\documentclass[11pt,a4paper]{article} \usepackage{epsfig}
\usepackage{latexsym}

\newcommand{\bi}{\begin{itemize}}
\newcommand{\ei}{\end{itemize}}
\newcommand{\be}{\begin{equation}}
\newcommand{\ee}{\end{equation}}

\newcommand{\bea}{\begin{eqnarray}}
\newcommand{\eea}{\end{eqnarray}}
\newcommand{\beastar}{\begin{eqnarray*}}
\newcommand{\eeastar}{\end{eqnarray*}}

\newcommand{\eq}[1]{(\ref{#1})}
\newcommand{\eqq}[2]{(\ref{#1},\ref{#2})}
\newcommand{\eqqq}[3]{(\ref{#1},\ref{#2},\ref{#3})}

\newcommand{\s}{\sigma}

\begin{document}

\title{Nonequilibrium fluctuations in small systems: From physics to biology}

\author{F Ritort\\
Department de Fisica Fonamental, Faculty of Physics, 
Universitat de Barcelona,\\
Diagonal 647, 08028 Barcelona, Spain; \\
E-Mail: {\tt ritort@ffn.ub.es}}

\date{December 2006}

\maketitle

\begin{abstract}
\noindent
In this paper I am presenting an overview on several topics related to
nonequilibrium fluctuations in small systems. I start with a
general discussion about fluctuation theorems and applications to
physical examples extracted from physics and biology: a bead in an
optical trap and single molecule force experiments. Next I present a
general discussion on path thermodynamics and consider distributions of
work/heat fluctuations as large deviation functions. Then I address the
topic of glassy dynamics from the perspective of nonequilibrium
fluctuations due to small cooperatively rearranging regions. Finally, I conclude with
a brief digression on future perspectives. 
\end{abstract}

\tableofcontents

\section{What are small systems?}
\label{intro}

Thermodynamics, a scientific discipline inherited from the 18th
century, is facing new challenges in the description of nonequilibrium
small (sometimes also called mesoscopic) systems. Thermodynamics is a
discipline built in order to explain and interpret energetic processes
occurring in macroscopic systems made out of a large number of
molecules on the order of the Avogadro number. Although
thermodynamics makes general statements beyond reversible processes
its full applicability is found in equilibrium systems where it can
make quantitative predictions just based on a few laws. The subsequent
development of statistical mechanics has provided a solid
probabilistic basis to thermodynamics and increased its predictive
power at the same time. The development of statistical mechanics goes
together with the establishment of the molecular hypothesis. Matter is
made out of interacting molecules in motion. Heat, energy and work are
measurable quantities that depend on the motion of molecules.  The
laws of thermodynamics operate at all scales.

Let us now consider the case of heat conduction
along polymer fibers. Thermodynamics applies at the microscopic
or molecular scale, where heat conduction takes place along molecules
linked along a single polymer fiber, up to the macroscopic scale where heat
is transmitted through all the fibers that make a piece
of rubber. The main difference between the two cases is the amount of
heat transmitted along the system per unit of time. In the first case the
amount of heat can be a few $k_BT$ per millisecond whereas in the
second can be on the order of $N_fk_BT$ where $N_f$ is the number of
polymer fibers in the piece of rubber. The relative amplitude of the
heat fluctuations are on the order of 1 in the molecular case and
$1/\sqrt{N_f}$ in the macroscopic case. Because $N_f$ is usually very large,
the relative magnitude of heat fluctuations is negligible for the
piece of rubber as compared to the single polymer fiber. We then
say that the single polymer fiber is a small system whereas the piece
of rubber is a macroscopic system made out of a very large collection of small
systems that are assembled together.

Small systems are those in which the energy exchanged with the
environment is a few times $k_BT$ and energy fluctuations are
observable. A few can be 10 or 1000 depending on the system. A small
system must not necessarily be of molecular size or contain a few
number of molecules. For example, a single polymer chain may behave as
a small system although it contains millions of covalently linked
monomer units. At the same time, a molecular system may not be small
if the transferred energy is measured over long times compared to the
characteristic heat diffusion time. In that case the average energy
exchanged with the environment during a time interval $t$ can be as
large as desired by choosing $t$ large enough. Conversely, a
macroscopic system operating at short time scales could deliver a tiny
amount of energy to the environment, small enough for fluctuations to
be observable and the system being effectively small.

Because macroscopic systems are collections of many molecules we expect
that the same laws that have been found to be applicable in macroscopic
systems are also valid in small systems containing a few number of
molecules \cite{Hill94,Gross01}. Yet, the phenomena that we will observe
in the two regimes will be different.  Fluctuations in large systems
are mostly determined by the conditions of the environment. Large
deviations from the average behavior are hardly observable and the
structural properties of the system cannot be inferred from the spectrum
of fluctuations.  In contrast, small systems will display large
deviations from their average behavior. These turn out to be
less sensitive to the conditions of the surrounding environment
(temperature, pressure, chemical potential) and carry information about
the structure of the system and its nonequilibrium behavior. We may then
say that information about the {\em structure} is carried in the tails
of the statistical distributions describing molecular properties.

The world surrounding us is mostly out of equilibrium, equilibrium
being just an idealization that requires of specific conditions to be
met in the laboratory. Even today we do not have a general theory about
nonequilibrium macroscopic systems as we have for equilibrium
ones. Onsager theory is probably the most successful attempt albeit its
domain of validity is restricted to the linear response regime. In small
systems the situation seems to be the opposite. Over the past years, a
set of theoretical results, that go under the name of fluctuation
theorems, have been unveiled . These theorems make specific predictions
about energy processes in small systems that can be scrutinized in the
laboratory.

The interest of the scientific community on small systems has been
boosted by the recent advent of micromanipulation techniques and
nanotechnologies. These provide adequate scientific instruments that can
measure tiny energies in physical systems under nonequilibrium
conditions.  Most of the excitement comes also from the more or less
recent observation that biological matter has successfully exploited the
smallness of biomolecular structures (such as complexes made out of nucleic acids and
proteins) and the fact that they are embedded in nonequilibrium environment to become
wonderfully complex and efficient at the same time
\cite{Ritort03,BusLipRit05}.

The goal of this review is to discuss these ideas from a physicsist
perspective by emphasizing the underlying common aspects in a broad
category of systems, from glasses to biomolecules. We aim to put
together some concepts in statistical mechanics that may become the
building blocks underlying a future theory of nonequilibrium small systems.  This is
not a review in the traditional sense but rather a survey of a few
selected topics in nonequilibrium statistical mechanics concerning
systems that range from physics to biology. The selection is biased by
my own particular taste and expertise. For this reason I have not tried
to cover most of the relevant references for each selected topic but
rather emphasize a few of them that make explicit connection with my
discourse. Interested readers are advised to look at other reviews that have
been recently written on related
subjects \cite{FreKro05,RegRubVil05,Qian05}

The outline of the review is as follows. Section \ref{small}
introduces two examples, one from physics and the other from biology,
that are paradigms of nonequilibrium behavior. Section~\ref{stoctherm:ft} is devoted to cover most
important aspects of fluctuation theorems whereas
Section~\ref{examples} presents applications of fluctuation theorems
to physics and biology. Section~\ref{path} presents the discipline of
path thermodynamics and briefly discusses large deviation
functions. Section~\ref{glassy} discusses the topic of glassy dynamics
from the perspective of nonequilibrium fluctuations in small
cooperatively rearranging regions. We conclude with a brief discussion
on future perspectives.

\section{Small systems in physics and biology}
\label{small}

\subsection{Colloidal systems}
\label{small:physics}
Condensed matter physics is full of examples where nonequilibrium
fluctuations of mesoscopic regions governs the nonequilibrium behavior
that is observed at the macroscopic level. A class of systems that have
attracted a lot of attention for many decades and that still remain
poorly understood are glassy systems, such as supercooled liquids and
soft materials \cite{CipRam05}. Glassy systems can be prepared in a
nonequilibrium state, e.g. by fast quenching the sample from high to low
temperatures, and subsequently following the time evolution of the
system as a function of time (also called the ``age'' of the system). Glassy systems display extremely
slow relaxation and aging behavior, i.e. an age dependent response to
the action of an external perturbation. Aging systems respond slower as
they {\em get older} keeping memory of their age for timescales that
range from picoseconds to years. The slow dynamics observed in glassy
systems is dominated by intermittent, large and rare fluctuations where
mesoscopic regions release some stress energy to the
environment. Current experimental evidence suggests that these events
correspond to structural rearrangements of clusters of molecules inside
the glass that release some energy through an activated and cooperative
process. These cooperatively rearranging regions are responsible of the
heterogeneous dynamics observed in glassy systems and lead to
a great disparity of relaxation times. The fact that, under appropriate conditions, 
the slow relaxation observed in glassy systems virtually takes forever, indicates that
the average amount of energy released in any rearrangement event must be small enough to
account for an overall net energy release of the whole sample (which is
not larger than the stress energy contained in the system in
the initial nonequilibrium state).

\begin{figure}
\begin{center}
\includegraphics[scale=0.8,angle=0]{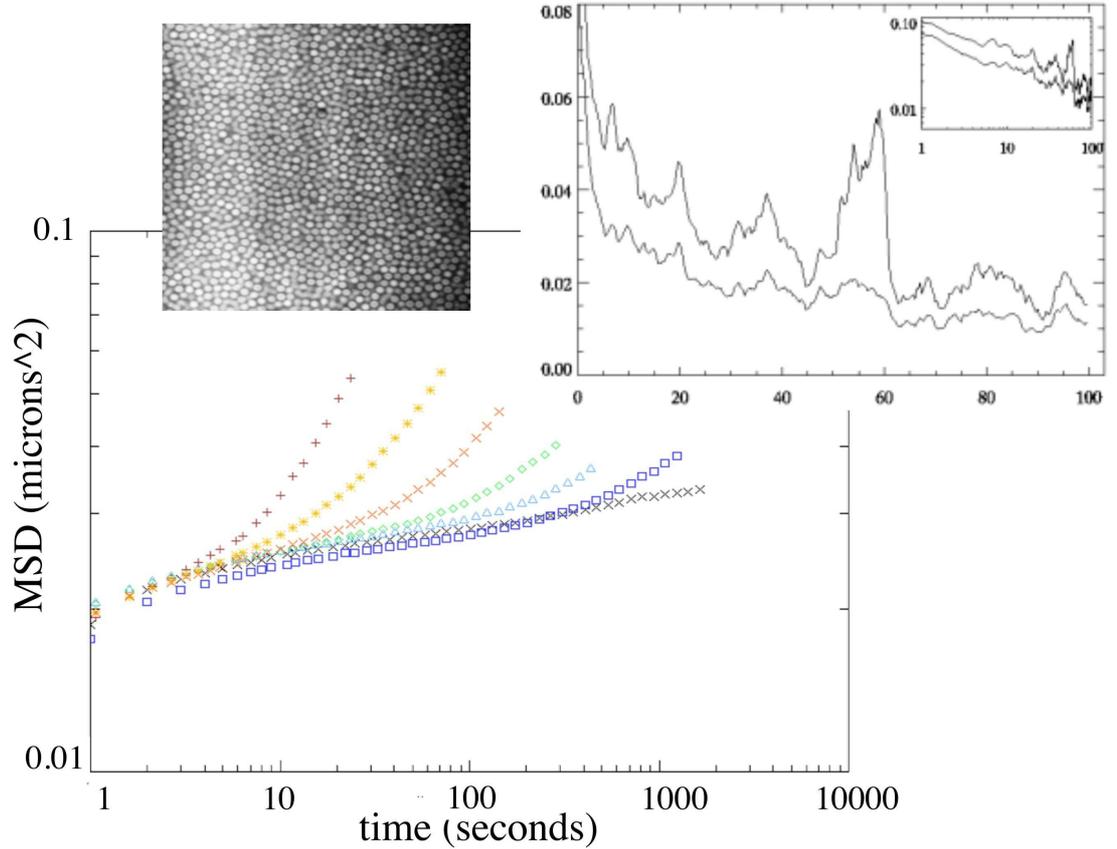}
\end{center}
\vspace{-0.4cm}
\caption{\em (Left inset) A snapshot picture of a colloidal system
obtained with confocal microscopy. (Left) Aging behavior observed in
the mean square displacement (MSD), $\langle\Delta x^2\rangle$, as a
function of time for different ages. The colloidal system reorganizes
slower as it becomes older. (Right) $\gamma=\sqrt{\langle\Delta
x^4\rangle/3}$ (upper curve) and $\langle\Delta x^2\rangle$ (lower
curve) as a function of the age measured over a fixed time window
$\Delta t=10 {\rm min}$. For a diffusive dynamics both curves should
coincide, however these measurements show deviations from diffusive
dynamics as well as intermittent behavior (Inset: the same as in main
panel but plotted in logarithmic timescale). Figures (A,B) taken from
http://www.physics.emory.edu/weeks
and Figure (C)
taken from \protect\cite{CouWee03}.}
\label{fig1}
\vspace{-0.2cm}
\end{figure}

In some systems, such as colloids, the free-volume (i.e. the volume of
the system that is available for motion to the colloidal particles) is
the relevant variable, and the volume fraction of colloidal particles
$\phi$ is the parameter governing the relaxation rate. Relaxation in
colloidal systems is determined by the release of tensional stress
energy and free volume in spatial regions that contain a few
particles. Colloidal systems offer great advantages to do experiments
for several reasons: 1) In colloids the control parameter is the volume
fraction, $\phi$, a quantity easy to control in experiments; 2) Under
appropriate solvent conditions colloidal particles behave as hard
spheres, a system that is pretty well known and has been theoretically
and numerically studied for many years; 3) The size of colloidal
particles is typically of a few microns making possible to follow the
motion of a small number particles using video microscopy and
spectroscopic techniques. This allows to detect cooperatively
rearranging clusters of particles and characterize their heterogeneous
dynamics. Experiments have been done with PMMA poly(methyl methacrylate)
particles of $\simeq 1\mu m$ radius suspended in organic solvents
\cite{PusMeg86,WeeCroLevSchWei00}. Confocal microscopy then allows to
acquire images of spatial regions of extension on the order of tens of
microns that contain a few thousand of particles, small enough to detect
the collective motion of clusters. In experiments carried by Weeks and
collaborators \cite{CouWee03} a highly stressed nonequilibrium state is
produced by mechanically stirring a colloidal system at volume fractions
$\phi\sim \phi_g$ where $\phi_g$ is the value of the volume fraction at
the glass transition where colloidal motion arrests. The subsequent
motion is then observed. A few experimental results are shown in Figure
\ref{fig1}. The mean square displacement of the particles inside the
confocal region show aging behavior. Importantly also, the region
observed is small enough to observe temporal heterogeneity, i.e. the
aging behavior is not smooth with the age of the system as usually
observed in light scattering experiments. Finally, the mean square
displacement for a single trajectory shows abrupt events characteristic
of collective motions involving a few tens of particles. By analyzing
the average number of particles belonging to a single cluster Weeks et
al.  \cite{CouWee03} find that no more than 40 particles participate in
the rearrangment of a single cluster suggesting that cooperatively
rearranging regions are not larger than a few particle radii in
extension. Large deviations, intermittent events and heterogeneous
kinetics are the main features observed in these experiments.

\subsection{Molecular machines}
\label{small:biology}
Biochemistry and molecular biology are scientific disciplines aiming to
describe the structure, organization and function of living matter
\cite{Alberts98,Watson04}. Both disciplines seek an understanding of
life processes in molecular terms. Their main object of study are
biological molecules and the function they play in the biological
process where they intervene. Biomolecules are small systems from
several points of view. First, from their size, where they span just a few
nanometers of extension. Second, from the energies they require to function
properly, which is determined by the amount of energy that can be
extracted by hydrolyzing one molecule of ATP (approximately 12
$k_BT$ at room temperature or $300K$). Third, from the typically short
amount of time that it takes to complete an intermediate step
in a biological reaction. Inside the cell many reactions that would take
an enormous long time under non-biological conditions are speeded up by
several orders of magnitude in the presence of specific enzymes.

\begin{figure}
\begin{center}
\includegraphics[scale=0.7,angle=0]{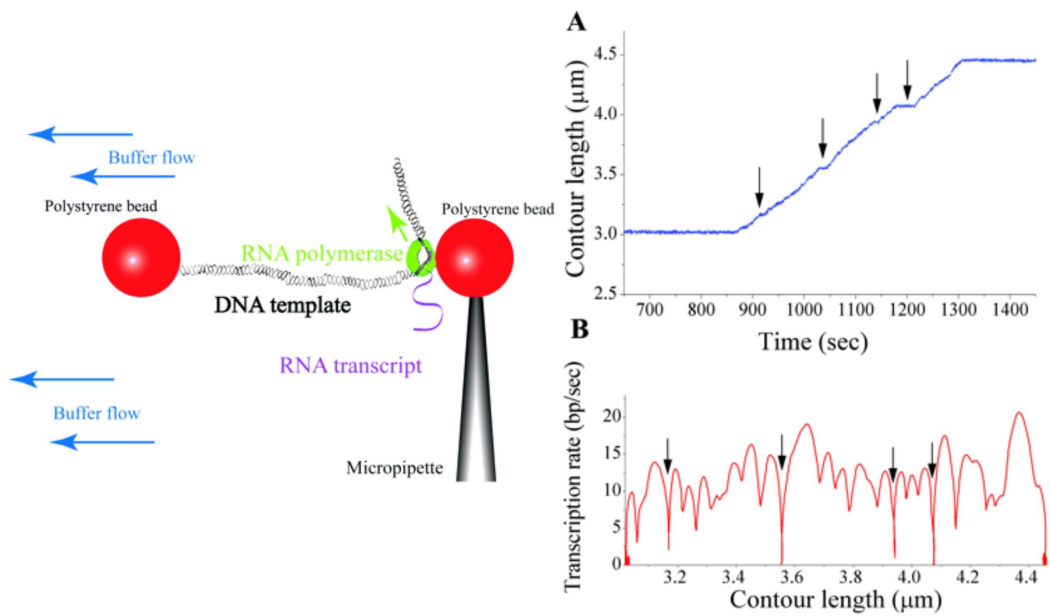}
\end{center}
\vspace{-0.4cm}
\caption{\em (Left) Experimental setup in force-flow measurements. Optical
tweezers are used to trap beads but forces are applied on the RNApol-DNA
molecular complex using the Stokes drag force acting on the left bead
immersed in the flow. In this setup force assists RNA transcription as
the DNA tether between beads increases in length as a function of
time. (A) The contour length of the DNA tether as a function of time
(blue curve) and (B) the transcription rate (red curve) as a function of
the contour length. Pauses (temporary arrests of transcription) are
shown as vertical arrows. Figure taken from
\protect\cite{ForIzhWooWuiBus02}.}
\label{fig2}
\vspace{-0.2cm}
\end{figure}

Molecular machines (also called molecular motors) are amazing complexes
made out of several parts or domains that coordinate their behavior to
perform specific biological functions by operating out of
equilibrium. Molecular machines hydrolyze energy carrier molecules such
as ATP to transform the chemical energy contained in the high energy
bonds into mechanical motion
\cite{EisHil85,CorErmOst92,JulAdjPro98,FisKol99}. An example of a
molecular machine that has been studied by molecular biologists and
biophysicists is the RNA polymerase \cite{GelLan98,Klug01}. This is an
enzyme that synthesizes the pre-messenger RNA molecule by translocating
along the DNA and reading, step by step, the sequence of bases along the
DNA backbone. The read out of the RNA polymerase is exported from the
nucleus to the cytoplasm of the cell to later be translated in the
ribosome, a huge molecular machine that synthesizes the protein coded
into the messenger RNA \cite{OrpRei02}. Using single molecule
experiments it is possible to grab one DNA molecule by both ends using
optical tweezers and follow the translocation motion of the RNA
polymerase \cite{WanSchYinLanGelBlo98,DavWuiLanBus00}. Current optical
tweezer techniques have even resolved the motion of the enzyme at the
level of a single base pair \cite{ShaAbbLanBlo03,AbbGreShaLanBlo05}. The
experiment requires flowing inside the fluidics chamber the enzymes and
proteins that are necessary to initiate the transcription reaction. The
subsequent motion and transcription by the RNA polymerase is called
elongation and can be studied under applied force conditions that assist
or oppose the motion of the enzyme \cite{ForIzhWooWuiBus02}. In Figure
\ref{fig2} we show the results obtained in the Bustamante group for the
RNA polymerase of {\em Escherichia Coli}, a bacteria found in the
intestinal tracts of animals.  In Figure \ref{fig2}A the polymerase
apparently moves at a constant average speed but is characterized by
pauses (black arrows) where motion temporarily arrests. In Figure
\ref{fig2}B we show the transcription rate (or speed of the enzyme) as a
function of time.  Note the large intermittent fluctuations in the
transcription rate, a typical feature of small systems embedded in a
noisy thermal environment. In contrast to the slow aging dynamics
observed in colloidal systems (Sec.~\ref{small:physics}) the kinetic
motion of the polymerase is steady and fast (translocation rates are
typically on the order of few tens of base pairs per second). We then
say that the polymerase is in a nonequilibrium steady state. As in the
previous case, large deviations, intermittent events (e.g. temporary
pauses in the translocation motion) and complex kinetics are the main
features observed in these experiments.

\section{Fluctuation theorems (FTs)}
\label{stoctherm:ft}
Fluctuation theorems (FTs) make statements about energy exchanges that
take place between a system and its surroundings under general
nonequilibrium conditions. Since their discovery in the mid 90's
\cite{EvaCohMor93,EvaSea94,GalCoh95} there has been a growing interest
to elucidate their importance and implications. FTs provide a fresh 
look to our understanding of old questions such as the origin of
irreversibility and the second law in statistical mechanics
\cite{Cohen02,Gallavotti03}. In addition, FTs provide statements about
energy fluctuations in small systems which, under generic conditions,
should be experimentally observable. FTs have been discussed in the
context of deterministic, stochastic and thermostatted systems. Although
the results obtained differ depending on the particular model of the
dynamics that is used, in the nutshell they look pretty equivalent.

FTs are related to the so called nonequilibrium work relations
introduced by Jarzynski \cite{Jarzynski97}. This fundamental relation
can be seen as consequence of the FTs \cite{ReiSevEva05,Evans03}. It represents
a new result beyond classical thermodynamics that shows the possibility to
recover free energy differences using irreversible processes. Several
reviews have been written on the subject
\cite{EvaSea02,Maes03,Jarzynski02,Kurchan05,Ritort03} with specific
emphasis on theory and/or experiments. In the next sections we review
some of the main results. Throughout the text we will take $k_B=1$.

\subsection{Nonequilibrium states}
\label{stoctherm:noneq}
An important concept in thermodynamics is the state variable. State
variables are those that, once determined, uniquely specify the
thermodynamic state of the system. Examples are the temperature, the
pressure, the volume and the mass of the different components in a given
system. To specify the state variables of a system it is common to put
the system in contact with a bath. The bath is any set of sources (of
energy, volume, mass, etc.) large enough to remain unaffected by the
interaction with the system under study. The bath ensures that a system
can reach a given temperature, pressure, volume and mass concentrations
of the different components when put in thermal contact with the bath
(i.e. with all the relevant sources).  Equilibrium states are then
generated by putting the system in contact with a bath and waiting until
the system properties relax to the equilibrium values. Under
such conditions the system properties do not change with time and the
average net heat/work/mass exchanged between the system and the bath is
zero.

Nonequilibrium states can be produced under a great variety of conditions,
either by continuously changing the parameters of the bath or by preparing
the system in an initial nonequilibrium state that slowly relaxes
toward equilibrium. In general a nonequilibrium state
is produced whenever the system properties change with time and/or the
net heat/work/mass exchanged by the system and the bath is non zero. We can
distinguish at least three different types of nonequilibrium states:

\begin{itemize}

\item{\bf Nonequilibrium transient state (NETS).} The system is initially
prepared in an equilibrium state and later driven out of equilibrium by
switching on an external perturbation. The system quickly returns to a new
equilibrium state once the external
perturbation stops changing.

\item{\bf Nonequilibrium steady-state (NESS).} The system is driven by
external forces (either time dependent or non-conservative) in a
stationary nonequilibrium state where its properties do not change with
time. The steady state is an irreversible nonequilibrium process, that
cannot be described by the Boltzmann-Gibbs distribution, where the
average heat that is dissipated by the system (equal to the entropy production
of the bath) is positive.

There are still other categories of NESS. For example
in nonequilibrium transient steady states the system starts in
a nonequilibrium steady state but is driven out of equilibrium by an external perturbation
to finally settle in a new nonequilibrium steady state.

\item{\bf Nonequilibrium aging state (NEAS).} The system is initially
prepared in a nonequilibrium state and put in contact with the
sources. The system is then let evolve alone but fails to reach thermal
equilibrium in observable or laboratory time scales. In this case the
system is in a non-stationary slowly relaxing nonequilibrium state
called {\em aging state} and characterized by a very small heat
dissipation to the sources. In the aging state two-times correlations
decay slower as the system becomes older. Two-time correlation functions
depend on both times and not just on the difference of times (time
translational invariance is said to be broken).

\end{itemize}

There are many examples of nonequilibrium states. A classic example of a
NESS is an electrical circuit made out of a battery and a
resistance. The current flows through the resistance and the chemical
energy stored in the battery is dissipated to the environment in the
form of heat; the average dissipated power, ${\cal P}_{\rm dis}=VI$, is
identical to the power supplied by the battery.  Another example is a
sheared fluid between two plates or coverslips where one of them is moved
relative to the other at a constant velocity $v$. To sustain such state
a mechanical power that is equal to ${\cal P}\propto \eta v^2$ has to be exerted
upon the moving plate where $\eta$ is the shear viscosity of the fluid. The
mechanical work produced is then dissipated in the form of heat through
the viscous friction between contiguous fluid layers.  Another examples
of NESS are chemical reactions in metabolic pathways that are sustained
by activated carrier molecules such as ATP. In such case, hydrolysis of
ATP is strongly coupled to specific oxidative reactions. For example,
ionic channels use ATP hydrolysis to transport protons against the
electromotive force.

A classic example of NETS is the case of a protein in its initial native
state that is mechanically pulled (e.g. using AFM) by exerting force at
the ends of the molecule. The protein is initially folded and in thermal
equilibrium with the surrounding aqueous solvent. By mechanically
stretching the protein is pulled away from equilibrium into a transient
state until it finally settles into the unfolded and extended new
equilibrium state. Another example of a NETS is a bead immersed in water
and trapped in an optical well generated by a focused laser beam. When
the trap is moved to a new position (e.g. by redirecting the laser beams) the
bead is driven into a NETS. After some time the bead reaches again
equilibrium at the new position of the trap. In another experiment the trap is suddenly put in motion at
a speed $v$ so the bead is transiently driven away from its equilibrium
average position until it settles into a NESS characterized by the speed
of the trap. This results in the average position of the bead lagging behind 
the position of the center of the trap.

The classic example of a NEAS is a supercooled liquid cooled below its
glass transition temperature. The liquid solidifies into an amorphous slowly
relaxing state characterized by huge relaxational times and anomalous low
frequency response. Other systems are colloids that can be prepared in a
NEAS by the sudden reduction of the volume fraction of the
colloidal particles or by putting the system under a strain/stress. 

\begin{figure}
\begin{center}
\includegraphics[scale=0.3,angle=-90]{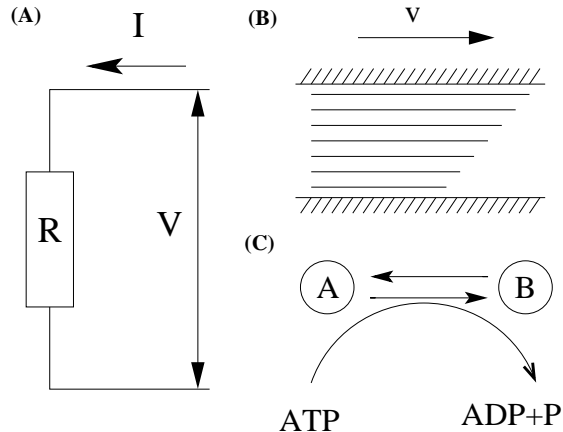}
\end{center}
\vspace{-0.cm}
\caption{\em Examples of NESS. (A) An electric current $I$ flowing through
a resistance $R$ and maintained by a voltage source or control parameter
$V$. (B) A fluid
sheared between two plates moving relative to each
other at speed $v$ (the control
parameter) . (C) A chemical reaction $A\rightarrow B$ coupled to ATP
hydrolysis. The control parameter here is relative concentration of ATP and
ADP.}
\label{fig3}
\vspace{-0.2cm}
\end{figure}

The classes of nonequilibrium states previously described do not make
distinctions whether the system is macroscopic or small. In small
systems, however, it is common to speak about the control parameter to
emphasize the importance of the constraints imposed by the bath that are
externally controlled and do not fluctuate. The control parameter
($\lambda$) represents a value (in general, a set of values) that
defines the state of the bath. Its value determines the equilibrium
properties of the system, e.g. the equation of state. In macroscopic
systems it is unnecessary to discern which value is externally
controlled because fluctuations are small and all equilibrium ensembles
give the same equivalent thermodynamic description, i.e. the same
equation of state. Differences arise only when including fluctuations in
the description. The nonequilibrium behavior of small systems is
then strongly dependent on the protocol used to drive them out of
equilibrium. The protocol is generally defined by the time evolution of
the control parameter $\lambda(t)$. As a consequence, the
characterization of the protocol $\lambda(t)$ is an essential step to
unambiguously define how the nonequilibrium state has been generated. Figure~\ref{fig3} shows a
representation of a few examples of NESS and control parameters.

\subsection{Fluctuation theorems (FTs) in stochastic dynamics}
In this section we are presenting a derivation of the FT based on stochastic
dynamics. In contrast to deterministic systems, stochastic dynamics
is naturally endowed of crucial assumptions that are needed for the derivation such as
the ergodicity hypothesis. The derivation we are presenting here follows the
approach introduced by Crooks-Kurchan-Lebowitz-Spohn \cite{Kurchan98,LebSpo99}
and includes also some results recently obtained by Seifert using
Langevin systems \cite{Seifert05a}.

\subsubsection{The master equation}

Let us consider a stochastic system described by a generic
variable ${\cal C}$. This variable may stand for the
position of a bead in an optical trap, the velocity field of a
fluid, the current passing through a resistance, the number of native
contacts in a protein, etc.  A trajectory or path $\Gamma$ in
configurational space is described by a discrete sequence of
configurations in phase space,
\be
\Gamma\equiv \lbrace {\cal C}_0,{\cal C}_1,{\cal C}_2,...,{\cal
C}_M\rbrace
\label{demo0}
\ee
where the system occupies configuration ${\cal C}_k$ at
time $t_k=k\Delta t$ and $\Delta t$ is the duration of the
discretized elementary time-step. In what follows we consider paths that start at
${\cal C}_0$ at time $t=0$ and end at the configuration ${\cal C}_M$ at
time $t=M\Delta t$. The continuous time limit is recovered by taking
$M\to\infty,\Delta t\to 0$ for a fixed value of $t$.

Let $\langle(...)\rangle$ denote the average over all paths that
start at $t=0$ at configurations ${\cal C}_0$ initially chosen from a
distribution $P_0({\cal C})$.  We also define $P_k({\cal C})$ as the
probability, measured over all possible dynamical paths, that the system is
in configuration ${\cal C}$ at time $t_k=k\Delta t$. Probabilities are
normalized for any $k$,
\be
\sum_{{\cal C}}P_k({\cal C})=1~~~~~.
\label{demo1}
\ee

The system is assumed to be in contact with a thermal bath at
temperature $T$. We also assume that the microscopic dynamics of the
system is
of the Markovian type: the probability for the system to be at a given
configuration and at a given time only depends on its previous
configuration. We then introduce the transition probability ${\cal W}_k({\cal
C}\to {\cal C'})$. This denotes the probability for the system to change
from ${\cal C}$ to ${\cal C'}$ at time-step $k$. According to Bayes
formula,
\be
P_{k+1}({\cal C})=\sum_{{\cal C'}}{\cal W}_k({\cal C'}\to {\cal C})P_k({\cal C'})
\label{demo2}
\ee
where the ${\cal W}'s$ satisfy the normalization condition,
\be
\sum_{{\cal C'}}{\cal W}_k({\cal C}\to {\cal C'})=1~~~~~.
\label{demo3}
\ee
Using \eq{demo1},\eq{demo2} we can write the following Master equation
for the probability $P_k({\cal C})$,
\be \Delta P_k({\cal C})=P_{k+1}({\cal C})-P_k({\cal C})=\sum_{{\cal
C'}\ne {\cal C}}{\cal W}_k({\cal C'}\to {\cal C})P_k({\cal C'})-\sum_{{\cal
C'}\ne {\cal C}}{\cal W}_k({\cal C}\to {\cal C'})P_k({\cal C})
\label{demo4}
\ee
where the terms ${\cal C}={\cal C'}$ have not been included 
as they cancel out in the first and second sums in the r.h.s.   
The first term in the r.h.s accounts for all transitions leading
to the configuration ${\cal C}$ whereas the second term counts all processes
leaving ${\cal C}$. 
It is convenient to introduce the rates $r_t({\cal C}\to {\cal C'})$ in
the continuous time limit $\Delta t\to 0$,
\be
r_t({\cal C}\to {\cal C'})=\lim_{\Delta t\to 0}\frac{{\cal W}_k({\cal
C}\to {\cal C'})}{\Delta t}~~~~;~~~\forall {\cal C}\ne {\cal C'}
\label{demo4b}
\ee
Equation \eq{demo4} becomes,
\be
\frac{\partial P_t({\cal C})}{\partial t}=\sum_{{\cal
C'}\ne {\cal C}}r_t({\cal C'}\to {\cal C})P_t({\cal C'})-\sum_{{\cal
C'}\ne {\cal C}}r_r({\cal C}\to {\cal C'})P_t({\cal C})~~~.
\label{demo4c}
\ee

\subsubsection{Microscopic reversibility}
\label{stochterm:mr}
We now introduce the concept of the control parameter $\lambda$ (see
Sec.~\ref{stoctherm:noneq}). In the present scheme the discrete-time
sequence $\lbrace\lambda_k;0\le k\le M\rbrace$ defines the perturbation
protocol. The transition probability ${\cal W}_k({\cal C}\to {\cal C'})$
now depends explicitly on time through the value of an external
time-dependent parameter $\lambda_k$. The parameter $\lambda_k$ may
indicate any sort of externally controlled variable that determines the
state of the system, for instance the value of the external magnetic
field applied on a magnetic system, the value of the mechanical force
applied to the ends of a molecule, the position of a piston containing a
gas or the concentrations of ATP and ADP in a molecular reaction coupled
to hydrolisis (see
Figure \ref{fig3}). The time variation of the control parameter,
$\dot{\lambda}=(\lambda_{k+1}-\lambda_k)/\Delta t$, is used as tunable
parameter which determines how strongly irreversible is the nonequilibrium process.
In order to emphasize the importance of the control parameter, in what
follows we will parametrize probabilities and transition probabilities
by the value of the control parameter at time-step $k$, $\lambda$ (rather
than by the time $t$). Therefore we will write $P_{\lambda}({\cal
C}),{\cal W}_{\lambda}({\cal C}\to {\cal C'})$ for the probabilities and
transition probabilities respectively at a given time $t$.

The transition probabilities ${\cal W}_{\lambda}({\cal C}\to {\cal C'})$
cannot be arbitrary but must guarantee that the equilibrium state
$P^{\rm eq}_{\lambda}(C)$ is a stationary solution of the master
equation \eq{demo4}. The simplest way to impose such condition is to
model the microscopic dynamics as ergodic and reversible for a fixed
value of ${\lambda}$,
\be \frac{{\cal W}_{\lambda}({\cal C}\to {\cal C'})}{{\cal W}_{\lambda}({\cal C'}\to {\cal C})}=\frac{P^{\rm
eq}_{\lambda}({\cal C'})}{P^{\rm eq}_{\lambda}({\cal C})}~~~.
\label{demo5}
\ee
The latter condition is commonly known as microscopic reversibility or
local detailed balance. This property is equivalent to time reversal
invariance in deterministic (e.g. thermostatted) dynamics. Microscopic
reversibility, a common assumption in nonequilibrium statistical
mechanics, is the crucial ingredient in the present derivation.

Equation \eq{demo5} has been criticized as a relation that is valid only
in the vicinity of equilibrium because the rates appearing in \eq{demo5} are
related to the equilibrium distribution $P^{\rm
eq}_{\lambda}(C)$. However, we must observe that the equilibrium
distribution evaluated at a given configuration depends only on the
Hamiltonian of the system at that configuration. Therefore, \eq{demo5}
must be read as a relation that only depends on the energy of
configurations. It should hold close but also far from equilibrium by
properly defining the configurational space.

Let us now consider all possible dynamical paths $\Gamma$ that
are generated starting from an ensemble of initial configurations at
time 0 (described by the initial distribution $P_{\lambda_0}({\cal C})$)
and which evolve according to \eq{demo5} until time $t$ ($t=M\Delta t$,
$M$ being the total number of discrete time-steps). Dynamical evolution takes
place according to a given protocol, $\lbrace \lambda_k,0\le k\le
M\rbrace$, the protocol defining the nonequilibrium
experiment. Different dynamical paths will be generated because
of the different initial conditions (all weighted with the probability
$P_{\lambda_0}({\cal C})$) and because of the stochastic nature of the
transitions between configurations at consecutive time-steps.

\subsubsection{The nonequilibrium equality}
Let us consider a generic observable ${\cal A}(\Gamma)$. The average
value of $A$ is given by,
\be
\langle{\cal A}\rangle=\sum_{\Gamma}P(\Gamma){\cal A}(\Gamma)
\label{ft1}
\ee
where $\Gamma$ denotes the path and $P(\Gamma)$ indicates the probability of that
path. Using the fact that the dynamics is Markovian together with the definition
\eq{demo0} we can write,
\be
P(\Gamma)=P_{\lambda_0}({\cal C}_0)\prod_{k=0}^{M-1}{\cal W}_{\lambda_{k}}({\cal
C}_k\to{\cal C}_{k+1})~~~~.
\label{demo8}
\ee
By inserting \eq{demo8} into \eq{ft1} we obtain,
\be 
\langle{\cal A}\rangle=\sum_{\Gamma}{\cal A}(\Gamma) 
P_{\lambda_0}({\cal C}_0)\prod_{k=0}^{M-1}{\cal W}_{\lambda_{k}}({\cal C}_k \to {\cal C}_{k+1})
~~~.
\label{demo9}
\ee
Using the detailed balance condition \eq{demo5} this expression
reduces to,
\bea
\langle{\cal A}\rangle=\sum_{\Gamma}
P_{\lambda_0}({\cal C}_0){\cal A}(\Gamma)\prod_{k=0}^{M-1}\Bigl[{\cal W}_{\lambda_{k}}({\cal C}_{k+1}\to {\cal C}_{k})
\frac{P^{\rm eq}_{\lambda_k}({\cal C}_{k+1})}{P^{\rm
eq}_{\lambda_{k}}({\cal C}_k)}\Bigr]\\
=\sum_{\Gamma} {\cal A}(\Gamma) P_{\lambda_0}({\cal C}_0)\exp\Bigl[\sum_{k=0}^{M-1}\log\Bigl(\frac{P^{\rm
eq}_{\lambda_k}({\cal C}_{k+1})}{P^{\rm eq}_{\lambda_{k}}({\cal
C}_k)}\Bigr)\Bigr]
\prod_{k=0}^{M-1}{\cal W}_{\lambda_{k}}({\cal C}_{k+1}\to{\cal C}_{k})~~~~.
\label{demo10}
\eea

This equation can not be worked out further. However, let us consider
the following observable ${\cal S}(\Gamma)$, defined by,
\be {\cal A}(\Gamma)=\exp(-{\cal S}(\Gamma))=\frac{b({\cal
C}_M)}{P_{\lambda_0}({\cal C}_0)}\prod_{k=0}^{M-1}\Bigl( \frac{P^{\rm
eq}_{\lambda_k}({\cal C}_k)}{P^{\rm eq}_{\lambda_k}({\cal
C}_{k+1})}\Bigr)
\label{ft2}
\ee
where $b({\cal C})$ is any positive definite and normalizable
function, 
\be
\sum_{{\cal C}} b({\cal C})=1~~~
\label{ft3}
\ee
and $P_{\lambda_0}({\cal C}_0)>0, \forall{\cal C}_0$. By inserting \eq{ft2} into \eq{demo10} we get,
\be
\langle \exp(-{\cal S})\rangle=\sum_{\Gamma}b({\cal
C}_M)\prod_{k=0}^{M-1}{\cal W}_{\lambda_{k}}({\cal C}_{k+1}\to{\cal C}_{k})=1
\label{ft4}
\ee
where we have applied a telescopic sum (we first summed over ${\cal C}_M$, used \eq{ft3} and
subsequently summed over the rest of variables
by using \eq{demo3}). We call ${\cal S}(\Gamma)$ the {\bf total dissipation}
of the system. It is given by,
\be
{\cal S}(\Gamma)=\sum_{k=0}^{M-1}\log\Bigl[ \frac{P^{\rm
eq}_{\lambda_k}({\cal C}_{k+1})}{P^{\rm
eq}_{\lambda_k}({\cal C}_k)}\Bigr] +\log(P_{\lambda_0}({\cal C}_0))-\log(b({\cal C}_M))~~~.
\label{ft5}
\ee
The equality \eq{ft4} immediately implies, by using Jensen's inequality,
the following inequality,
\be
\langle {\cal S}\rangle\ge 0
\label{ft6}
\ee
which is reminiscent of second law of thermodynamics for nonequilibrium
systems: the entropy of the universe (system plus the environment)
always increases.  Yet, we have to identify the different terms
appearing in \eq{ft5}. It is important to stress that entropy production
in nonequilibrium systems can be defined just in terms of the
work/heat/mass transferred by the system to the external sources which
represent the bath. The definition of the total dissipation \eq{ft5} is
arbitrary because it depends on an undetermined function $b({\cal C})$
\eq{ft3}. Therefore, the total dissipation ${\cal S}$ may not
necessarily have a general physical meaning and could be interpreted in
different ways depending on the specific nonequilibrium context.

Equation \eq{ft4} has appeared in the past in the literature
\cite{BocKuz81} and is mathematically related to the
Jarzynski equality \cite{Jarzynski97}. We analyze this connection below
in Sec.~\ref{applications:NETS}.

\subsubsection{The fluctuation theorem}

A physical insight on the meaning of the total dissipation ${\cal S}$
can be obtained by deriving the fluctuation theorem. We start by
defining the reverse path $\Gamma^*$ of a given path $\Gamma$. Let us
consider the path, $\Gamma\equiv {\cal C}_0 \rightarrow {\cal C}_1
\rightarrow ...  \rightarrow {\cal C}_M$ corresponding to the forward (F) protocol
which is described by the sequence of values of $\lambda$ at different time-steps $k$,
$\lambda_k$. Every transition occurring at time-step $k$ , ${\cal C}_k
\rightarrow {\cal C}_{k+1}$, is governed by the transition probability
${\cal W}_{\lambda_k}({\cal C}_k\to{\cal C}_{k+1})$. The reverse path of
$\Gamma$ is defined as the time reverse sequence of configurations,
$\Gamma^*\equiv {\cal C}_M \rightarrow {\cal C}_{M-1} \rightarrow ...
\rightarrow {\cal C}_0$ corresponding to the reverse (R) protocol
described by the time-reversed sequence of values of $\lambda$,
$\lambda^R_k=\lambda_{M-k-1}$.

The probability of a given path and its reverse are given by,
\bea
{\cal P}_F(\Gamma)=\prod_{k=0}^{M-1} {\cal W}_{\lambda_k}({\cal C}_k\to{\cal
C}_{k+1})\label{ft7a}\\
{\cal P}_R(\Gamma^*)=\prod_{k=0}^{M-1} {\cal W}_{\lambda^R_k}({\cal C}_{M-k}\to{\cal
C}_{M-k-1})=\prod_{k=0}^{M-1} {\cal W}_{\lambda_k}({\cal C}_{k+1}\to{\cal
C}_k)\label{ft7b}
\eea
where in the last line we have shifted variables $k\to M-1-k$. We use the
notation ${\cal P}$ for the path probabilities rather than the usual
letter $P$. This difference in notation is introduced to stress the fact
that path probabilities \eqq{ft7a}{ft7b} are non-normalized conditional
probabilities, i.e. $\sum_{\Gamma} {\cal P}_{F(R)}(\Gamma)\ne 1$.  By
using \eq{demo5} we get,
\be
\frac{{\cal P}_F(\Gamma)}{{\cal P}_R(\Gamma^*)}=\prod_{k=0}^{M-1}\frac{P^{\rm
eq}_{\lambda_k}({\cal C}_{k+1})}{P^{\rm
eq}_{\lambda_k}({\cal C}_k)}=\exp(S_p(\Gamma))
\label{ft8}
\ee
where we defined the {\bf entropy production} of the system,
\be
S_p(\Gamma)=\sum_{k=0}^{M-1}\log\Bigl (\frac{P^{\rm
eq}_{\lambda_k}({\cal C}_{k+1})}{P^{\rm
eq}_{\lambda_k}({\cal C}_k)}\Bigr)~~~~.
\label{ft9}
\ee

Note that $S_p(\Gamma)$ is just a part of the total dissipation
introduced in \eq{ft5},
\be
{\cal S}(\Gamma)=S_p(\Gamma)+B(\Gamma)
\label{ft10}
\ee
where $B(\Gamma)$ is the so-called {\bf boundary term},
\be
B(\Gamma)=\log\bigl( P_{\lambda_0}({\cal C}_0)\bigr) -\log\bigl(b({\cal C}_M)\bigr)~~~.
\label{ft11}
\ee
We tend to identify $S_p(\Gamma)$ as the entropy production in a
nonequilibrium system whereas $B(\Gamma)$ is a term that contributes
just at the beginning and end of the nonequilibrium process. Note that
the entropy production $S_p(\Gamma)$ is antisymmetric under time
reversal, $S_p(\Gamma^*)=-S_p(\Gamma)$, expressing the fact that the
entropy production is a quantity related to irreversible
motion. According to \eq{ft8} paths that produce a given amount of
entropy are much more probable than those that consume the same amount
of entropy. How much improbable is entropy consumption depends
exponentially on the amount of entropy consumed. The larger the system
is, the larger the probability to produce (rather than consume) a given amount of
entropy $S_p$.

Equation \eq{ft8} has already the form of a fluctuation
theorem. However, in order to get a proper fluctuation theorem we still need to
specify relations between probabilities for physically measurable
observables rather than paths. From \eq{ft8} it is straightforward to
derive a fluctuation theorem for the total dissipation ${\cal S}$. Let us take
$b({\cal C})=P_{\lambda_M}({\cal C})$. With this choice we get,
\be
{\cal S}(\Gamma)=S_p(\Gamma)+B(\Gamma)=\sum_{k=0}^{M-1}\log\Bigl( \frac{P^{\rm
eq}_{\lambda_k}({\cal C}_{k+1})}{P^{\rm
eq}_{\lambda_k}({\cal C}_k)}\Bigr) +\log(P_{\lambda_0}({\cal C}_0))-\log(P_{\lambda_M}({\cal C}_M))~~.
\label{ft12}
\ee
The physical motivation behind this choice is that ${\cal S}$ now becomes an
antisymmetric observable under time reversal. Albeit $S_p(\Gamma)$ is always antisymmetric,
the choice \eq{ft12} is the only one that guarantees that the total
dissipation ${\cal S}$ changes sign upon reversal of the path, ${\cal
S}(\Gamma^*)=-{\cal S}(\Gamma)$. The symmetry property of observables
under time reversal and the possibility to consider boundary terms where
${\cal S}$ is symmetric (rather than antisymmetric) under time reversal
has been discussed in \cite{MaeWie06}.

The probability to produce a total dissipation ${\cal S}$ along the forward
protocol is given by,
\bea
P_F({\cal S})=\sum_{\Gamma}P_{\lambda_0}({\cal C}_0){\cal P}_F(\Gamma)\delta({\cal
S}(\Gamma)-{\cal S})=\sum_{\Gamma}P_{\lambda_0}({\cal C}_0){\cal P}_R(\Gamma^*)\exp(S_p(\Gamma))\delta({\cal
S}(\Gamma)-{\cal S})=\nonumber\\
\sum_{\Gamma}P_{\lambda_M}({\cal C}_M){\cal P}_R(\Gamma^*)\exp({\cal S}(\Gamma))\delta({\cal
S}(\Gamma)-{\cal S})=\nonumber\\
\exp({\cal S})\sum_{\Gamma^*}P_{\lambda_M}({\cal C}_M){\cal P}_R(\Gamma^*)\delta({\cal
S}(\Gamma^*)+{\cal S})=\exp({\cal S})P_R(-{\cal S})~~~.
\eea
In the first line of the derivation we used \eq{ft8}, in the second we
used \eq{ft12} and in the third we took into account the antisymmetric
property of ${\cal S}(\Gamma)$ and the unicity of the assignment
$\Gamma\to\Gamma^*$. This result is known under the generic name of
{\bf fluctuation theorem},
\be
\frac{P_F({\cal S})}{P_R(-{\cal S})}=\exp({\cal S})~~~~.
\label{ft13}
\ee
It is interesting to observe that this relation is not satisfied by the
entropy production because the inclusion of boundary term \eq{ft11} in
the total dissipation is required to respect the fluctuation symmetry. 
In what follows we discuss some of its consequences in some specific
situations. 

\begin{itemize}

\item{\bf Jarzynski equality.} The nonequilibrium equality \eq{ft4} is just a consequence
of \eq{ft13} that is obtained by rewriting it as $P_R(-{\cal
S})=P_F({\cal S})\exp(-{\cal S})$ and integrating both sides of the
equation from  ${\cal S}=-\infty$ to  ${\cal S}=\infty$.

\item{\bf Linear response regime.} Equation \eq{ft13} is trivially satisfied for ${\cal S}=0$ if
$P_F(0)=P_R(0)$. The process where $P_{F(R)}({\cal S})=\delta({\cal S})$
is called quasistatic or reversible. When ${\cal S}$ is different from
zero but small (${\cal S}< 1$) we can expand \eq{ft13} around ${\cal S}=0$ to obtain,
\bea
{\cal S}P_F({\cal S})={\cal S}\exp({\cal S})P_R(-{\cal S})\nonumber\\
\langle {\cal S}\rangle_F=\langle (-{\cal S}+{\cal S}^2)\rangle_R+{\cal
O}({\cal S}^3)\nonumber\\
\langle ({\cal S}^2)\rangle_{F(R)}=2\langle {\cal S}\rangle_{F(R)}
\label{ft14}
\eea
where we used $\langle {\cal S}\rangle_F=\langle {\cal S}\rangle_R$,
valid up to second order in ${\cal S}$. Note the presence of the subindex $F(R)$ for the
expectation values in the last line of \eq{ft14} which emphasizes the
equality of these averages along the forward and reverse process. The
relation \eq{ft14} is a version of the fluctuation-dissipation theorem
(FDT) valid in the linear response region and equivalent to the Onsager
reciprocity relations \cite{Gallavotti96}.

\end{itemize}

\subsection{Applications of the FT to nonequilibrium states}
\label{applications}
The FT \eq{ft13} finds application in several nonequilibrium
contexts. Here we describe specific results for transient and steady
states. 

\subsubsection{Nonequilibrium transient states (NETS)}
\label{applications:NETS}
We will assume a system initially in thermal equilibrium that is
transiently brought to a nonequilibrium state.  We are going to show that, under
such conditions, the entropy production \eq{ft9} is equal to the heat
delivered by the system to the sources. We rewrite \eq{ft9} by
introducing the potential energy function $G_{\lambda}({\cal C})$,
\be
P_{\lambda}^{\rm eq}({\cal C})=\frac{\exp(-G_{\lambda}({\cal C}))}{{\cal
Z}_{\lambda}}=\exp(-G_{\lambda}({\cal C})+{\cal G}_{\lambda})
\label{applications:NETS:1}
\ee
where ${\cal Z}_{\lambda}=\sum_{{\cal C}}\exp(-G_{\lambda}({\cal
C}))=\exp(-{\cal G}_{\lambda})$ is the partition function and ${\cal
G}_{\lambda}$ is the thermodynamic potential. The existence of the
potential $G_{\lambda}({\cal C})$ and the thermodynamic potential ${\cal
G}_{\lambda}$ is guaranteed by Boltzmann-Gibbs ensemble theory. For
simplicity we will consider here the canonical ensemble where the volume
$V$, the number of particles $N$ and the temperature $T$ are
fixed. Needless to say that the following results can be generalized to
arbitrary ensembles. In the canonical case $G_{\lambda}({\cal C})$ is
equal to $E_{\lambda}({\cal C})/T$ where $E_{\lambda}({\cal C})$ is the
total energy function (that includes the kinetic plus the potential
energy terms). ${\cal G}_{\lambda}$ is equal to $F_{\lambda}(V,T,N)/T$
where $F_{\lambda}$ stands for the Helmholtz free energy.

With these definitions the entropy production \eq{ft9} is given by,
\be
S_p(\Gamma)=\sum_{k=0}^{M-1}\Bigl(G_{\lambda_k}({\cal
C}_k)-G_{\lambda_k}({\cal C}_{k+1})\Bigr)=\\
\frac{1}{T}\sum_{k=0}^{M-1}\Bigl(E_{\lambda_k}({\cal C}_k)-E_{\lambda_k}({\cal C}_{k+1})\Bigr)~~~.
\label{applications:NETS:3}
\ee
For the boundary term \eq{ft11} let us take $b({\cal C})=P_{\lambda_M}^{\rm eq}({\cal C})$,
\bea B(\Gamma)=\log\bigl( P_{\lambda_0}^{\rm eq}({\cal C}_0)\bigr)
-\log\bigl( P_{\lambda_M}^{\rm eq}({\cal C}_M)\bigr)=\nonumber\\
=G_{\lambda_M}({\cal C}_M)-G_{\lambda_0}({\cal C}_0)-{\cal G}_{\lambda_M}+{\cal G}_{\lambda_0}=\nonumber\\
=\frac{1}{T}(E_{\lambda_M}({\cal C}_M)-E_{\lambda_0}({\cal C}_0)-F_{\lambda_M}+F_{\lambda_0})
\label{applications:NETS:4}
\eea
The total dissipation \eq{ft12} is then equal to,
\be
{\cal S}(\Gamma)=S_p(\Gamma)+\frac{1}{T}(E_{\lambda_M}({\cal C}_M)-E_{\lambda_0}({\cal C}_0)-F_{\lambda_M}+F_{\lambda_0})
\label{applications:NETS:5}
\ee
which can be rewritten as a balance equation for the variation of the
energy $E_{\lambda}({\cal C})$ along a given path,
\be
\Delta E(\Gamma)=E_{\lambda_M}({\cal C}_M)
-E_{\lambda_0}({\cal C}_0)=T{\cal S}(\Gamma)+\Delta F-TS_p(\Gamma)~~~.
\label{applications:NETS:6}
\ee
where $\Delta F=F_{\lambda_M}-F_{\lambda_0}$. This is the first law of
thermodynamics where we have identified the term in the lhs with the
total variation of the internal energy $\Delta E(\Gamma)$. Whereas
$T{\cal S}(\Gamma)+\Delta F$ and $TS_p(\Gamma)$ are identified with the
mechanical work exerted on the system and the heat delivered to the bath
respectively,
\bea
\Delta E(\Gamma)=W(\Gamma)-Q(\Gamma)\label{applications:NETS:7a}\\
W(\Gamma)=T{\cal S}(\Gamma)+\Delta F\label{applications:NETS:7b}\\
Q(\Gamma)=TS_p(\Gamma)\label{applications:NETS:7c}~~~.
\eea
By using \eq{applications:NETS:3} we obtain the following expressions
for work and heat,
\bea
W(\Gamma)=\sum_{k=0}^{M-1}\Bigl(E_{\lambda_{k+1}}({\cal C}_{k+1})-E_{\lambda_k}({\cal C}_{k+1})\Bigr)\label{applications:NETS:7aa}\\
Q(\Gamma)=\sum_{k=0}^{M-1}\Bigl(E_{\lambda_k}({\cal C}_k)-E_{\lambda_k}({\cal C}_{k+1})\Bigr)\label{applications:NETS:7ab}~~~.
\eea
The physical meaning of both entropies is now clear. Whereas $S_p$
stands for the heat transferred by the system to the sources
\eq{applications:NETS:7c}, the total dissipation term $T{\cal S}$
\eq{applications:NETS:7b} is just the difference between the total
mechanical work exerted upon the system, $W(\Gamma)$, and the reversible
work, $W_{\rm rev}=\Delta F$. It is customary to define this quantity as
the dissipated work, $W_{\rm diss}$,
\be
W_{\rm diss}(\Gamma)=T{\cal S}(\Gamma)=W(\Gamma)-\Delta
F=W(\Gamma)-W_{\rm rev}~~~.
\label{applications:NETS:8}
\ee
The nonequilibrium equality \eq{ft4} becomes the nonequilibrium work relation
originally derived by Jarzynski using Hamiltonian dynamics \cite{Jarzynski97}, 
\be
\langle \exp\Bigl(-\frac{W_{\rm diss}}{T}\Bigr)\rangle=1~~~~{\rm or}~~~~ \langle
\exp\Bigl(-\frac{W}{T}\Bigr)\rangle=\exp\Bigl(-\frac{\Delta F}{T}\Bigr)~~~.
\label{applications:NETS:9}
\ee
This relation is called the Jarzynski equality (hereafter referred as
JE) and can be used to recover free energies from nonequilibrium
simulations or experiments (see Sec.~\ref{bio:free}).  The FT
\eq{ft13} becomes the Crooks fluctuation theorem (hereafter referred as CFT) 
\cite{Crooks99,Crooks00},
\be
\frac{P_F(W_{\rm diss})}{P_R(-W_{\rm diss})}=\exp\Bigl(\frac{W_{\rm
diss}}{T}\Bigr)~~~{\rm or}~~~~~~
\frac{P_F(W)}{P_R(-W)}=\exp\Bigl(\frac{W-\Delta F}{T}\Bigr)~~~~.
\label{applications:NETS:10}
\ee
The second law of thermodynamics $\overline{W}\ge \Delta F$ also follows
naturally as a particular case of \eq{ft6} by using
\eqq{applications:NETS:8}{applications:NETS:9}. Note that for the heat
$Q$ a general relation equivalent to \eq{applications:NETS:10} does not
exist. We mention three aspects of the JE and the CFT.

\begin{itemize} 

\item{\bf The fluctuation-dissipation parameter $R$.} In the limit of
small dissipation $W_{\rm diss}\to 0$ the linear response result
\eq{ft14} holds. It is then possible to introduce a parameter
,$R$, that measures deviations from the linear response
behavior\footnote{Sometimes $R$ is called fluctuation-dissipation ratio,
not to be confused with the identically called but different quantity
introduced in glassy systems, see Sec.\ref{glassy:noneqtemp}, that quantifies
deviations from the fluctuation-dissipation theorem that is valid in equilibrium.}. It is defined as,
\be
R=\frac{\sigma_W^2}{2TW_{\rm diss}}
\label{applications:NETS:11}
\ee
where $\sigma_W^2=<W^2>-<W>^2$ is the variance of the work
distribution. In the limit $W_{\rm diss}\to 0$, a second order cumulant
expansion in \eq{applications:NETS:9} shows that $R$ is equal to 1 and
\eq{ft14} holds. Deviations from $R=1$ are interpreted as
deviations of the work distribution from a Gaussian. When the work
distribution is non-Gaussian the system is far from the linear response
regime and \eq{ft14} is not satisfied anymore.

\item{\bf The Kirkwood formula.} A particular case of the JE \eq{applications:NETS:9} is the Kirkwood formula
\cite{Kirkwood35,Zwanzig54}.  It corresponds to
the case where the control parameter only takes two values $\lambda_0$
and $\lambda_1$. The system is initially in equilibrium at the
value $\lambda_0$ and, at an arbitrary later time $t$, the value of
$\lambda$ instantaneously switches to $\lambda_1$. In this case \eq{applications:NETS:7aa}
reads,
\be
W(\Gamma)=\Delta E_{\lambda}({\cal C})=E_{\lambda_1}({\cal C})-E_{\lambda_0}({\cal C})~~~.
\label{applications:NETS:12}
\ee
In this case a path corresponds to a single configuration,
$\Gamma\equiv{\cal C}$, and \eq{applications:NETS:9} becomes,
\be
\overline{\exp\Bigl(-\frac{\Delta E_{\lambda}({\cal C})}{T}\Bigr)}=\exp\Bigl(-\frac{\Delta F}{T}\Bigr)
\label{applications:NETS:13}
\ee
the average $\overline{(..)}$ is taken over all configurations ${\cal C}$ sampled according to
the equilibrium distribution taken at $\lambda_0$, $P_{\lambda_0}^{\rm eq}({\cal C})$. 

\item{\bf Heat exchange between two bodies.} Suppose that we take two bodies
initially at equilibrium at temperatures $T_H,T_C$ where $T_H,T_C$ stand
for a hot and a cold temperature. At time $t=0$ we put them in contact and
ask about the probability distribution of heat flow between them. In
this case, no work is done between the two bodies and the heat
transferred is equal to the energy variation of each of the bodies. Let
$Q$ be equal to the heat transferred from the hot to the cold body in
one experiment. It
can be shown \cite{JarWoj04} that in this case the total dissipation
${\cal S}$ is given by,
\be
{\cal S}=Q(\frac{1}{T_c}-\frac{1}{T_H})
\label{applications:NETS:14}
\ee
and the equality \eq{applications:NETS:9} reads,
\be
\langle\exp\Bigl(-Q(\frac{1}{T_c}-\frac{1}{T_H})\Bigr)=1\rangle
\label{applications:NETS:15}
\ee
showing that, in average, net heat is always transferred from the hot to the cold
body. Yet, sometimes, we also expect some heat to flow from the cold to
the hot body. Again, the probability of such events will be
exponentially small with the size of the system.

\end{itemize}

\subsubsection{Nonequilibrium steady states (NESS)}
\label{applications:NESS}
Most investigations in nonequilibrium systems were initially
carried out in NESS.  It is widely believed that NESS are among the
best candidate nonequilibrium systems to possibly extend Boltzmann-Gibbs
ensemble theory beyond equilibrium \cite{OonPan98,SasTas04}. 

We can distinguish two types of NESS: Time-dependent conservative (C)
systems and non-conservative (NC) systems. In the C case the system is
acted by a time-dependent force that derives from an external
potential. In the NC case the system is driven by (time-dependent or
not) non-conservative forces. In C systems the control parameter
$\lambda$ has the usual meaning: it specifies the set of external
parameters that, once fixed, determine an equilibrium state. Examples
are: a magnetic dipole in an oscillating field ($\lambda$ is the value
of the time-dependent magnetic field); a bead confined in an moving
optical trap and dragged through water ($\lambda$ is the position of the
center of the moving trap); a fluid sheared between two plates
($\lambda$ is the time-dependent relative position of the upper and lower
plates). In C systems we assume local detailed balance so \eq{demo5} still holds.

In contrast to the C case, in NC systems the local detailed balance property, in the form of
\eq{demo5}, does not hold because the system does not reach thermal equilibrium but a stationary
or steady state.  It is then customary to characterize the NESS by the parameter
$\lambda$ and the stationary distribution by $P_{\lambda}^{\rm ss}({\cal
C})$. NESS systems in the linear regime (i.e. not driven arbitrarily far
from equilibrium) satisfy the Onsager reciprocity relations where the
fluxes are proportional to the forces. NESS can be maintained, either by
keeping constant the forces or the fluxes. Examples of NC systems are:
the flow of a current in an electric circuit (e.g. the control parameter
$\lambda$ is either the constant current, $I$, flowing through the circuit or the
constant voltage difference, $\Delta V$); a Poiseuille fluid flow inside a cylinder
($\lambda$ could be either the constant fluid flux, $\Phi$, or the
pressure difference, $\Delta P$); heat flowing between two sources kept at two
different temperatures ($\lambda$ could be either the heat
flux, $J_Q$, or the temperature gradient, $\Delta T$); the particle exclusion process
($\lambda=\mu^{+,-}$ are the rates of inserting and removing particles
at both ends of the chain). In NESS of NC type the local detailed
balance property \eq{demo5} still holds but replacing $P_{\lambda}^{\rm
eq}({\cal C})$ by the corresponding stationary distribution,
$P_{\lambda}^{\rm ss}({\cal C})$,
\be \frac{{\cal W}_{\lambda}({\cal C}\to {\cal C'})}{{\cal W}_{\lambda}({\cal C'}\to {\cal C})}=\frac{P^{\rm
ss}_{\lambda}({\cal C'})}{P^{\rm ss}_{\lambda}({\cal C})}~~~.
\label{applications:NESS:1}
\ee
In a steady state in a NC system $\lambda$ is maintained
constant. Because the local form of detailed balance
\eq{applications:NESS:1} holds the main results of
Sec.~\ref{stoctherm:ft} follow. In particular, the nonequilibrium
equality \eq{ft4} and the FT \eq{ft13} are still true. However there is
an important difference. In steady states the reverse process is
identical to the forward process $P_F({\cal S})=P_R({\cal S})$ because
$\lambda$ is maintained constant. Therefore \eq{ft4} and the FT
\eq{ft13} become,
\bea
\langle{\exp(-{\cal S})}\rangle=1
\label{applications:NESS:2a}\\
\frac{P({\cal S})}{P(-{\cal S})}=\exp({\cal S})~~~~~.\label{applications:NESS:2b}
\eea
We can now extract a general FT for the entropy
production $S_p$ in NESS. Let us assume that, in average, $S_p$ grows
linearly with time, i.e. $S_p\gg B$ for large $t$. Because ${\cal
S}=S_p+B$ \eq{ft10}, in the large $t$ limit fluctuations in ${\cal S}$ are asymptotically
dominated by fluctuations in $S_p$. In average, fluctuations in $S_p$ grow
like $\sqrt{t}$ whereas fluctuations in the boundary term are
finite. By taking the logarithm of \eq{applications:NESS:2b} and
using \eq{ft10} we obtain,
\be
S=\log(P(S))-\log(P(-S))\rightarrow S_p+B=\log(P(S_p+B))-\log(P(-S_p-B))~~~~~.
\label{applications:NESS:3}
\ee
In NESS the entropy produced, $S_p(\Gamma)$, along paths of
duration $t$ is a fluctuating quantity. 
Expanding \eq{applications:NESS:3} around $S_p$ we get,
\be
S_p=\log\Bigl(\frac{P(S_p)}{P(-S_p)}\Bigr)+B\Bigl(\frac{P'(S_p)}{P(S_p)}+\frac{P'(-S_p)}{P(-S_p)}-1 \Bigr)
\label{applications:NESS:4a}
\ee
The average entropy production $\langle
S_p\rangle$ is defined by averaging $S_p$ along an infinite number of
paths. Dividing \eq{applications:NESS:4a} by $\langle
S_p\rangle$ we get,
\be
\frac{S_p}{\langle S_p \rangle}=\frac{1}{\langle S_p
\rangle}\log\Bigl(\frac{P(S_p)}{P(-S_p)}\Bigr)+\frac{B}{\langle S_p \rangle}\Bigl(\frac{P'(S_p)}{P(S_p)}+\frac{P'(-S_p)}{P(-S_p)}-1 \Bigr)
\label{applications:NESS:4b}
\ee
We introduce a quantity $a$ that is equal to the ratio between the entropy
production and its average value, $a=\frac{S_p}{\langle S_p \rangle}$. We
can define the function,
\be
f_t(a)=\frac{1}{\langle S_p \rangle}\log\Bigl(\frac{P(a)}{P(-a)}\Bigr)~~~~.
\label{applications:NESS:5}
\ee
Equation \eq{applications:NESS:4b} can be rewritten as,
\be
f_t(a)=a-\frac{B}{\langle S_p
\rangle}\Bigl(\frac{P'(S_p)}{P(S_p)}+\frac{P'(-S_p)}{P(-S_p)}-1 \Bigr)
\label{applications:NESS:6}
\ee
In the large time limit, assuming that $\log(P(S_p))\sim t$, and because
$B$ is finite, the second term vanishes relative to the first and
$f_t(a)=a+{\cal O}(1/t)$. Substituting this result into
\eq{applications:NESS:5} we find that an FT holds in the large $t$
limit.  However, this is not necessarily always true. Even for very
large $t$ there can be strong deviations in the initial and final state
that can make the boundary term $B$ large enough to be comparable to
$\langle S_p\rangle$. In other words, for certain initial and/or final
conditions, the second term in the rhs of \eq{applications:NESS:6} can
be on the same order and comparable to the first term, $a$. The boundary term can
be neglected only if we restrict the size of such large deviations, i.e.
if we require $|a|\le a^*$ where $a^*$ is a maximum given value. With this
proviso, the FT in NESS reads,
\be
\lim_{t\to\infty}\frac{1}{\langle S_p
\rangle}\log\Bigl(\frac{P(a)}{P(-a)}\Bigr)=a~~~~~~;~~~~~~|a|\le a^*
\label{applications:NESS:7}
\ee
In general it can be very difficult to determine the nature of the
boundary terms. A specific result in an exactly solvable case is
discussed below in
Sec.~\ref{example:bead:entropy}. Eq.\eq{applications:NESS:7} is the
Gallavotti-Cohen FT derived in the context of deterministic Anosov
systems \cite{GalCoh95}. In that case $S_p$ stands for the so called
phase space compression factor. It has been experimentally tested by
Ciliberto and coworkers in Rayleigh-Bernard convection \cite{CilLar98}
and turbulent flows \cite{CilGarHerLacPinRui04}. Similar relations have
been also tested in athermal systems, e.g. in fluidized granular media
\cite{FeiMen04} or the case of two-level systems in fluorescent diamond
defects excited by light \cite{SchSpeTieWraSei05}.

The FT \eq{ft13} also describes fluctuations in the total dissipation
for transitions between steady states where $\lambda$ varies according
to a given protocol. In that case, the system starts at time $0$ in a
given steady state, $P^{\rm ss}_{\lambda_0}({\cal C})$, and evolves away
from that steady state at subsequent times. The boundary term for steady state
transitions is then given by,
\be
B(\Gamma)=\log(P^{\rm ss}_{\lambda_0}({\cal C}_0))-\log(P^{\rm ss}_{\lambda_M}({\cal
C}_M))
\label{applications:NESS:8}
\ee
where we have chosen the boundary function $b({\cal C})=P^{\rm
ss}_{\lambda_M}({\cal C})$. In that case the total dissipation is
antisymmetric under time-reversal and \eq{ft13} holds. Only in cases
where the reverse process is equivalent to the forward process
\eq{applications:NESS:2b} remains an exact result.  Transitions between
nonequilibrium steady states and expressions for the function ${\cal S}$
have been considered by Hatano and Sasa in the context of Langevin
systems \cite{HatSas01}.
 
\section{Examples and applications}
\label{examples}
In this section we analyze in detail two cases where analytical
calculations can be carried out and FTs have been experimentally tested. We
have chosen two examples: one extracted from physics, the other from
biology. We first analyze the bead in a trap to later consider single
molecule pulling experiments. These examples show that there are lots of
interesting observations that can be made by comparing theory and
nonequilibrium experiments in simple systems.

\subsection{A physical system: a bead in an optical trap}
\label{example:bead}
%
%
It is very instructive to work out in detail the fluctuations of a bead
trapped in a moving potential. This case is of great interest for at
least two reasons. First, it provides a simple example of both a NETS
and a NESS that can be analytically solved in detail. Second, it can be
experimentally realized by trapping micron-sized beads using optical
tweezers. The first experiments studying nonequilibrium fluctuations in
a bead in a trap were carried out by Evans and collaborators
\cite{WanSevMitSeaEva02} and later on extended in a series of works
\cite{CarReiWanSevSeaEva04,ReiCarWanSevSeaEva04}. Mazonka and Jarzynski
\cite{MazJar99} and later Van Zon and Cohen
\cite{ZonCoh03a,ZonCoh03b,ZonCoh04} have carried out detailed
theoretical calculations of heat and work fluctuations. Recent experiments have
also analyzed the case of a particle in a non-harmonic optical potential
\cite{BliSpeHelSeiBec06}. These results have greatly contributed to clarify the
general validity of the FT and the role of the boundary terms appearing
in the total dissipation ${\cal S}$.

The case of a bead in a trap is also equivalent to the power
fluctuations in a resistance in an RC electrical circuit
\cite{ZonCilCoh04} (see Figure \ref{fig4}).  The experimental setup is
shown in Figure \ref{fig5}. A micron-sized bead is immersed in water and
trapped in an optical well. In the simplest case the trapping potential
is harmonic.  Here we will assume that the potential well can have an
arbitrary shape and carry out specific analytical computations for the
harmonic case.

Let $x$ be the position of the bead in the laboratory frame and
$U(x-x^*)$ the trapping potential of a laser focus that is centered at a
reference position $x^*$. For harmonic potentials we will take
$U(x)=(1/2){\kappa}x^2$. By changing the value of $x^*$ the trap is
shifted along the $x$ coordinate. A nonequilibrium state can be
generated by changing the value of $x^*$ according to a protocol
$x^*(t)$. In the notation of the previous sections, $\lambda\equiv x^*$
is the control parameter and ${\cal C}\equiv x$ is the configuration. A
path $\Gamma$ starts at $x(0)$ at time 0 and ends at $x(t)$ at time $t$,
$\Gamma\equiv\lbrace x(s);0\le s\le t\rbrace$.

At low Reynolds number the motion of the bead can be described by a
one-dimensional Langevin equation that contains only the overdamping term,
\be
\gamma\dot{x}=f_{x^*}(x)+\eta~~~~~~;~~~~~~\langle \eta(t)\eta(s)\rangle=2T\gamma\delta(t-s)
\label{example:1}
\ee
where $x$ is the position of the bead in the laboratory frame, $\gamma$ is the friction
coefficient, $f_{x^*}(x)$ is a conservative force deriving from the trap potential $U(x-x^*)$, 
\be
f_{x^*}(x)=-(U(x-x^*))'=-\Bigl (\frac{\partial U(x-x^*)}{\partial x}\Bigr)
\label{conservative}
\ee
and $\eta$ is a stochastic white noise.  

In equilibrium $x^*(t)=x^*$ is constant in time. In this case, the
stationary solution of the master equation is the equilibrium solution
\be
P^{\rm eq}_{x^*}(x)=\frac{\exp(-\frac{U(x-x^*)}{T})}{\int
dx\exp(-U(x-x^*)/T)}=\frac{\exp(-\frac{U(x-x^*)}{T})}{{\cal Z}}
\label{example:2}
\ee
where ${\cal Z}=\int dx\exp(-U(x)/T)$ is the partition function that is
independent of the reference position $x^*$. Because the free energy
$F=-T\log({\cal Z})$ does not depend on the control parameter $x^*$, the
free energy change is always zero for arbitrary translations of the trap.

Let us now consider a NESS where the trap is moved at constant velocity,
$x^*(t)=vt$. It is not possible to solve the Fokker-Planck
equation to find the probability distribution
in the steady state for arbitrary potentials.   Only for harmonic potentials,
$U(x)=\kappa x^2/2$, the Fokker-Planck equation can be solved exactly. The
result is,
\be
P^{\rm ss}_{x^*}(x)=\Bigl(\frac{2\pi
T}{\kappa}\Bigr)^{-\frac{1}{2}}\exp\Bigl(-\frac{\kappa (x-x^*(t)+\frac{\gamma v}{\kappa})^2}{2T} \Bigr)
\label{example:2b}
\ee
Note that the steady-state solution \eq{example:2b} depends explicitly
on time through $x^*(t)$. To obtain a time-independent solution we must change variables
$x\to x-x^*(t)$ and describe the motion of the bead in the reference
frame that is solidary and moves with the trap. We will come back to this problem later in
Sec.~\ref{example:ss}.

\subsubsection{Microscopic reversibility}
\label{example:mr}
In this section we show that the Langevin dynamics \eq{example:1}
satisfies the microscopic reversibility assumption or local detailed
balance \eq{demo5}. We recall that $x$ is the position of the bead in
the laboratory frame. The transition rates ${\cal W}_{x^*}(x\to x')$ for
the configuration $x$ at time $t$ to change to $x'$ at a later time $t+\Delta t$
can be computed from \eq{example:1}. We discretize the Langevin equation
\cite{ZinnJustin96} by writing,
\be
x'=x+\frac{f(x-x^*)}{\gamma}\Delta t+\sqrt{\frac{2T\Delta
t}{\gamma}}r+{\cal O}\Bigl ((\Delta t)^2\Bigr)
\label{example:3}
\ee
where $r$ is a random Gaussian number of zero mean and unit
variance. For a given value of $x$, the distribution of values $x'$ is
also a Gaussian with average and variance given by,
\bea
\overline{x'}=x+\frac{f(x-x^*(t))}{\gamma}\Delta
t+{\cal O}\Bigl ((\Delta t)^2\Bigr)\label{example:4a}\\
\sigma^2_{x'}=\overline{(x')^2}-(\overline{(x')})^2=\frac{2T\Delta t}{\gamma}+{\cal O}\Bigl ((\Delta t)^2\Bigr)
\label{example:4b}
\eea
and therefore,
\be
{\cal W}_{x^*}(x\to
x')=(2\pi\sigma^2_{x'})^{-\frac{1}{2}}\exp\Bigl(-\frac{(x'-x+\frac{f(x-x^*)\Delta
t}{\gamma})^2}{2\sigma^2_{x'}} \Bigl)~~~~.
\label{example:5}
\ee
From \eq{example:5} we compute the ratio between the transition
probabilities to first order in $\Delta t$,
\be
\frac{{\cal W}_{x^*}(x\to x')}{{\cal W}_{x^*}(x'\to x)}=\exp\Bigl(-\frac{(x'-x)(f(x-x^*)+f(x'-x^*))}{2T}\Bigr)~~~.
\label{example:7}
\ee
We can now use the Taylor expansions, 
\bea
U(x'-x^*)=U(x-x^*)-f(x-x^*)(x'-x)+{\cal O}\bigl((x'-x)^2\bigr)\\
U(x-x^*)=U(x'-x^*)-f(x'-x^*)(x-x')+{\cal O}\bigl((x'-x)^2\bigr)
\label{example:8}
\eea
and subtract both equations to finally obtain,
\be
(x'-x)(f(x-x^*)+f(x'-x^*))=2(U(x'-x^*)-U(x-x^*))
\label{example:9}
\ee
which yields,
\be
\frac{{\cal W}_{x^*}(x\to x')}{{\cal W}_{x^*}(x'\to x)}=\exp\Bigl(-\frac{U(x'-x^*)-U(x-x^*)}{T}\Bigr)=\frac{P^{\rm eq}_{x^*}(x')}{P^{\rm eq}_{x^*}(x)}
\label{example:10}
\ee
which is the local detailed balance assumption \eq{demo5}. 

\subsubsection{Entropy production, work and total dissipation}
\label{example:bead:entropy}
Let us consider an arbitrary nonequilibrium protocol $x^*(t)$ where
$v(t)=\dot{x}^*(t)$ is the velocity of the moving trap. The
entropy production for a given path, $\Gamma\equiv\lbrace x(s);0\le
s\le t\rbrace$ can be computed using \eq{ft9},
\be
S_p(\Gamma)=\int_0^t ds \dot{x}(s)\Bigl(\frac{\partial \log P^{\rm
eq}_{x^*(s)}(x)}{\partial x}\Bigr)_{x=x(s)}~~~~~.
\label{example:11}
\ee
We now define the variable $y(t)=x(t)-x^*(t)$. From \eq{example:2} we
get \footnote{Note that $\dot{x}$, the velocity of the bead, is not well
defined in \eqq{example:11}{example:12}. However, $ds\dot{x}(s)=dx$ it
is. Yet we prefer to use the notation in terms of velocities just to
make clear the identification between the time integrals in
\eqq{example:11}{example:12} and the discrete time-step sum in \eq{ft9}.},
\bea
S_p(\Gamma)=\frac{1}{T}\int_0^t ds
\dot{x}(s)f(x(s)-x^*(s))=\frac{1}{T}\int_0^t ds(\dot{y}(s)
+v(s))f(y(s))=\\
\frac{1}{T}\Bigl(\int_{y(0)}^{y(t)}dy f(y)+\int_0^tds
v(s)f(s)\Bigr)=\frac{-\Delta U+W(\Gamma)}{T}
\label{example:12}
\eea
with
\be 
\Delta U=U(x(t)-x^*(t))-U(x(0)-x^*(0))~~~;~~~ 
W(\Gamma)=\int_0^tds v(s)f(s)~~~.
\label{example:13}
\ee 
where we used \eq{conservative} in the last equality of
\eq{example:12}. $\Delta U$ is the variation of internal energy between
the initial and final positions of the bead and $W(\Gamma)$ is the
mechanical work done on the bead by the moving trap. Using the first law
$\Delta U=W-Q$ we get,
\be
S_p(\Gamma)=\frac{Q(\Gamma)}{T}
\label{example:14}
\ee
and the entropy production is just the heat transferred from the bead to
the bath divided by the temperature of the bath.

The total dissipation ${\cal S}$ \eq{ft10} can be evaluated by adding the
boundary term \eq{ft11} to the entropy production. For the boundary term
we have some freedom as to which function $b$ we use in the rhs of
\eq{ft11},
\be
B(\Gamma)=\log\Bigl(P_{x^*(0)}(x(0))\Bigr)-\log\Bigl(b(x(t))\Bigr)~~~~.
\label{example:15}
\ee

\begin{figure}
\begin{center}
\includegraphics[scale=0.4,angle=0]{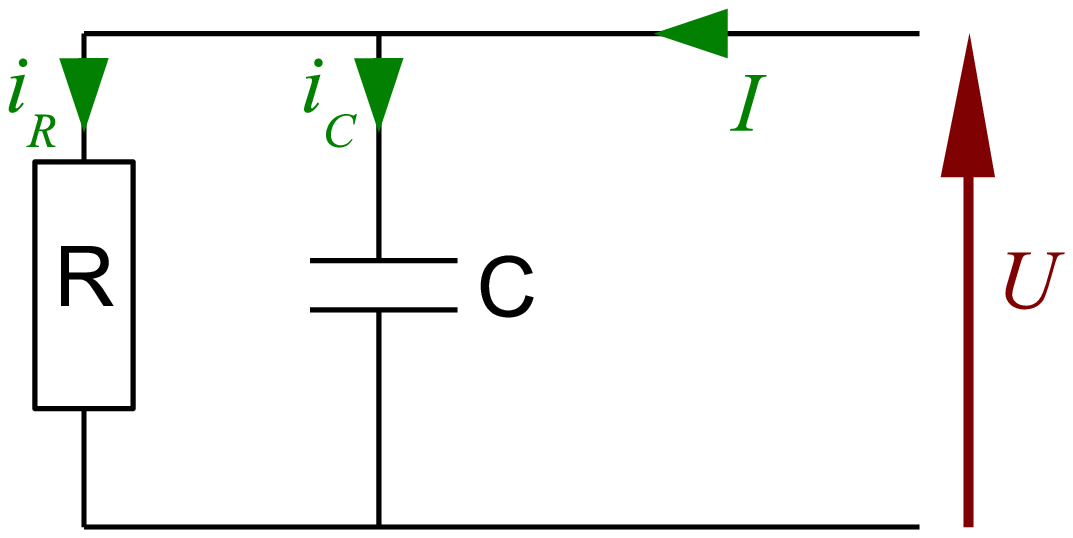}\includegraphics[scale=0.3,angle=0]{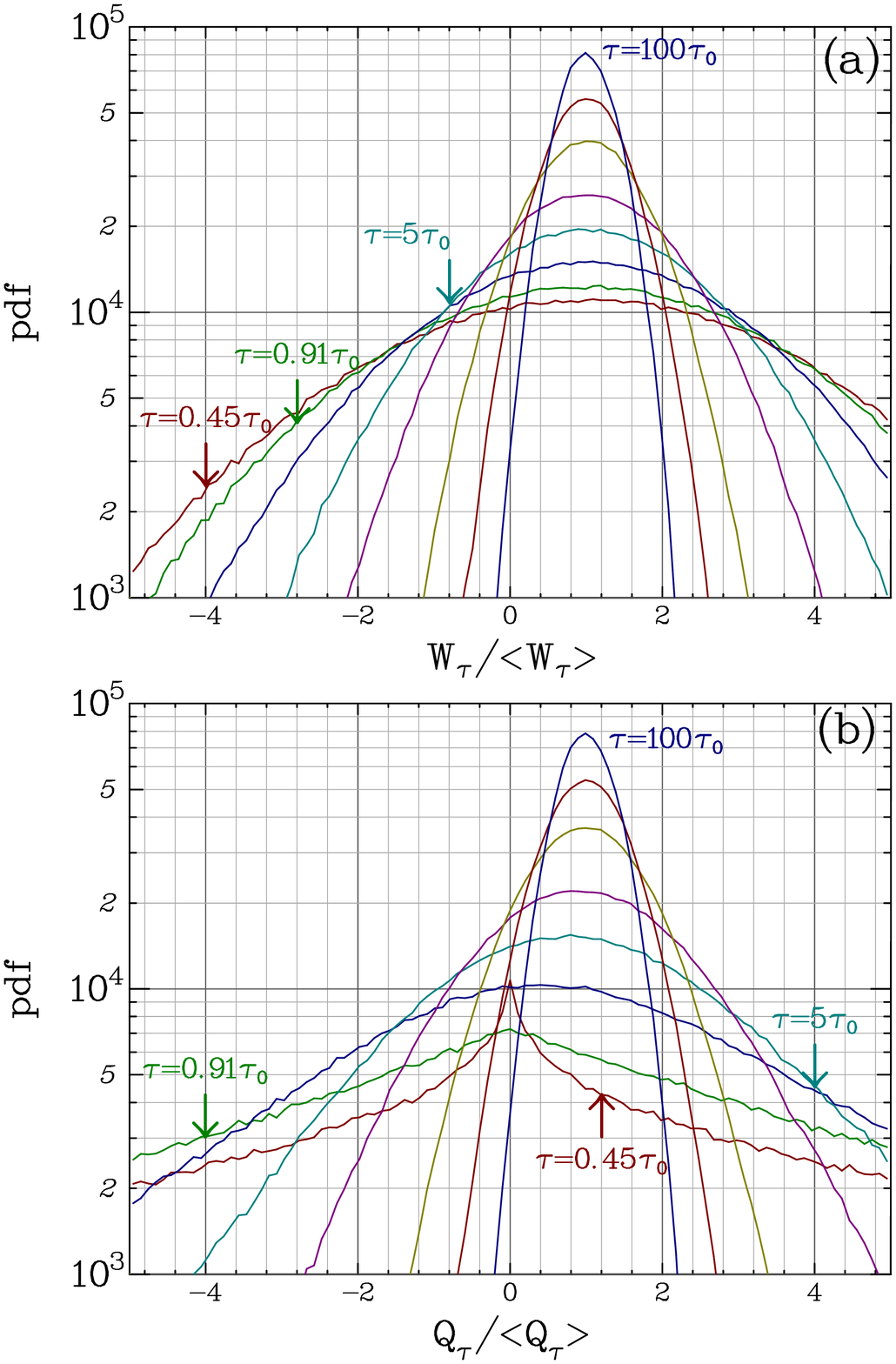}\includegraphics[scale=0.3,angle=0]{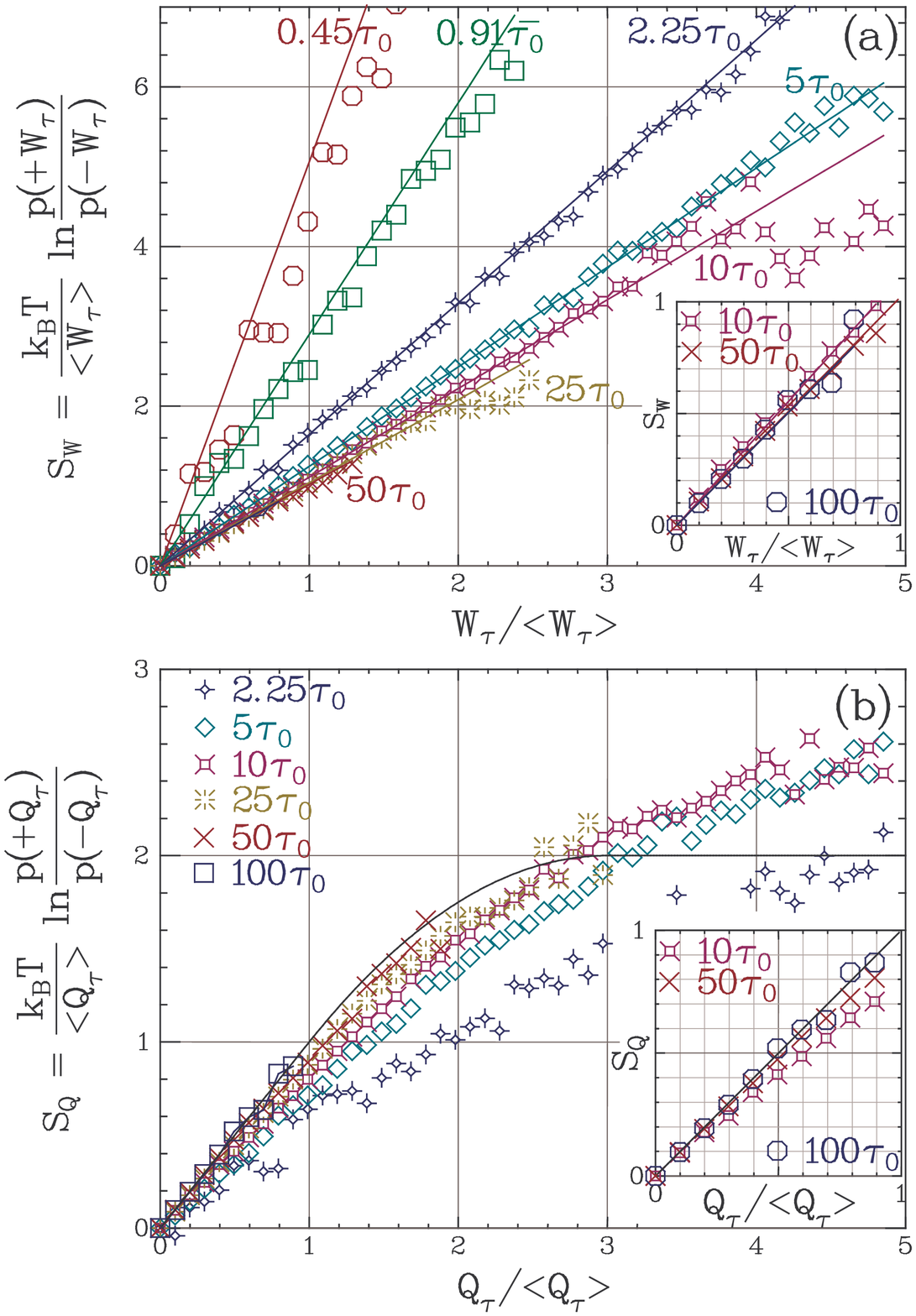}
\end{center}
\vspace{-0.4cm}
\caption{\em Heat and work fluctuations in an electrical circuit
(left). PDF distributions (center) and verification of the FTs
\eqq{example:22}{example:23} (right). Figure
taken from \protect\cite{GarCil04}.}
\label{fig4}
\vspace{-0.2cm}
\end{figure}

Because we want ${\cal S}$ to be antisymmetric against
time-reversal there are two possible choices for the function $b$ depending on the initial
state,

\begin{itemize}

\item{\bf Nonequilibrium transient state (NETS).} Initially the bead is in
equilibrium and the trap is at rest in a given position $x^*(0)$. Suddenly the trap is set in
motion. In this case we choose $b(x)=P^{\rm eq}_{x^*(t)}(x)$ and the boundary term
\eq{ft11} reads,
\be
B(\Gamma)=\log\Bigl(P^{\rm eq}_{x^*(0)}(x(0))\Bigr)-\log\Bigl(P^{\rm eq}_{x^*(t)}(x(t))\Bigr)~~~.
\label{example:16}
\ee
By inserting \eq{example:2} we obtain,
\be
B(\Gamma)=\frac{1}{T}\bigl(U(x(t)-x^*(t))-U(x(0)-x^*(0))\bigr)=\frac{\Delta U}{T}
\label{example:17}
\ee
and ${\cal S}=S_p+B=(Q+\Delta U)/T=W/T$ so the work satisfies the
nonequilibrium equality \eq{ft4} and the FT \eq{ft13},
\be
\frac{P_F(W)}{P_R(-W)}=\exp\Bigl(\frac{W}{T}\Bigr)
\label{example:17b}
\ee
Note that in the
reverse process the bead starts in equilibrium at the final position
$x^*(t)$ and the motion of the trap is reversed $(x^*)^R(s)=x^*(t-s)$. 
The result \eq{example:17b} is valid for arbitrary potentials $U(x)$. In general, the
reverse work distribution $P_R(W)$ will differ from the forward
distribution $P_F(W)$. Only for symmetric potentials $U(x)=U(-x)$ both
work distributions are identical \cite{BaiJacMaeSka06}. Under this additional assumption
\eq{example:17b} reads,
\be
\frac{P(W)}{P(-W)}=\exp(\frac{W}{T})
\label{example:18}
\ee
Note that this is a particular case of the CFT \eq{applications:NETS:10} with $\Delta F=0$. 

\item{\bf Nonequilibrium steady state (NESS).} If the initial state is a steady
state, $P_{\lambda_0}({\cal C}_0)\equiv P^{\rm ss}_{x^*(0)}(x)$, then we choose $b(x)=P^{\rm ss}_{x^*(t)}(x)$. The boundary term
reads,
\be
B(\Gamma)=\log\Bigl(P^{\rm ss}_{x^*(0)}(x(0))\Bigr)-\log\Bigl(P^{\rm ss}_{x^*(t)}(x(t))\Bigr)
\label{example:19}
\ee
Only for harmonic potentials we exactly know the steady state solution
\eq{example:2b} so we can write down an explicit expression
for $B$,
\be
B(\Gamma)=\frac{\Delta U}{T}-\frac{v\gamma\Delta f}{\kappa T}
\label{example:20}
\ee
where $\Delta U$ is defined in \eq{example:13} and $\Delta
f=f_{x^*(t)}(x(t))-f_{x^*(0)}(x(0))$. The total dissipation is given by,
\be
{\cal S}=S_p+B=\frac{Q+\Delta U}{T}-\frac{v\gamma\Delta f}{\kappa
T}=\frac{W}{T}-\frac{v\gamma\Delta f}{\kappa T}~~~.
\label{example:21}
\ee
It is important to stress that \eq{example:21} does not satisfy
\eqq{applications:NESS:2a}{applications:NESS:2b} because the last
boundary term in the rhs of \eq{example:21} ($v\gamma\Delta f/\kappa T$) is not antisymmetric against time
reversal. Van Zon and Cohen \cite{ZonCoh03a,ZonCoh03b,ZonCoh04} have
analyzed in much detail work and heat fluctuations in the NESS. They
find that work fluctuations satisfy the exact relation,
\be
\frac{P(W)}{P(-W)}=\exp\Bigl( \frac{W}{T}\frac{1}{1+\frac{\tau}{t}(\exp(-\frac{t}{\tau})-1)}\Bigr)
\label{example:22}
\ee
where $t$ is the time window over which work is measured and $\tau$ is
the relaxation time of the bead in the trap, $\tau=\gamma/\kappa$. Note
that the FT \eq{example:18} is satisfied in the limit $\tau/t\to
0$. Corrections to the FT are on the order of $\tau/t$ as expected (see
the discussion in the last part of
Sec.~\ref{applications:NESS}). Computations can be also carried out for
heat fluctuations. The results are expressed in terms of the relative
fluctuations of the heat, $a=\frac{S_p}{\langle S_p \rangle}$. The large
deviation function $f_t(a)$ \eq{applications:NESS:5} has been shown to be given by,

\bea
\lim_{t\to\infty}f_t(a)=a~~~~(0\le a\le 1);\nonumber\\ 
\lim_{t\to\infty}f_t(a)=a-(a-1)^2/4~~~~(1\le a <3);\nonumber\\
\lim_{t\to\infty}f_t(a)= 2~~~~(3\le a);
\label{example:23}
\eea
and $f_t(-a)=-f_t(a)$. Very accurate experiments to test
\eqq{example:22}{example:23} have been carried out by Garnier and
Ciliberto who measured the Nyquist noise in an electric resistance \cite{GarCil04}. Their
results are in very good agreement with the theoretical predictions
which include corrections in the convergence of \eq{example:23} on the order of $1/t$
as expected. A few results are shown in Figure \ref{fig4}.

\end{itemize}

\subsubsection{Transitions between steady states} 
\label{example:ss}
Hatano and Sasa \cite{HatSas01} have derived an interesting result for nonequilibrium
transitions between steady states. Despite of the generality of Hatano
and Sasa's approach, explicit computations can be worked out only for
harmonic traps. In the present example the system starts in a steady
state described by the stationary distribution \eq{example:2b} and is
driven away from that steady state by varying the speed of the trap,
$v$.  The stationary distribution can be written in the frame system
that moves solidary with the trap. If we define $y(t)=x-x^*(t)$ then
\eq{example:2b} becomes,
\be
P^{\rm ss}_{v}(y)=(2\pi
T/\kappa)^{-\frac{1}{2}}\exp\Bigl(-\frac{\kappa (y+\frac{\gamma v}{\kappa})^2}{2T} \Bigr)~~~~.
\label{example:ss:1}
\ee
Note that, when expressed in terms of the reference moving frame, the
distribution in the steady state becomes stationary or time
independent. The transition rates \eq{example:5} can also be expressed in the
reference system of the trap,
\be
{\cal W}_{v}(y\to
y')=(\frac{4\pi T\Delta t}{\gamma})^{-\frac{1}{2}}\exp\bigl(-\frac{\gamma(y'-y+(v+\frac{\kappa}{\gamma}y)\Delta
t)^2}{4T\Delta t} \bigl)
\label{example:ss:2}
\ee
where we have used $f(x-x^*)=f(y)=-\kappa y$. The transition rates
${\cal W}_{v}(y\to y')$ now depend on the velocity of the trap. This
shows that, for transitions between steady states, $\lambda\equiv v$
plays the role of the control parameter, rather than the value of $x^*$.
A path is then defined by the evolution $\Gamma\equiv\lbrace y(s); 0\le
s\le t\rbrace$ whereas the perturbation protocol is specified by the
time evolution of the speed of the trap $\lbrace v(s);0\le s\le t\rbrace$.

The rates ${\cal W}_{v}(y\to y')$ satisfy the local detailed balance property
\eq{applications:NESS:1}. From \eq{example:ss:2} and \eq{example:ss:1}
we get (in the limit $\Delta t\to 0$),
\bea
\frac{{\cal W}_{v}(y\to y')}{{\cal W}_{v}(y'\to y)}=\frac{P^{\rm ss}_{v}(y')}{P^{\rm
ss}_{v}(y)}=\\
=\exp\Bigl(-\frac{\kappa}{2T}(y'^2-y^2)-\frac{\gamma v}{T}(y'-y)
\Bigr)=\exp\Bigl( -(\frac{\Delta U}{T}-\frac{v\gamma \Delta f}{\kappa T})\Bigr)
\label{example:ss:3}
\eea
Note that the exponent in the rhs of \eq{example:ss:3} is equal to the
boundary term \eq{example:20}. In the reference system of the trap we
can then compute the entropy production $S_p$ and the total dissipation
${\cal S}$. From either \eq{ft9} or \eq{example:ss:3} and using
\eq{example:ss:1} we get,
\bea
S_p(\Gamma)=\int_0^tds\dot{y}(s)\Bigl(\frac{\log(P_{v}^{\rm
ss}(y)}{\partial
y}\Bigr)_{y=y(s)}=-\frac{\Delta U}{T}+\frac{\gamma}{\kappa T}\int_0^tds v(s)\dot{F}(s)=\\
=-\frac{1}{T}\Bigl(\Delta
U-\frac{\gamma}{\kappa}(\Delta(vF))+\frac{\gamma}{\kappa}\int_0^tds \dot{v}(s)F(s)\Bigr)
\label{example:ss:4}
\eea
where we integrated by parts in the last step of the
derivation. For the boundary term \eq{applications:NESS:8} we get,
\bea
B(\Gamma)=\log\Bigr(P^{\rm ss}_{v(0)}(y(0))\Bigr)-\log\Bigl(P^{\rm ss}_{v(t)}(y(t))\Bigr)=\\
\frac{1}{T}\Bigl( \Delta
U-\frac{\gamma}{\kappa}(\Delta(vF))+\frac{\gamma^2\Delta(v^2)}{2\kappa} \Bigr)
\label{example:ss:5}
\eea
where we used \eq{example:ss:1}. By adding \eq{example:ss:4} and
\eq{example:ss:5} we obtain the total dissipation,
\bea
{\cal S}=S_p+B=\frac{1}{T}\Bigl(\Delta(\frac{\gamma^2v^2}{2\kappa})-
\frac{\gamma}{\kappa}\int_0^tds \dot{v}(s)F(s)\Bigr)=\\
=-\frac{\gamma}{\kappa}\int_0^tds \dot{v}(s)(F(s)-\gamma v(s))
\label{example:ss:6}
\eea
The quantity $S$ (called $Y$ by Hatano and Sasa) satisfies the
nonequilibrium equality \eq{ft4} and the FT \eq{ft13}. Only for time
reversal invariant protocols, $v^R(s)=v(t-s)$, we have $P_F({\cal S})=P_R({\cal S})$, and the FT
\eq{applications:NESS:2b} is also valid.  We emphasize two aspects of \eq{example:ss:6},
\begin{itemize}

\item{\bf Generalized second law for steady state transitions.} From the inequality
\eq{ft6} and \eq{example:ss:6} we obtain,
\be
\frac{\gamma}{\kappa}\int_0^tds \dot{v}(s)F(s)\le \Delta(\frac{\gamma^2v^2}{2\kappa})
\label{example:ss:7}
\ee
which is reminiscent of the Clausius inequality $Q\ge -T\Delta S$ where
the average dissipation rate $\overline{P}_{\rm diss}=\gamma v^2$ plays
the role of a state function similar to the equilibrium entropy. In
contrast to the Clausius inequality the transition now occurs between
steady states rather than equilibrium states \cite{TreJarRitCroBusLip04}.

\item{\bf Non-invariance of entropy production under Galilean
transformations}. In steady states where $\dot{v}=0$, $S_p$
\eq{example:ss:4} becomes a boundary term and ${\cal S}=0$
\eq{example:ss:6}. However we saw in
\eq{example:14} that $S_p$ is equal to the heat delivered to the
environment (and therefore proportional to the time elapsed $t$) whereas now
$S_p$ is a boundary term that does not grow with $t$. This important difference
arises from the fact that the entropy production is not invariant under
Galilean transformations. In the reference of the moving trap the bath
is moving at a given speed which impedes to define the heat in a proper
way. To evaluate the entropy production for transitions between steady
states one has to resort to the description where $x^*$ is the control
parameter and $x^*(t)=\int_0^tds v(s)$ is the perturbation protocol. In
such description the results \eq{example:13} and \eq{example:14} are still
valid.

\end{itemize}

\begin{figure}
\begin{center}
\includegraphics[scale=0.7,angle=0]{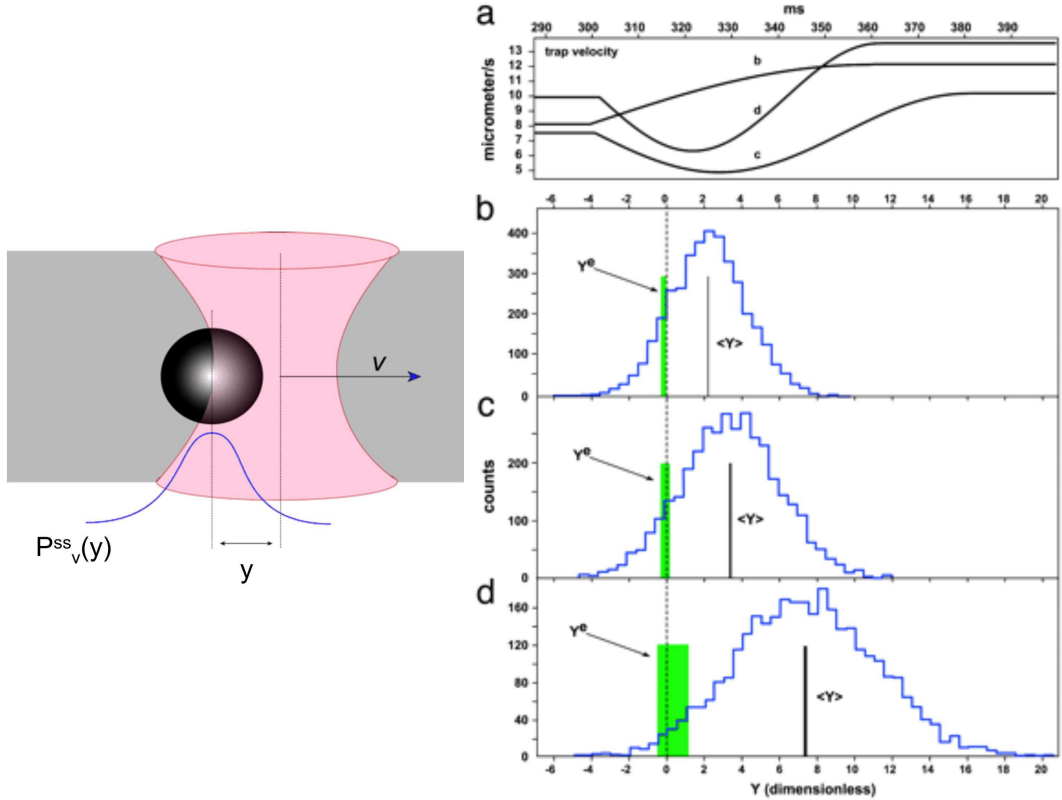}
\end{center}
\vspace{-0.4cm}
\caption{\em (Left) Bead confined in a moving optical trap. (Right) Total
entropy ${\cal S}$ distributions (b,c,d) for the velocity protocols shown
in (a). Figure taken from \protect\cite{TreJarRitCroBusLip04}.}
\label{fig5}
\vspace{-0.2cm}
\end{figure}

These results have been experimentally tested for trapped beads
accelerated with different velocity protocols
\cite{TreJarRitCroBusLip04}. Some results are shown in Figure \ref{fig5}.

\subsection{A biological system: pulling biomolecules}
\label{pullbio}
The development of accurate instruments capable of measuring forces in
the piconewton range and extensions on the order of the nanometer gives
access to a wide range of phenomena in molecular biology and
biochemistry where nonequilibrium processes that involve small energies
on the order of a few $k_BT$ are measurable (see Sec.~\ref{small}). From
this perspective the study of biomolecules is an excellent playground to
explore nonequilibrium fluctuations. The most successful investigations
in this area have been achieved in single molecule experiments using
optical tweezers \cite{LanBlo03}.  In these experiments biomolecules can
be manipulated one at a time by applying mechanical force at their
ends. This allows us to measure small energies under varied conditions
opening new perspectives in the understanding of fundamental
problems in biophysics, e.g. the folding of biomolecules
\cite{TinBus99,ThiHye05,Finkelstein03}. The field of single molecule
research is steadily growing with new molecular systems being explored
that show nonequilibrium behavior characteristic of small systems. The reader interested
in a broader view of the area of single molecules research is suggested
to have a look at reference \cite{Ritort06}.

\subsubsection{Single molecule force experiments}
\label{bio:sme}
\begin{figure}
\begin{center}
\includegraphics[scale=0.5,angle=0]{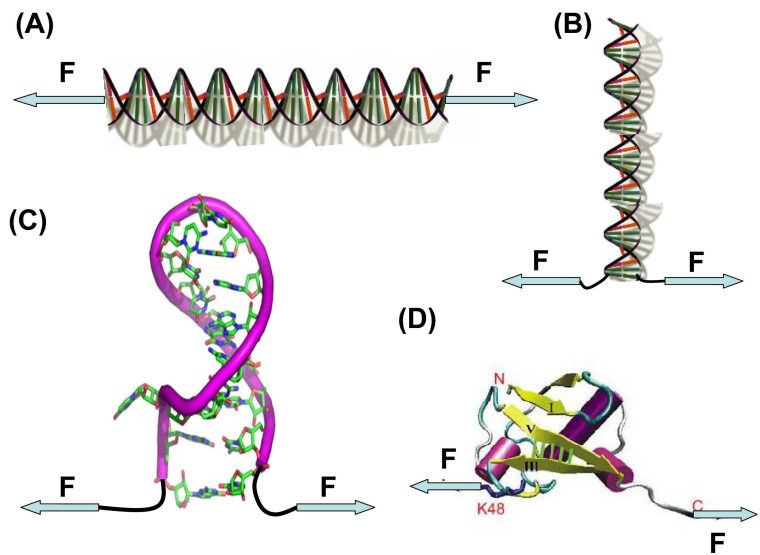}
\end{center}
\vspace{-0.4cm}
\caption{\em Pulling single molecules. (A) Stretching DNA; (B) Unzipping
DNA; (C) Mechanical unfolding of RNA and (D) Mechanical unfolding of proteins.}
\label{fig6}
\vspace{-0.2cm}
\end{figure}

In single molecule force experiments it is possible to apply force on
individual molecules by grabbing their ends and pulling them apart
\cite{BusMacWui00,StrAllCroBen01,StrDesChaDekAllBenCro03,BusChemForIzh04}. Examples
of different ways in which mechanical force is applied to single molecules are
shown in Figure \ref{fig6}.  In what follows we will consider single
molecule force experiments using optical tweezers, although everything we
will say extends to other force techniques (AFM, magnetic tweezers or
biomembrane force probe, see \cite{Ritort06}) In these experiments, the
ends of the molecule (for example DNA \cite{CalDre97}) are labeled with
chemical groups (e.g. biotin or digoxigenin) that can bind specifically
to their complementary molecular partners (e.g. streptavidin or
anti-digoxigenin respectively). Beads are then coated with the
complementary molecules and mixed with the DNA in such a way that a
tether connection can be made between the two beads through specific
attachments. One bead is in the optical trap and used as a force
sensor. The other bead is immobilized on the tip of a micropipette that
can be moved by using a piezo-controlled stage to which the
pipette is attached. The experiment consists in measuring force-extension curves
(FECs) by moving the micropipette relative to the trap
position \cite{SmiCuiBus03}. In this way it is possible to investigate
the mechanical and elastic properties of the DNA molecule
\cite{BusSmiLipSmi00,MarCoc03}.

Many experiments have been carried out by using this setup: the stretching
of single DNA molecules, the unfolding of RNA molecules or proteins and the
translocation of molecular motors (Figure \ref{fig2}). Here we focus our attention in force
experiments where mechanical work can be exerted on the molecule and
nonequilibrium fluctuations measured. The most successful studies along
this line of research are the stretching of small domain molecules such
as RNA \cite{LipOnoSmiTinBus01} or protein
motifs\cite{CecShaBusMar05}. Small RNA domains consist of a few tens of
nucleotides folded into a secondary structure that is further stabilized by
tertiary interactions. Because an RNA molecule is too small to be
manipulated with micron-sized beads, it has to be inserted between
molecular handles. These act as polymer spacers that avoid
non-specific interactions between the bead and the molecule as well as
the contact between the two beads.

\begin{figure}
\begin{center}
\includegraphics[scale=0.7,angle=0]{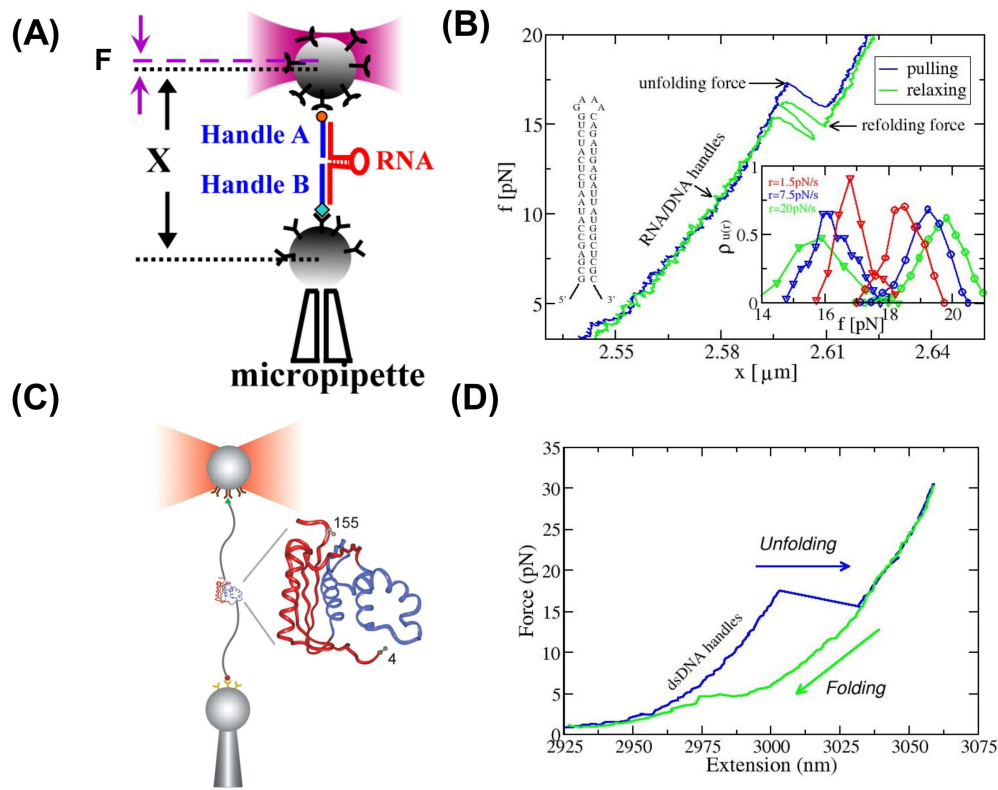}
\end{center}
\vspace{-0.4cm}
\caption{\em Mechanical unfolding of RNA molecules (A,B) and proteins
(C,D) using optical tweezers. (A) Experimental setup in RNA pulling experiments. (B) Pulling cycles
in the caconical hairpin CD4 and force rip distributions during the
unfolding and refolding at three pulling
speeds. (C) Equivalent setup in proteins. (D) Force extension curve when
pulling the protein RNAseH. Figure (B) taken from
\protect\cite{ManColRit06}. Figures (A,D) are a courtesy from C. Cecconi \protect\cite{CecShaBusMar05}.}
\label{fig7}
\vspace{-0.2cm}
\end{figure}

The basic experimental setup is shown in Figure \ref{fig7}. We also show
a typical FEC for an RNA hairpin and a protein. Initially the FEC shows
an elastic response due to the stretching of the molecular
handles. Then, at a given value of the force, the molecule under study
unfolds and a rip is observed in both force and extension. The rip
corresponds to the unfolding of the small RNA/protein molecule. The
molecule is then repeatedly stretched and relaxed starting from the
equilibrated native/extended state in the pulling/relaxing process.  In
the pulling experiment the molecule is driven out of equilibrium to a
NETS by the action of a time dependent force. The unfolding/refolding
reaction is stochastic, the dissociation/formation of the molecular
bonds that maintain the native structure of the molecule is overdamped
by the Brownian motion of the surrounding water molecules
\cite{CocMarMon02}.  Every time the molecule is pulled different
unfolding and refolding values of the force are observed (inset of
Figure~\ref{fig7}B). The average value of the force at which the
molecule unfolds during the pulling process increases with the loading
rate (roughly proportional to the pulling speed) in a logarithmic way as
expected for a two-state process (see the discussion at the end of
Sec.~\ref{pw:example} and equation \eq{pw:10a}).

Single molecule pulling experiments can be described with the formalism
developed in Sec.~\ref{applications:NETS}. In the simplest setting the
configurational variable ${\cal C}$ corresponds to the molecular
extension of the complex (handles plus inserted molecule) and the
control parameter $\lambda$ is either the force $f$ measured in the bead
or the molecular extension of the system, $x$. For small enough systems
the thermodynamic equation of state is dependent on what is the variable
that is externally controlled \cite{KelSwiBus03}. In the actual
experiments, the assumption that either the force or the extension are
controlled is just an approximation. Neither the molecular extension or
the force can be really controlled in optical tweezers
\cite{ManRit05}. For example, in order to control the force a feedback
mechanism must operate at all times. This feedback mechanism has a time
delay response so the force is never really constant
\cite{Man06a,Man06b}.  By assuming a constant force we are neglecting
some corrections in the analysis\footnote{By using two traps, it is
possible to maintain a constant force \cite{GreWooAbbBlo05}. This is
also possible with magnetic tweezers. However, because of the low
stiffness of the magnetic trap, the spatial resolution along individual
trajectories is limited by thermal fluctuations to 10 nanometers or
more.}. Under some conditions these corrections are shown to be
unimportant (see below). Let us now consider that the force acting on
the inserted molecule is controlled (the so called isotensional
ensemble).  For a molecule that is repeatedly pulled from a minimum
force value, $f_{\rm min}$, up to a maximum force, $f_{\rm max}$, the
work \eq{applications:NETS:7aa} along a given path is given by,
\be
W_f(\Gamma)=\int_{f_{\rm min}}^{f_{\rm max}}df\frac{\partial
E(x,f)}{\partial f}=-\int_{f_{\rm min}}^{f_{\rm max}}x(f)df
\label{bio:sm:1}
\ee
where the energy function is given by $E(x,f)=E(x,0)-fx$, $E(x,0)$
being the energy function of the molecule at zero force.  The subindex $f$
in $W_f$ is written to underline the fact that we are considering the
isotensional case where $f$ is the control parameter. The Jarzynski
equality \eq{applications:NETS:9} and the FT \eq{applications:NETS:10}
hold with $\Delta F$ equal to the free energy difference between the
initial and final equilibrium states. We assume that the molecule is
immersed in water at constant temperature $T$, pressure $p$ and acted
by a force $f$. The thermodynamic free energy $F(T,p,f)$ in this
description is the Legendre transform of the Gibbs free energy at zero
force, ambient temperature $T$ and pressure $p$, $G(T,p)$ \cite{TinBus02,Tin04},
\be
F(T,p,f)=G(T,p)-fx(T,p,f)\rightarrow x(T,P,f)=-\frac{\partial
F(T,p,f)}{\partial f}~~~~.
\label{bio:sm:2}
\ee
We are interested in knowing the Gibbs free energy difference at zero
force, $\Delta G$, rather than the free energy difference $\Delta F$ between
the folded state at $f_{\rm min}$ and the unfolded-extended state at $f_{\rm max}$. We can express
\eqq{applications:NETS:9}{applications:NETS:10} in terms of $G$ (rather
than $F$) and define the corrected work $W^c_f(\Gamma)$ along a path,
\be W^c_f(\Gamma)=W_f(\Gamma)+\Delta(xf)=W_f(\Gamma)+(x_{\rm max}f_{\rm
max}-x_{\rm min}f_{\rm min})
\label{bio:sm:3a}
\ee
where the extensions $x_{\rm min},x_{\rm max}$ are now fluctuating
quantities evaluated at the initial and final times along each pulling.  The corrected work
$W^c_f(\Gamma)$ includes an additional boundary term and therefore does
not satisfy neither the JE and the CFT. If we now consider that $x$ is
the control parameter then we can define the equivalent of \eq{bio:sm:1}
(the so-called isometric ensemble),
\be
W_x(\Gamma)=W_f(\Gamma)+\Delta(xf)=\int_{x_{\rm min}}^{x_{\rm
max}}f(x)dx
\label{bio:sm:3b}
\ee
where now $x$ is controlled and $x_{\rm min},x_{\rm max}$ are fixed by
the pulling protocol.  Equations \eqq{bio:sm:3a}{bio:sm:3b} look
identical, however they refer to different experimental
protocols. Note that the term $W_f(\Gamma)$ appearing in \eq{bio:sm:3b} is now evaluated between the
initial and final forces at fixed initial and final times. Both works
$W_x,W_f$ satisfy the relations
\eqq{applications:NETS:9}{applications:NETS:10}. For a reversible
process where $f$ is controlled we have $W_f^{\rm rev}=\Delta F$ whereas
if $x$ is controlled we have $W_x^{\rm rev}=\Delta G$.  In experiments
it is customary to use \eq{bio:sm:3b} for the work. First, because that
quantity is more easily recognized as the mechanical work. Second,
because it gives the free energy difference between the folded and the
unfolded states at zero force, a quantity that can be compared with the
value obtained from thermal denaturation experiments.

\begin{figure}
\begin{center}
\includegraphics[scale=0.3,angle=-90]{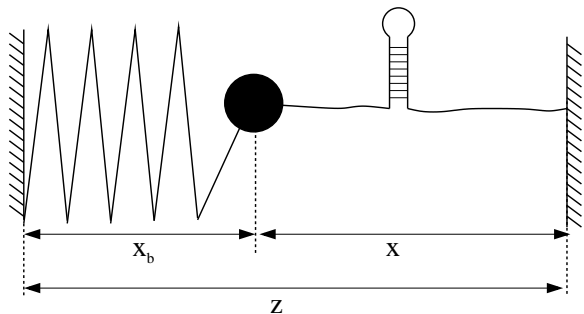}
\end{center}
\vspace{-0.4cm}
\caption{\em A molecular system of extension $x$ is connected at its
leftmost end to a bead trapped in an optical well (or to the tip of an
AFM cantilever) and at its rightmost end to an immobilized surface (or a
bead fixed to the tip of micropipette). The position of the bead
relative to the center of the trap, $x_b$, gives a read out of the
acting force $f=\kappa x_b$. The control parameter in this setup is
$z=x_b+x$ whereas both $x_b$ and $x$ are fluctuating quantities.}
\label{fig8}
\vspace{-0.2cm}
\end{figure}

In general neither the force or the molecular extension can be
controlled in the experiments so both definitions
\eqqq{bio:sm:1}{bio:sm:3a}{bio:sm:3b} result into approximations to the
{\em true} mechanical work that satisfies
\eqq{applications:NETS:9}{applications:NETS:10}. The control parameter
in single molecule experiments using optical tweezers is the distance
between the center of trap and the immobilized bead
\cite{ManRit05}. Both the position of the bead in the trap and extension
of the handles are fluctuating quantities. It has been observed
\cite{HumSza00,Jarzynski01,SchFuj03} that in pulling experiments the
proper work that satisfies the FT includes some corrections to the
expressions \eqq{bio:sm:2}{bio:sm:3b} mainly due to the effect of the
trapped bead. There are two considerations to take into account when
analyzing experimental data:

\begin{itemize}

\item{\bf $W_x$ or $W_f$?} Let us suppose that $f$ is the control
parameter. In this case the JE and CFT
\eqq{applications:NETS:9}{applications:NETS:10} are valid for the work
\eq{bio:sm:1}. How large is the error that we make when we apply the JE
using $W_x$ instead? This question has been experimentally addressed by
Ciliberto and coworkers \cite{Douarche05a,Douarche05b} who measured the
work in an oscillator system with high precision (within tenths of
$k_BT$). As shown in \eq{bio:sm:3b} the difference between both works is
mainly a boundary term, $\Delta(xf)$. Fluctuations of this term can be a
problem if they are on the same order than fluctuations of $W_x$
itself. For an harmonic oscillator of stiffness constant equal to
$\kappa$, the variance of fluctuations in $fx$ are equal to $\kappa
\delta(x^2)$, i.e.  approximately on the order of $k_BT$ due to the
fluctuation-dissipation relation. Therefore, for experimental
measurements that do not reach such precision, $W_x$ or $W_f$ are
equally good.

\item{\bf The effect of the bead or cantilever.} Hummer and Szabo
\cite{HumSza00} have analyzed the effect of a force sensor attached to
the system (i.e. the bead in the optical trap or the cantilever in the
AFM) in the work measurements. To this end they consider a simplified
model of the experimental setup (Figure~\ref{fig8}). In such model the
molecular system (that includes the molecule of interest -RNA or
protein- and the handles) is connected to a spring (that models the
trapped bead or the AFM cantilever) and the whole system is embedded in
a thermal bath. The total extension of the molecular system is $x$ but
the control parameter is $z=x+x_b$ where $x_b$ is the position of the
bead respect to the center of the trap. The total free energy of the
system is given by $F(x)+(1/2)\kappa x_b^2$ where $F(x)$ is the free
energy of the molecular system alone and $\kappa$ is the stiffness of
the trap. The molecular extension $x$ and the distance $x_b$ are related
by the force balance equation,
\be
f=\kappa x_b=\frac{\partial F(x)}{\partial x}
\label{bio:sm:4}
\ee
where we assume that the bead is locally equilibrated at all values of
the nonequilibrium molecular extension $x$ (this is a good approximation
if, as often occurs, the bead relaxes fast enough compared to the
typical time for the unfolding/refolding of the molecule). The
mechanical work \eq{applications:NETS:7aa} is then given by,
\be
W(\Gamma)=\int_{z_{\rm min}}^{z_{\rm max}}fdz=W_x(\Gamma)+\Delta\Bigl(\frac{f^2}{2\kappa} \Bigr)
\label{bio:sm:5}
\ee
where we used $dz=dx_b+dx$ and \eq{bio:sm:4}. The difference between
the proper work $W$ and $W_x$ is again a boundary term. Because $z$ is
the control parameter the JE 
and the CFT are valid for the work $W$ but not 
for $W_x$. Again, the FT will not hold if fluctuations in the
boundary term are important.  The variance of these fluctuations
is given by,
\be
\langle\delta\Bigl(\frac{f^2}{2\kappa}\Bigr)\rangle\sim
\frac{k_BT\kappa}{\kappa_x+\kappa}\le k_BT 
\label{bio:sm:6}
\ee
where $\kappa_x$ is the stiffness of the molecular system \cite{GerBunHwa01,ManRit05}. Usually, $\kappa_x\gg
\kappa$ so fluctuations in the boundary term are again smaller than $k_BT$.
In general, as a rule of thumb, we can say that it does not matter
much which mechanical work we measure if we do not seek for free energy
estimates with an accuracy less than $k_BT$. This is true unless the bead (cantilever)
does not equilibrate within the timescale of the experiments. This
may be the case when $\kappa$ is too low and \eq{bio:sm:4} is not applicable.

\end{itemize}

\subsubsection{Free energy recovery}
\label{bio:free}
As we already emphasized the JE \eq{applications:NETS:9} and the FT
\eq{applications:NETS:10} can be used to predict free energy
differences. In single molecule experiments it is usually difficult to
pull molecules in a reversible way due to drift effects in the
instrument. It is therefore convenient to devise nonequilibrium methods
(such as the JE or the CFT) to extract equilibrium free energies
differences from data obtained in irreversible processes. The first
experimental test of the JE was carried out by pulling RNA hairpins that
are obtained as derivatives of the L21 Tetrahymena ribozyme
\cite{LipDumSmiTinBus02}. In these experiments RNA molecules were pulled
not too fast, the average dissipated work in such experiments was less
$6k_BT$ and the work distributions turned out to be approximately
Gaussian. Recent experiments have studied RNA molecules that are driven
farther away from equilibrium in the non-linear regime. In such
non-linear regime the average dissipated work is non-linear with the
pulling speed \cite{RitBusTin02} and the work distribution strongly
deviates from a Gaussian \cite{ColRitJarSmiTinBus05}. In addition, these
experiments have provided the first experimental test of the CFT
\eq{applications:NETS:10}. These measurements have also shown the
possibility to recover free energy
differences by using the CFT with larger accuracy than that obtained by using the
JE. There are two main predictions of the CFT \eq{applications:NETS:10}
that have been scrutinized in these experiments,
\begin{itemize}

\item{\bf Forward and reverse work distributions cross at
$W=\Delta G$.} In order to obtain $\Delta G$ we can measure the forward
and reverse work distributions $P_F(W),P_R(-W)$ and look at the work
value $W^*$ where they cross, $P_F(W^*)=P_R(-W^*)$. According
to \eq{applications:NETS:10} both distributions should cross at
$W^*=\Delta G$ independently of how far the system is driven out of
equilibrium (i.e. independently of the pulling speed). Figure \ref{fig9} shows
experiments on a short canonical RNA hairpin CD4 (i.e. just containing
Watson-Crick complementary base pairs) at three different pulling speeds which agree
very well with the FT prediction.

\begin{figure}
\begin{center}
\includegraphics[scale=0.7,angle=0]{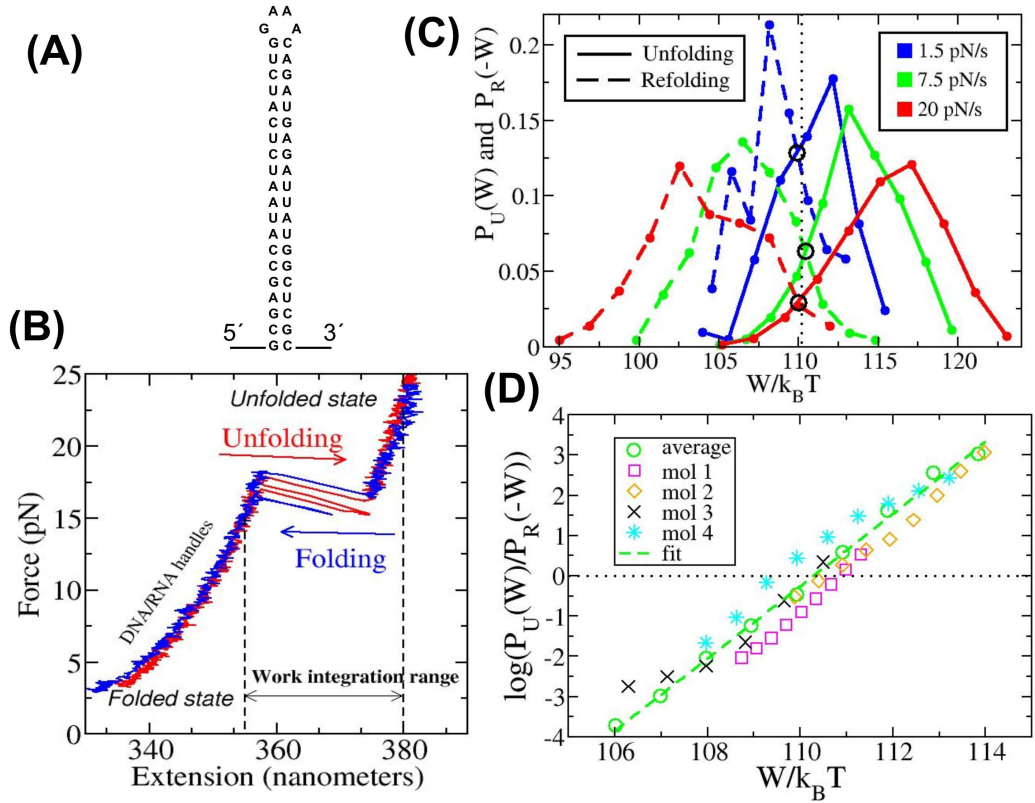}
\end{center}
\vspace{-0.4cm}
\caption{\em (A) Structure of the canonical CD4 hairpin. (B) FECs at a
loading rate of $1.7$pN/s. (C) Unfolding and refolding work
distributions at three loading rates (see inset). The unfolding and
refolding work distributions cross at a value $\Delta G$ independent of
the pulling speed as predicted by the CFT. Data correspond to 100,400
and 700 pulls for the lowest, medium and highest pulling speeds
respectively. (D) Test of the CFT at the intermediate loading rate
$7.5$pN/s for 4 different tethers. The trend of the data is reproducible
from tether to tether and consistent with the CFT prediction. Figures
taken from \protect\cite{ColRitJarSmiTinBus05}.}
\label{fig9}
\vspace{-0.2cm}
\end{figure}

\item{\bf Verification of the CFT.} The CFT \eq{applications:NETS:10}
can be tested by plotting $\log(P_F(W)/P_R(-W))$ as a function of
$W$. The resulting points should fall in a straight line of slope 1 (in
$k_BT$ units) that intersects the work axis at $W=\Delta G$. Of course,
this relation can be tested only in the region of work values along the
work axis where both distributions (forward and reverse) overlap. An
overlap between the forward and reverse distributions is hardly observed
if the molecules are pulled too fast or if the number of pulls is too
small. In such cases, other statistical methods (Bennet's acceptance
ratio or maximum likelihood methods, Sec.~\ref{bio:accep}) can be
applied to get reliable estimates of $\Delta G$.  The validity of the
CFT has been tested in the case of the RNA hairpin CD4 previously mentioned and
the three way junction RNA molecule as well. Figures~\ref{fig9}C,D and
\ref{fig10}C show results for these two molecules.

\end{itemize}

\begin{figure}
\begin{center}
\includegraphics[scale=0.7,angle=0]{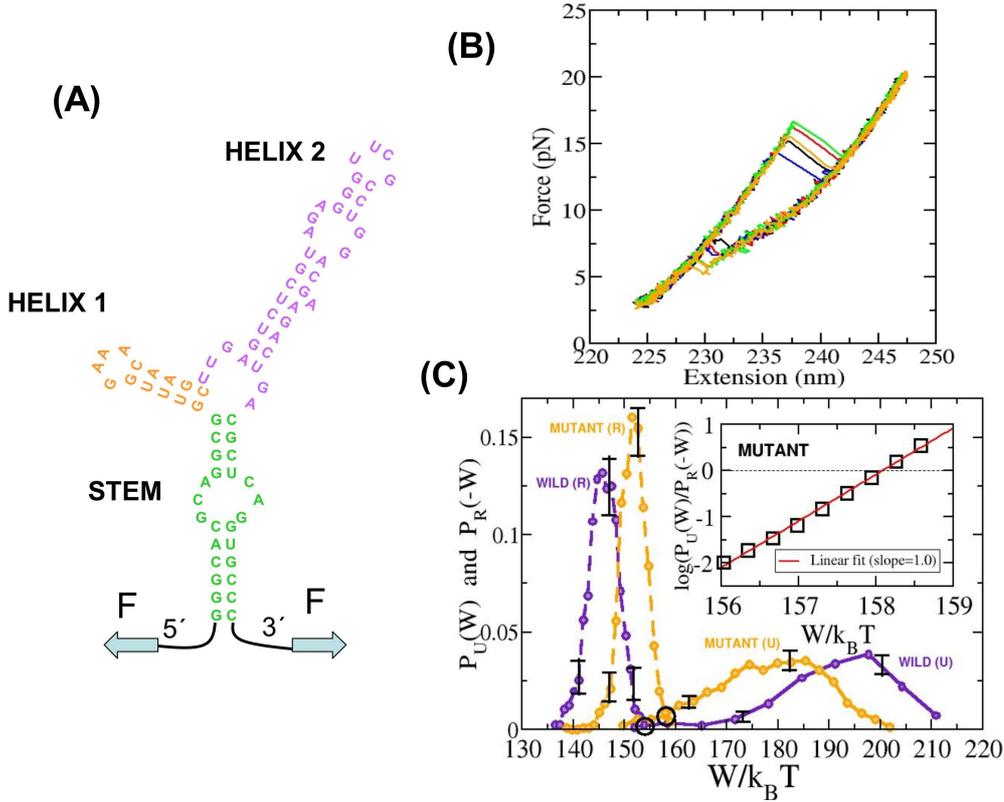}
\end{center}
\vspace{-0.4cm}
\caption{\em (A) Secondary structure of the three way junction S15. (B) A few FECs
for the wild type. (C) Unfolding/refolding work distributions for the
wild type and the mutant. (Inset) Experimental verification of the
validity of the CFT for the mutant where unfolding and refolding
distributions overlap each other over a limited range of work
values. Data correspond to 900 pulls for the wild type and 1200 pulls
for the mutant. Figure
taken from \protect\cite{ColRitJarSmiTinBus05}.}
\label{fig10}
\vspace{-0.2cm}
\end{figure}

In general, both the JE and the CFT are only valid in the limit
of an infinite number of pulls. For a finite number
of pulls, $N$, the estimated value for $\Delta G$ that is obtained by applying the JE is
biased \cite{GorRitBus03}. The free-energy estimate $F^{\rm JE}_k$ for a
given set, $k$, of $N$ work
values $W_1^k,W_2^k,...,W_N^k$ is defined as, 
\be
F^{\rm JE}_k=-T\log\Bigl(\frac{1}{N}\sum_{i=1}^N \exp(-\frac{W_i^k}{T})\Bigr)~~~~.
\label{bio:sm:7}
\ee
The free-energy bias is defined by averaging the estimator $F^{\rm
JE}_k$ over an infinite number of sets, 
\be
B(N)=\lim_{M\to \infty}\Bigl[\frac{1}{M}\sum_{k=1}^MF^{\rm JE}_k\Bigr]-\Delta F
\label{bio:sm:8}
\ee
where $\Delta F$ is the true free energy difference. The bias $B(N)$
converges to 0 for $N\to\infty$. However, it is of practical importance
to devise methods to estimate how many pulls are required to obtain the
Jarzynski free energy estimate $F^{\rm JE}$ within a reasonable error
far from the true value \cite{WooMuhTho91,RitBusTin02}. The bias is a
complicated mathematical object because the Jarzynski average catches
important contributions from large deviations of the work.  As we will
see later in Sec.~\ref{pw:bias}, the bias is a large deviation function
that requires specific mathematical methods to analyze its finite $N$
behavior and large $N$ convergence. There we proof that, for large $N$,
the bias decreases as $1/N$, a result that is known as Woods formula
\cite{WooMuhTho91}. In the intermediate $N$ regime the behavior of the
bias is more complicated \cite{ZucWol02b}. Free-energy recovery
techniques are also used in numerical simulations to evaluate
free-energy differences
\cite{IsrGaoSch01,JenParTajSch02,ParKhaTajSch03,AndDinKar03} and reconstruct
free-energy profiles or potentials of mean-field force
\cite{ParSch04,HumSza05}.

\subsubsection{Efficient strategies and numerical methods}
\label{bio:accep}
An important question is to understand the optimum nonequilibrium
protocol to recover free-energies using the JE given specific constraints in
experiments and simulations. There are several considerations to take
into account,

\begin{itemize}
\item{\bf Faster or slower pulls?.} In single molecule experiments
tethers break often so it is not possible to repeatedly pull the same
tether an arbitrary number of times. Analogously, in numerical
simulations only a finite amount of computer time is available and only
a limited number of paths can be simulated. Given these limitations, and
for a given amount of available time, is it better to perform many fast
pulls or a few number of slower pulls to recover the free energy
difference using the JE?. In experiments, drift effects in the
instrument always impose severe limitations to the minimum speed at
which molecules can be pulled. To obtain good quality data it is
advisable to carry out pulls as fast as possible. In numerical
simulations the question about the best strategy for free energy
recovery has been considered in several papers
\cite{Hummer01,HenJar01,ZucWol02a,GorRitBus03}. The general conclusion
that emerges from these studies is that, in systems that are driven far
away from equilibrium, it is preferable to carry out many pulls at high
speed than a few number of them at slower speeds. The reason can be
intuitively understood. Convergence in the JE is dominated by the
so-called outliers, i.e. work values that deviate a lot from the average
work and are smaller than $\Delta F$. The outliers contribute a lot to
the exponential average \eq{applications:NETS:9}. For higher pulling
speeds we can perform more pulls so there are more chances to catch a
large deviation event, i.e. to catch an outlier. At the same time,
because at higher speeds the pulling is more irreversible, the average
dissipated work becomes larger making the free energy estimate less
reliable. However the contribution of the outliers that is required to recover
the right free-energy largely compensates the slow convergence due to
the increase of the average dissipated work. We should also mention that
periodically oscillating pulls have been also considered, however it is
unclear whether they lead to improved free energy estimates
\cite{BraHanSei04,ImpPel06}.

\item{\bf Forward or reverse process?.} Suppose we want to evaluate the
free energy difference between two states, $A$ and $B$, by using the
JE. Is it better to estimate $\Delta F$ by carrying out irreversible
experiments from $A$ to $B$, or is it better to do them from $B$ to $A$?
Intuitively it seems natural that the less irreversible process among
the two (forward and reverse), i.e. the one with smaller dissipated work
$W_{\rm diss}$, is also the most convenient in order to
extract the free energy difference. However this
is not the case. In general, a larger average dissipate work implies a
larger work variance \eq{applications:NETS:11}, i.e. larger
fluctuations.  The larger are the fluctuations, the larger the
probability to catch a large deviation that contributes to the
exponential average. It seems reasonable that, if outliers contribute
much more to obtain the correct free energy difference than proper tuning of the average
value of the work does, then the process that more fluctuates, i.e. the more
dissipative one, is the preferred process in order to efficiently
recover $\Delta F$. This result was anticipated in \cite{Ritort04} and
analyzed in more detail in \cite{Jarzynski06}. For Gaussian work
distributions, the minimum number of pulls, $N^*$, required to efficiently recover free
energy differences within $1k_BT$ by using the JE grows exponentially
with the dissipated work along the nonequilibrium process \cite{RitBusTin02}. However, for general
work distributions, the value of $N^*_{F(R)}$ along the forward (reversed)
process depends on the
average dissipated work along the reverse (forward) process \cite{Jarzynski06}. This
implies that,
\be
N^*_{F(R)}\sim \exp\Bigl(\frac{W_{\rm diss}^{R(F)}}{T} \Bigr)
\label{bio:accep:0}
\ee
and the process that dissipates most between the forward and the reverse
is the most convenient to efficiently recover $\Delta F$.

\end{itemize}

\begin{figure}
\begin{center}
\includegraphics[scale=0.28,angle=-90]{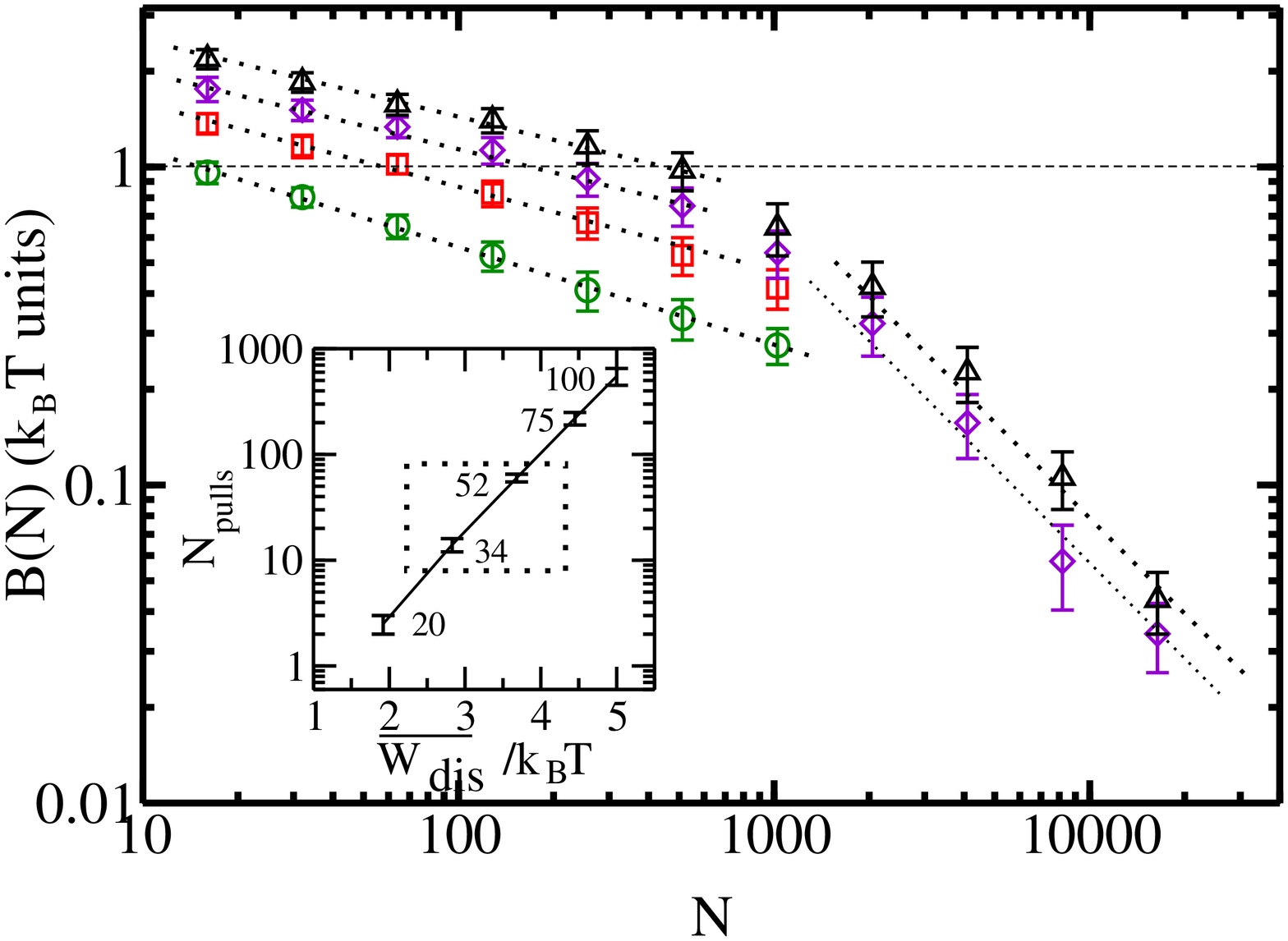}\includegraphics[scale=0.28,angle=-90]{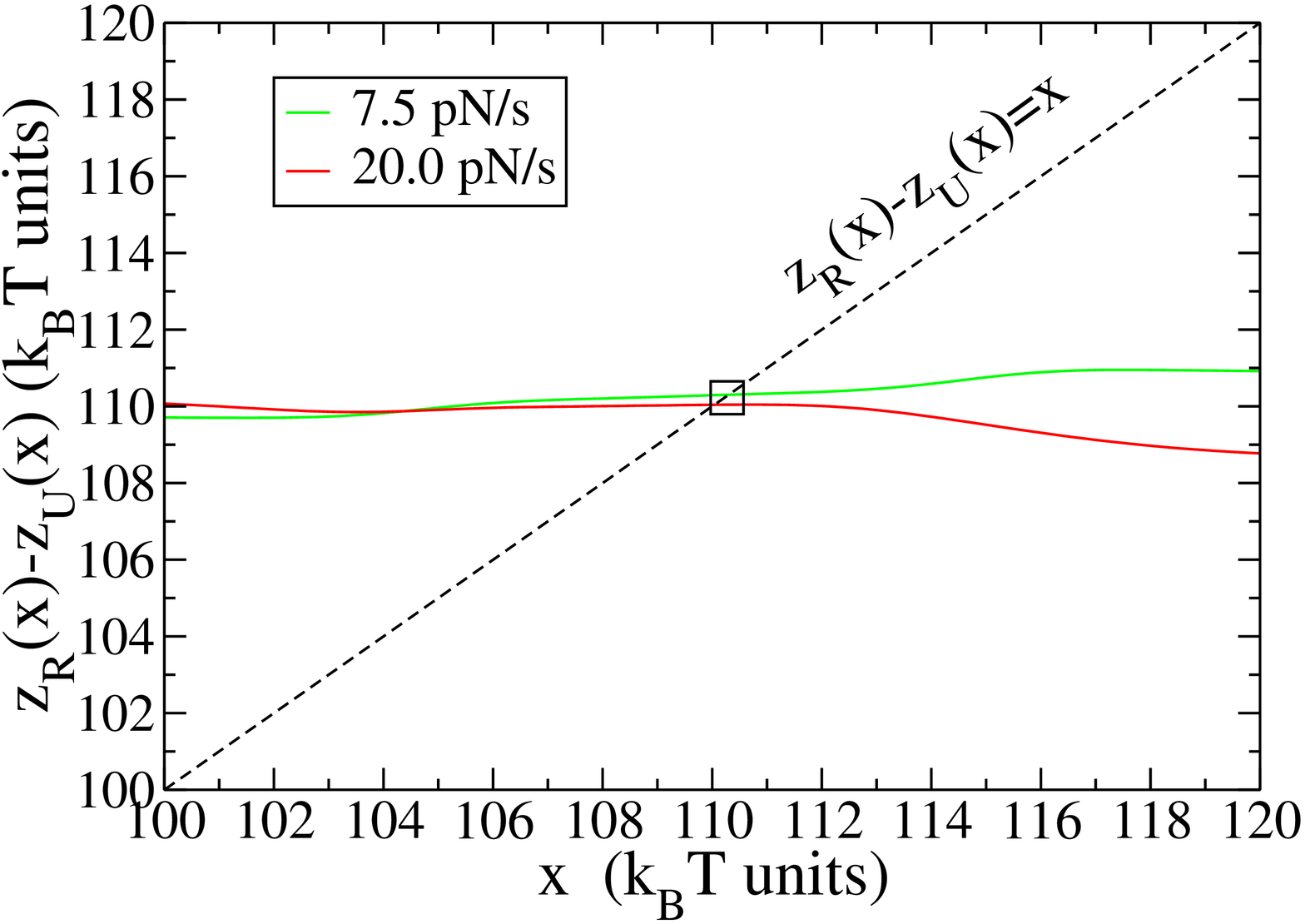}
\end{center}
\vspace{-0.4cm}
\caption{\em (Left) Bias as a function of the number of pulls $N$ for a two-states model. The
inset shows the number of pulls as a function of the dissipated work
required to recover the free-energy with an error within
$1k_BT$. (Right) Function $z_R-z_F$ for the data shown in
Figure~\ref{fig9}C at the two largest pulling speeds. Left figure 
taken from \protect\cite{RitBusTin02,Ritort03}. Right figure taken from
the supplementary material in Ref.\cite{ColRitJarSmiTinBus05}.}
\label{fig11}
\vspace{-0.2cm}
\end{figure}

Until now we discussed strategies to recover free energy differences
using the JE. We might be interested in free energy recovery by
combining the forward and reverse distributions at the same time that we
use the CFT.  This is important in either experiments
\cite{ColRitJarSmiTinBus05} or simulations
\cite{KosBarJan05,DelFabCov06} where it is convenient and natural to use
data from the forward and reverse process. The best strategy to
efficiently recover free energies using the forward and reverse
processes was proposed by C. Bennett long time ago in the context of
equilibrium sampling \cite{Bennett76}. The method has been later
extended by Crooks to the nonequilibrium case \cite{Crooks00} and is
known as Bennett's acceptance ratio method. The basis of the method is
as follows. Let us multiply both sides of $\eq{applications:NETS:10}$ by
the function $g_{\mu}(W)$,
\be
g_{\mu}(W)\exp\Bigl(-\frac{W}{T}\Bigr) P_F(W)=g_{\mu}(W)P_R(-W)\exp\Bigl(-\frac{\Delta F}{T}\Bigr)
\label{bio:accep:1}
\ee
where $g_{\mu}(W)$ is an arbitrary real function that depends on the
parameter $\mu$. Integrating both sides between $W=-\infty$ and
$W=\infty$ gives,
\be
\langle  g_{\mu}(W)\exp\Bigl(-\frac{W}{T}\Bigr)\rangle_F=\langle
g_{\mu}(W)\rangle_R\exp\Bigl(-\frac{\Delta F}{T}\Bigr)
\label{bio:accep:2}
\ee
where $\langle...\rangle_{(F,R)}$ denote averages over the forward and
reverse process. Taking the logarithm at both sides we have,
\be
z_R(\mu)-z_F(\mu)=\frac{\Delta F}{T}
\label{bio:accep:2b}
\ee
where we have defined,
\bea
z_R(\mu)=\log\Bigl(\langle g_{\mu}(W)\rangle_R\Bigr)\label{bio:accep:2c1}\\
z_F(\mu)=\log \Bigl( \langle  g_{\mu}(W)\exp\Bigl(-\frac{W}{T}\Bigr)\rangle_F\Bigr)~~~~~.
\label{bio:accep:2c2}
\eea
Equation \eq{bio:accep:2b} implies that the difference between both
functions $z_F,z_R$ must be a constant over all $\mu$ values. The question we
would like to answer is the following. Given a finite number of forward
and reverse pulls, what is the optimum choice for $g_{\mu}(W)$ that
gives the best estimate \eq{bio:accep:2b} for $\Delta F$? For a finite
number of experiments $N_F,N_R$ along the forward and reverse process we
can write for any observable $A$,
\be
\langle A(W) \rangle_{F(R)}=\frac{1}{N_{F(R)}}\sum_{i=1}^{N_{F(R)}}A(W_i)~~~·
\label{bio:accep:3}
\ee
Equation \eq{bio:accep:2} yields an estimate
for $\Delta F$,
\be
(\Delta F)^{\rm est}=T\Bigl(\log\Bigl( \langle
g_{\mu}(W)\rangle_R\Bigr)-\log\Bigl(\langle  g_{\mu}(W)\exp\Bigl(-\frac{W}{T}\Bigr)\rangle_F\Bigr)\Bigr)~~~~.
\label{bio:accep:4}
\ee
Minimization of the variance, 
\be
\sigma_{\Delta F}^2=\langle\Bigl( (\Delta F)^{\rm est}-\Delta F\Bigr)^2\rangle
\label{bio:accep:5}
\ee
($\langle..\rangle$ denotes the average over the distributions $P_F,P_R$)
respect all possible functions $g_{\mu}(W)$ shows \cite{Bennett76,Crooks00} that the
optimal solution is given by,
\be
g_{\mu}(W)=\frac{1}{1+\frac{N_F}{N_R}\exp\Bigl(\frac{W-\mu}{T}\Bigr)}
\label{bio:accep:6}
\ee
with $\mu=\Delta F$. The same result has been obtained by Pande and
coworkers by using maximum likelihood methods \cite{ShiBaiHooPan03}. In this case, one starts from a
whole set of work data encompassing $N_F$ forward and $N_R$ reversed
values. One then defines the likelihood function of distributing all
work values between the forward and reverse sets. Maximization of the
likelihood leads to Bennett's acceptance ratio formula. To
extract $\Delta F$ is then customary to plot the difference in
the lhs of \eq{bio:accep:2b}, $z_R(\mu)-z_F(\mu)$, as a function of $\mu$
by using \eqq{bio:accep:2c1}{bio:accep:2c2}. The
intersection with the line $z_R(\mu)-z_F(\mu)=\mu$ gives the best
estimate for $\Delta F$. An example of this method is shown in
Figure \ref{fig11}. Recently, the maximum likelihood method has been generalized
to predict free energy differences between more than two states \cite{MarSpiKar05}.

\section{Path thermodynamics}
\label{path}
\subsection{The general approach}
\label{path:strategy}
The JE \eq{applications:NETS:9} indicates a way to
recover free energy differences by measuring the work along all possible paths that start
from an equilibrium state.  Its mathematical form reminds a lot of the
partition function in the canonical ensemble used to compute free energies in  statistical
mechanics. The formulae for the two cases read,
\bea
\sum_{{\cal C}}\exp\bigl(-\frac{E({\cal
C})}{T}\bigr)=\exp(-\frac{F}{T})~~~~{\rm (partition~~function)}
\label{path:1a}\\
\sum_{\Gamma}\exp\bigl(-\frac{W(\Gamma)}{T}\bigr)=\exp(-\frac{\Delta
F}{T})~~~~{\rm (Jarzynski~~equality)}\label{path:1b}
\eea
where $F$ is the equilibrium free energy of the system at temperature
$T$.  Throughout this section we take $k_B=1$.  In the canonical
ensemble the entropy $S(E)$ is equal to the logarithm of the density
of states with a given energy $E$. That density is proportional to the
number of configurations with energy equal to $E$. Therefore
\eq{path:1a} becomes,
\be
\exp\Bigl(-\frac{F}{T}\Bigr)=\sum_{{\cal C}}\exp\Bigl(-\frac{E({\cal
C})}{T}\Bigr)=\sum_{E}\exp\Bigl(S(E)-\frac{E}{T}\Bigr)=\sum_{E}\exp\Bigl(-\frac{\Phi(E)}{T}\Bigr)
\label{path:2}
\ee
where $\Phi(E)=E-TS$ is the free-energy functional. In the large volume
limit the sum in \eq{path:2} is dominated by the value $E=E^{\rm eq}$ where
$F(E)$ is minimum. The value $E^{\rm eq}$ corresponds to the equilibrium energy
of the system and $\Phi(E^{\rm eq})$ is the equilibrium free
energy. The following chain of relations hold,
\be
F=\Phi(E^{\rm eq})~~~;~~~\Bigl(\frac{\partial \Phi(E)}{\partial
E}\Bigr)_{E=E^{\rm eq}}=0\rightarrow \Bigl(\frac{\partial S(E)}{\partial E}\Bigr)_{E=E^{\rm eq}}=\frac{1}{T}
\label{path:3}
\ee
The equilibrium energy $E^{\rm eq}$ is different from the most probable
energy, $E^{\rm mp}$, defined by $S'(E=E^{\rm mp})=0$. $E^{\rm mp}$ is
the average energy we would find if we were to randomly select
configurations all with identical {\em a priori} probability. The
equilibrium energy, rather than the most probable energy, is the
thermodynamic energy for a system in thermal equilibrium.

Proceeding in an similar way for the JE we can define
the {\em path entropy} $S(W)$ as the logarithm of the density of paths with
work equal to $W$, $P(W)$,
\be
P(W)=\exp(S(W))~~~~.
\label{path:3b}
\ee
We can rewrite \eq{path:1b} in the following way,
\bea
\exp\Bigl(-\frac{\Delta
F}{T}\Bigr)=\sum_{\Gamma}\exp\Bigl(-\frac{W(\Gamma)}{T}\Bigr)=\int dW
P(W)\exp\Bigl(-\frac{W}{T}\Bigr)=\nonumber\\
\int dW \exp\Bigl(S(W)-\frac{W}{T}\Bigr)=\int dW \exp \Bigl(-\frac{\Phi(W)}{T}\Bigr)
\label{path:4}
\eea
where $\Phi(W)=W-TS(W)$ is the {\em path free energy}. In the large volume
limit the sum in \eq{path:4} is dominated by the work value, $W^{\dag}$,
where $\Phi(W)$ is minimum. Note that the value $W^{\dag}$ plays the
role of the equilibrium energy in the canonical case \eq{path:3}. From
\eq{path:4} the path free energy $\Phi(W^{\dag})$ is equal to the free energy difference
$\Delta F$. The following chain of relations hold,
\bea
\Delta F=\Phi(W^{\dag})=W^{\dag}-TS(W^{\dag})\label{path:5a}\\
\Bigl(\frac{\partial \Phi(W)}{\partial
W}\Bigr)_{W=W^{\dag}}=0\rightarrow \Bigl(\frac{\partial S(W)}{\partial W}\Bigr)_{W=W^{\dag}}=\frac{1}{T}
\label{path:5b}
\eea
At the same time, $W^{\dag}$ is different from the most probable
work, $W^{\rm mp}$, defined as the work value at which $S(W)$ is
maximum, 
\be \Bigl(\frac{\partial S(W)}{\partial W}\Bigr)_{W=W^{\rm
mp}}=0\rightarrow \Bigl(\frac{\partial \Phi(W)}{\partial
W}\Bigr)_{W=W^{\rm mp}}=1~~~~.
\label{path:6}
\ee
The role of $W^{\rm mp},W^{\dag}$ in the case of the JE \eq{path:1a} and
$E^{\rm mp},E^{\rm eq}$ in the partition function case \eq{path:1b}
appear exchanged. $W^{\rm mp}$ is the work value typically observed upon
repetition of the same experiment a large number of times. In contrast,
in the partition function case \eq{path:1a}, the typical energy is
$E^{\rm eq}$ rather than $E^{\rm mp}$. In addition,
$W^{\dag}$ is not the typical work but the work that must be sampled
along paths in order to be able to extract the free energy difference
using the JE. As we have already emphasized, as the system size
increases, less and less paths can sample the region of work values
around $W^{\dag}$. Therefore, although both formalisms (partition
function and JE) are mathematically similar, the physical meaning of the
quantities $W^{\dag}$ and $E^{\rm eq}$ is different. In the large volume
limit $E^{\rm eq}$ is almost {\em always} observed whereas $W^{\dag}$ is
almost {\em never} observed.

\begin{figure}
\begin{center}
\includegraphics[scale=0.35,angle=-90]{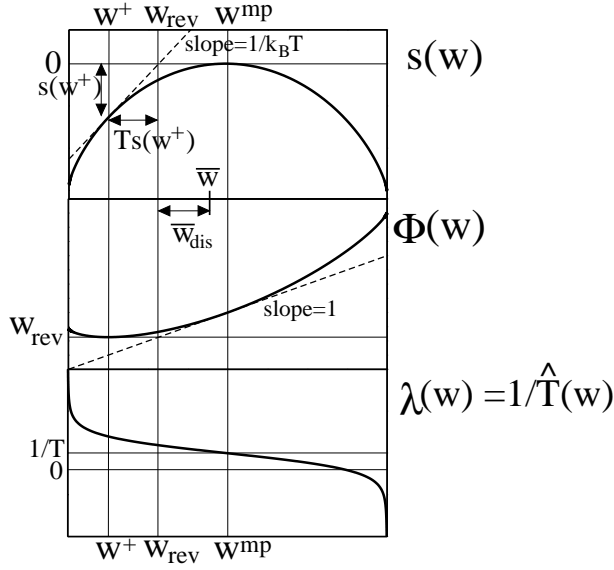}
\end{center}
\vspace{-0.4cm}
\caption{\em Upper panel: Path entropy $s(w)$. Middle panel: Path
free-energy $\Phi(w)=w-Ts(w)$. Lower panel: Lagrange multiplier
$\lambda(w)$ equal to the inverse of the path temperature
$1/\hat{T}(w)$.   $w^{\rm mp}$ is the most probable work value given by $s'(w^{\rm
mp})=\lambda(w^{\rm mp})=0$ or $\Phi'(w^{\rm mp})=1$; $w^{\dag}$ is the
value of the work that has to be sampled to recover free energies from
nonequilibrium work values using the JE. This is
given by $s'(w^{\dag})=1/T$ or $\Phi'(w^{\dag})=0$; 
$w_{\rm rev},\overline{w_{\rm dis}}$ are the reversible and average
dissipated work respectively. Figure taken from
\protect\cite{Ritort04}.}
\label{fig12}
\vspace{-0.2cm}
\end{figure}

In general, from the path entropy we can also define a {\em path temperature}, $\hat{T}(W)$,
\be
\frac{\partial S(W)}{\partial W}=\lambda(W)=\frac{1}{\hat{T}(W)} \rightarrow
\hat{T}(W^{\dag})=T~~~~,
\label{path:7}
\ee
where $\lambda(W)$ is a Lagrange multiplier that transforms the path
entropy $S(W)$ into the path free energy $\Phi(W)$, equation
\eq{path:5a}. The mathematical relations between the new quantities
$W^{\dag}$ and $W^{\rm mp}$ can be graphically represented for a given
path entropy $S(W)$. This is shown in Figure \ref{fig12}.

The path thermodynamics formalism allows us to extract some general
conclusions on the relation between $W^{\dag}$ and $W^{\rm mp}$. Let us
consider the CFT \eq{applications:NETS:10}. In terms of the path
entropies for the forward and reverse processes, $S_F(W)$ and $S_R(W)$, \eq{applications:NETS:10} can
be written as,
\be
S_F(W)-S_R(-W)=\frac{W-\Delta F}{T}\rightarrow (S_F)'(W)+(S_R)'(-W)=\frac{1}{T}
\label{path:7a}
\ee
where we used the definition \eq{path:3b} and later derived with respect
to $W$. By inserting $W=W_F^{\dag}$ and $-W_R^{\dag}$ in the rhs of
\eq{path:7a} and using \eqq{path:5b}{path:6} we obtain the following
chain of relations,
\bea
(S_F)'(W_F^{\dag})+(S_R)'(-W_F^{\dag})=\frac{1}{T}\rightarrow
(S_R)'(-W_F^{\dag})=0\rightarrow W_F^{\dag}=-W_R^{\rm mp}\label{path:7ba}\\
(S_F)'(-W_R^{\dag})+(S_R)'(W_R^{\dag})=\frac{1}{T}\rightarrow
(S_F)'(-W_R^{\dag})=0\rightarrow W_R^{\dag}=-W_F^{\rm mp}~~~.\label{path:7bb}
\eea
The rightmost equalities in \eqq{path:7ba}{path:7bb} imply that the most
probable work along the forward (reverse) process is equal to the work
value ($W^{\dag}$) that must be sampled, in a finite number of
experiments, along the reverse (forward) process for the JE to be
satisfied. This result has been already
discussed in Sec.~\ref{bio:accep}: the process that dissipates most
between the forward and the reverse is the one that samples more
efficiently the region of values in the vicinity of $W^{\dag}$. This conclusion that
may look counterintuitive can be rationalized by noting that larger
dissipation implies also larger fluctuations and therefore more chances
to get rare paths that sample the vicinity of $W^{\dag}$.  The
symmetries \eqq{path:7ba}{path:7bb} were originally discussed in
\cite{Ritort04} and analyzed in detail for the case of the gas contained
in a piston \cite{Jarzynski06}.

We close this section by analyzing the case where the work distribution
is Gaussian. The Gaussian case describes the linear response regime
usually (but not necessarily) characterized by small deviations from equilibrium. Let us
consider the following distribution,
\be
P(W)=(2\pi\sigma_W^2)^{-\frac{1}{2}}\exp\Bigl(-\frac{(W-W^{\rm mp})^2}{2\sigma_W^2}\Bigr)
\label{path:8}
\ee
where the average value of the work, $\langle W\rangle$, is just equal to the
most probable value $W^{\rm mp}$. The path entropy is given by $S(W)=-(W-W^{\rm mp})^2/(2\sigma_W^2)+{\rm
constant}$, so \eq{path:6} is satisfied. From \eq{path:7} we get,
\be
\hat{T}(W)=-\frac{\sigma_W^2}{W-W^{\rm mp}}\rightarrow W^{\dag}=W^{\rm mp}-\frac{\sigma_W^2}{T}~~~.
\label{path:9}
\ee
From \eq{path:5b} and \eq{path:9} we get
$W^{\dag}=\Delta F-(\sigma_W^2/2T)$. Therefore,
\bea
W_{\rm diss}^{\dag}=W^{\dag}-W_{\rm rev}=W^{\dag}-\Delta
F=-\frac{\sigma_W^2}{2T}\label{path:10a}\\
W_{\rm diss}^{\rm mp}=W^{\rm mp}-W_{\rm rev}=W^{\rm mp}-\Delta
F=\frac{\sigma_W^2}{2T}\label{path:10b}
\eea
leading to the final result $W_{\rm diss}^{\dag}=-W_{\rm diss}^{\rm
mp}=-\langle W_{\rm diss}\rangle$. Therefore, in order to recover the
free energy using the JE, paths with negative dissipated work and of
magnitude equal to the average dissipated work must be
sampled. Sometimes the paths with negative dissipated work are referred
as {\em transient violations of the second law}. This name has raised
strong objections among some rows of physicists. Of course the second
law remains inviolate. The name just stresses the fact that paths with
negative dissipated work must be experimentally accessible to
efficiently recover free energy differences.  Note that, for the
specific Gaussian case, we get $\langle W_{\rm
diss}\rangle=\frac{\sigma_W^2}{2T}$ and therefore the
fluctuation-dissipation parameter $R$ \eq{applications:NETS:11} is equal
to 1 as expected for systems close to equilibrium. The result $R=1$ has
been shown to be equivalent to the validity of the
fluctuation-dissipation theorem \cite{SchFuj03}.

\subsection{Computation of the work/heat distribution}
\label{stoctherm:pw}
The JE and the CFT describe relations between work distributions
measured in NETS. However, they do not imply a specific form for the
work distribution. In small systems fluctuations of the work relative
to the average work are large so work distributions can strongly
deviate from Gaussian distributions and be highly non trivial. In
contrast, as the system size increases, deviations of the work respect
to the average value start to become rare and exponentially suppressed
with the size of the system. To better characterize the pattern of
nonequilibrium fluctuations it seems important to explore analytical
methods that allow us to compute, at least approximately, the shape of
the energy distributions (e.g. heat or work) along nonequilibrium
processes. Of course there is always the possibility to carry out
exact calculations in specific solvable cases. In general, however,
the exact computation of the work distribution turns out to be a very difficult
mathematical problem (solvable examples are
\cite{LuaGro05,BenBroKaw05,SpeSei05b,CroJar06,CleBroKaw06}) that is
related to the evaluation of large deviation functions
(Sec~\ref{pw:tails}). This problem has traditionally received a lot of
attention by mathematicians and we foresee it may become a central
area of research in statistical physics in the next years.

\subsubsection{An instructive example}
\label{pw:example}

To put the problem in proper perspective let us consider an instructive
example: an individual magnetic dipole of moment $\mu$ subject to a magnetic
field $H$ and embedded in a thermal bath. The dipole can switch between the up and
down configurations, $\pm \mu$. The transition rates between the up and
down orientations are of the Kramers type
\cite{HanTalBor90,Melnikov91},
\bea
k_{-\mu\to\mu}(H)=k_{\rm up}(H)=k_0\frac{\exp(\frac{\mu
H}{T})}{2\cosh(\frac{\mu H}{T})}\label{pw:1a}\\
k_{\mu\to-\mu}(H)=k_{\rm down}(H)=k_0\frac{\exp(\frac{-\mu
H}{T})}{2\cosh(\frac{\mu H}{T})}\label{pw:1b}
\eea
with $k_0=k_{\rm up}(H)+k_{\rm down}(H)$ independent of $H$. The rates
\eqq{pw:1a}{pw:1b} satisfy detailed balance \eq{demo5},
\be
\frac{k_{\rm up}(H)}{k_{\rm down}(H)}=\frac{P^{\rm eq}(\mu)}{P^{\rm eq}(-\mu)}=\exp\Bigl(\frac{2\mu
H}{T}\Bigr)
\label{pw:2}
\ee
with $P^{\rm eq}(+(-)\mu)=\exp(-(+)\mu H/T)/{\cal Z}$ where ${\cal
Z}=2\cosh(\mu H/T)$ is the equilibrium partition function. In this system
there are just two possible configurations, ${\cal C}={-\mu,\mu}$. We
consider a nonequilibrium  protocol where the control parameter $H$ is varied as a
function of time, $H(t)$. The dynamics of the dipole is a continuous
time Markov process, and a path is specified by the time-sequence
$\Gamma\equiv \lbrace \mu(t)\rbrace$. 

\begin{figure}
\begin{center}
\includegraphics[scale=0.3,angle=-90]{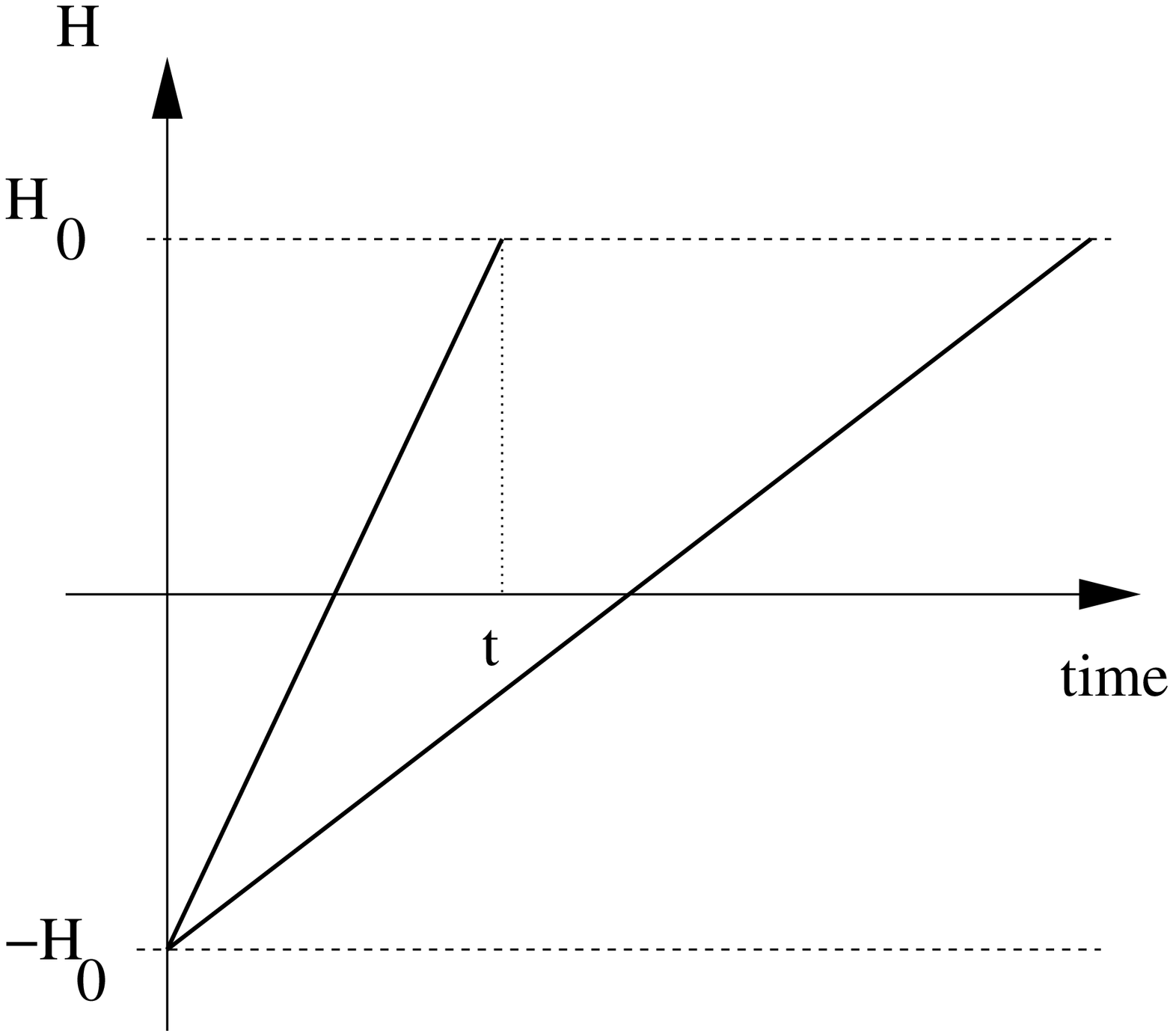}\includegraphics[scale=0.3,angle=-90]{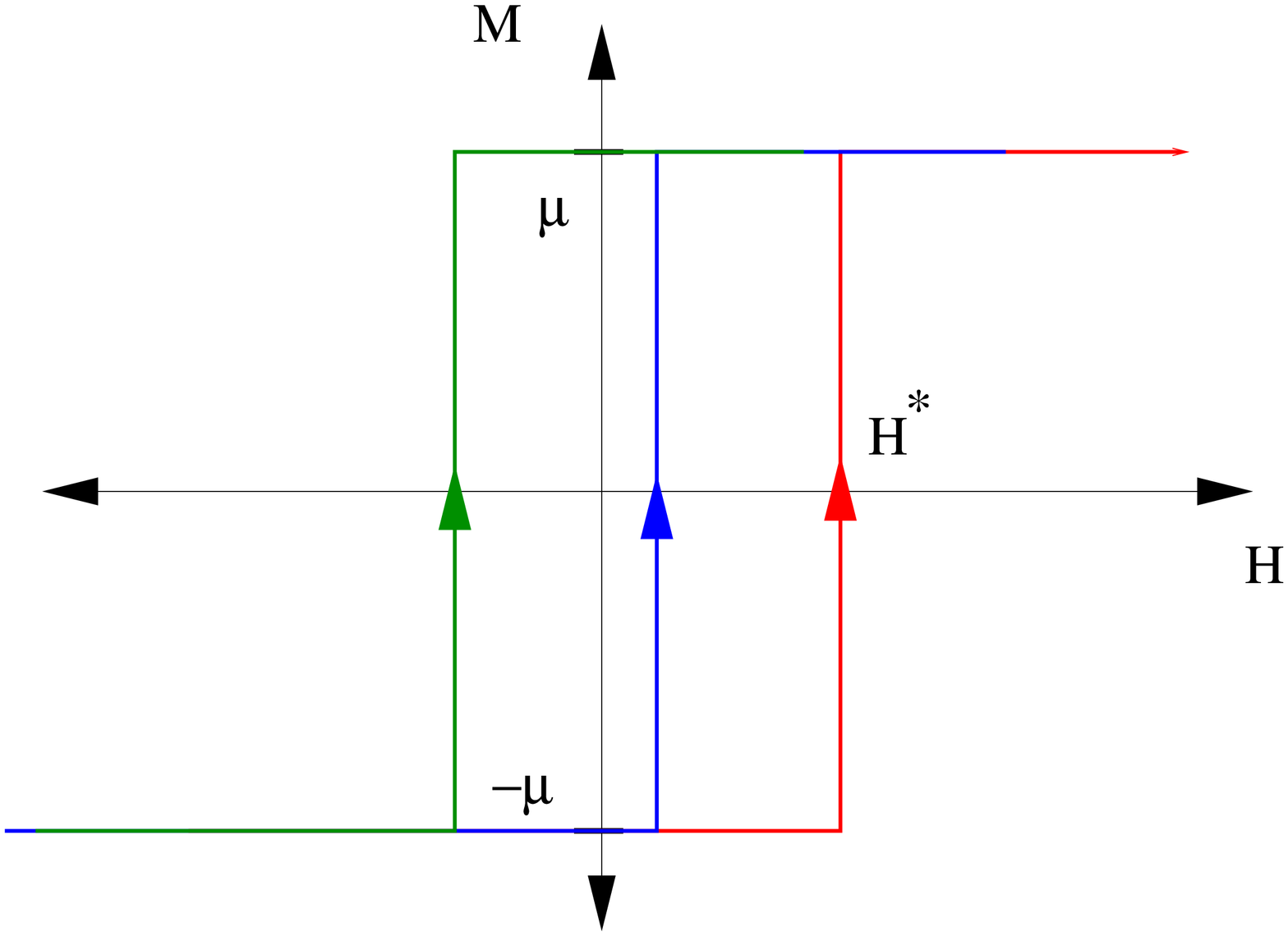}
\end{center}
\vspace{-0.4cm}
\caption{\em (Left) Ramping protocol. The ramping speed is defined by
$r=2H_0/t$ where $t$ is the duration of the ramp. (Right) Three examples of paths where
the down dipole reverses orientation at different values of the field, $H^*$.}
\label{fig13}
\vspace{-0.2cm}
\end{figure}

Let us consider the following protocol: the dipole starts in the down
state $-\mu$ at $H=-H_0$. The field is then ramped from $-H_0$ to $+H_0$
at a constant speed $r=\dot{H}$, so $H(t)=rt$ (Figure~\ref{fig13}A). The
protocol lasts for a time $t_{\rm max}=2H_0/r$ and the field stops
changing when it has reached the value $H_0$. The free energy difference
between the initial and final state is 0 because the free energy is an
even function of $H$. To ensure that the dipole initially points down
and that this is an equilibrium state, we take the limit $H_0\to \infty$
but we keep the ramping speed $r$ finite. In this way we generate paths
that start at $H=-\infty$ at $t=-\infty$ and end up at $H=\infty$ at
$t=\infty$. We can now envision all possible paths followed by the
dipole. The up configuration is statistically preferred for $H>0$
whereas the down configuration is preferred for $H<0$. Therefore, in a
typical path the dipole will stay in the down state until the field is
reversed. At some point, after the field changes sign, the dipole will
switch from the down to the up state and remain in the up state for the
rest of the protocol. In average there will be always a time lag between
the time at which the field changes sign and the time at which the
dipole reverses orientation.  In another paths the dipole will reverse
orientation before the field changes sign, i.e. when $H<0$. These sort
of paths become more and more rare as the ramping speed
increases. Finally, in the most general case, the dipole can reverse
orientation more than once. The dipole will always start in the down
orientation and end in the up orientation with multiple transitions
occurring along the path.

The work along a given path is given by \eq{applications:NETS:7aa},
\be
W(\Gamma)=-\int_{-\infty}^{\infty}dt \dot{H}(t) \mu(t)=-r\int_{-\infty}^{\infty}dt\mu(t)
\label{pw:3}
\ee
Note that because $E_{H}(\mu)=E_{-H}(-\mu)$ then $\Delta E=0$ and
$Q(\Gamma)=W(\Gamma)$ (i.e. the first law, $\Delta E=W-Q)$) so heat and
work distributions are identical in this example. Moreover, due to the
time reversal symmetry of the ramping protocol, the work distribution
$P(W)$ is identical along the forward and reverse process. Therefore we
expect that the JE \eq{applications:NETS:9} and the CFT
\eq{applications:NETS:10} are both satisfied with $\Delta F=0$,
\be
\frac{P(W)}{P(-W)}=\exp\Bigl(\frac{W}{T}\Bigr) ~~~~; ~~~~\langle \exp\Bigl(-\frac{W}{T}\Bigr)\rangle =1
\label{pw:3a}
\ee
The exact computation of $P(W)$ in this simple one-dipole model is already a
very arduous task that, to my knowledge, has not yet been exactly solved. We can
however consider a limiting case and try to elucidate the properties of
the work (heat) distribution. Here we consider the limit of
large ramping speed $r$ where the dipole executes just one transition
from the down to the up orientation. A few of these paths are depicted in
Figure \ref{fig13}B. This is also called a first order Markov process because it
only includes transitions that occur along one direction (from down to
up). In this reduced and oversimplified description a path is fully specified by the value of the field
$H^*$ at which the dipole reverses orientation. The work along one of
these paths is given by,
\be
W(\Gamma\equiv H^*)=-\lim_{H_0\to \infty}\int_{-H_0}^{H_0}dH
\mu(H)=((H^*+H_0)-(H_0-H^*))\mu=2\mu H^*
\label{pw:4}
\ee
According to the second law, $\langle W\rangle=\langle Q\rangle\ge 0$, which
implies that the average switching field is positive, $\langle
H^*\rangle\ge 0$ (as expected due to the time lag between
the reversal of the field and the reversal of the dipole). The work distribution is
just given by the switching field distribution $p(H^*)$. This is
a quantity easy to compute. The probability that the dipole is in the down state
at field $H$ satisfies a master equation that only includes the death process,
\be
\frac{\partial p_{\rm down}(H)}{\partial H}=-\frac{k_{\rm up}(H)}{r}p_{\rm
down}(H)~~~.
\label{pw:5}
\ee
This equation can be exactly solved,
\be
p_{\rm down}(H)=\exp\Bigl( -\frac{1}{r}\int_{-\infty}^H dH k_{\rm up}(H) \Bigr)
\label{pw:6}
\ee
where we have inserted the initial condition $p_{\rm
down}(-\infty)=1$. The integral in the exponent can be easily evaluated
using \eqq{pw:1a}{pw:1b}. We get,
\be
p_{\rm down}(H)=\Bigl(1+\exp(\frac{2\mu H}{T})\Bigr)^{-\frac{Tk_0}{2\mu r}}
\label{pw:7}
\ee
The switching field probability distribution $p(H^*)$ is given by $p(H^*)=-(p_{\rm
down})'(H^*)$. From \eq{pw:4} we get,
\be
P(W)=\frac{k_0}{4\mu r}\Bigl(1+\exp(\frac{W}{T})\Bigr)^{-\frac{Tk_0}{2\mu r}} \frac{\exp(\frac{W}{2T})}{\cosh(\frac{W}{2T})}
\label{pw:8}
\ee
and from this result we obtain the path entropy,
\be
S(W)=\log(P(W))=-\frac{Tk_0}{2\mu
r}\log\Bigl(\exp(\frac{W}{T})+1\Bigr)+\frac{W}{2T}-\log\Bigl(\cosh(\frac{W}{2T})\Bigr)+{\rm constant}
\label{pw:9}
\ee
It is important to stress that \eq{pw:8} does not satisfy \eq{pw:3a} except in the limit $r\to\infty$ where this
approximation becomes exact. We now compute $W^{\rm mp},W^{\dag}$ in the
large $r$ limit. We obtain to the leading order,
\bea
S'(W^{\rm mp})=0 \rightarrow W^{\rm mp}=T\log(\frac{2\mu r}{k_0T})+{\cal
O}\bigl(\frac{1}{r} \bigr)\label{pw:10a}\\
S'(W^{\dag})=\frac{1}{T} \rightarrow W^{\dag}=-T\log(\frac{2\mu r}{k_0T})+{\cal
O}\bigl(\frac{1}{r} \bigr)\label{pw:10b}
\eea
so the symmetry \eq{path:7ba} (or \eq{path:7bb}) is satisfied to the
leading order (yet it can be shown how the $1/r$ corrections appearing
in $W^{\rm mp},W^{\dag}$ \eqq{pw:10a}{pw:10b} are different). We can also
compute the leading behavior of the fluctuation-dissipation parameter
$R$ \eq{applications:NETS:11} by observing that the average work $\langle
W\rangle$ is asymptotically equal to the most probable work. The
variance of the work, $\sigma_W^2$, is found by expanding $S(W)$ around
$W^{\rm mp}$,
\bea
S(W)=S(W^{\rm mp})+\frac{S''(W^{\rm mp})}{2}(W-W^{\rm mp})^2+{\rm (higher~~order~~terms)}\label{pw:11a}\\
\sigma_W^2=-\frac{1}{S''(W^{\rm mp})}\label{pw:11b}
\eea
A simple computation shows that $\sigma_W^2=2T$ and therefore,
\be
R=\frac{\sigma_W^2}{2TW_{\rm diss}}\to \frac{1}{\log(\frac{2\mu r}{k_0T})}
\label{pw:12}
\ee
so $R$ decays logarithmically to zero. The logarithmic increase of the
average work with the ramping speed \eq{pw:10a} is just a consequence of
the logarithmic increase of the average value of the switching field
$\langle H^*\rangle$ with the ramping speed. This result has been also
predicted for the dependence of the average breakage force of molecular
bonds in single molecule pulling experiments. This phenomenology,
related to the technique commonly knownn as dynamic force spectroscopy,
allows to explore free energy landscapes by varying the pulling speed
over several orders of magnitude \cite{Evans01,EvaWil00}.

\subsubsection{A mean-field approach}
\label{pw:mf}
We now focus our attention in describing an analytical
method useful to compute work distributions, $P(W)$, in mean-field systems. The method
has been introduced in \cite{Ritort04} and developed in full generality by
A. Imparato and L. Peliti \cite{ImpPel05b,ImpPel05c}. This section is a
bit technical. The reader not interested in the details can just skip
this section and go to Sec~\ref{pw:tails}.

The idea behind the method is the following. We express the probability
distribution $P(W)$ as a sum over all paths that start from a given
initial state. This sum results in a path integral that can be
approximated by its dominant solution or classical path in the large $N$
limit, $N$ being the number of particles. The present approach exploits
the fact that, as soon as $N$ becomes moderately large, the contribution
to the path integral is very well approximated by the classical path.
In addition, the classical path exactly satisfies the FT.
Here we limit ourselves to show in a very sketchy way how the method applies to solve the
specific example shown in Sec.~\ref{pw:example}. A detailed and more
complete derivation of the method can be found in \cite{Ritort04,ImpPel05c}

We come back to the original model \eqq{pw:1a}{pw:1b} and include all
possible paths where the dipole reverses orientation more than once. The problem now gets too complicated so
what we do is to modify the original model by considering an ensemble of
non-interacting $N$ identical dipoles. A configuration in the system is
specified by the $N$-component vector ${\cal C}\equiv
\lbrace \vec{\mu}=(\mu_i)_{1\le i\le N}\rbrace$ with $\mu_i=\pm \mu$ the two possible
orientations of each dipole. A path is specified by the time-sequence
$\Gamma\equiv \lbrace \vec{\mu}(s);0\le s\le t\rbrace$. The energy of the system is
given by,
\be
E({\cal C})=-hM({\cal C})=-h\sum_{i=1}^N\mu_i
\label{pw:mf:1}
\ee
where $M=\sum_i\mu_i$ is the total magnetisation. The equilibrium free
energy is $F=-N\log(2\cosh(\mu H/T))$ and the kinetic rules are
the same as given in \eqq{pw:1a}{pw:1b} and are identical for each dipole. The work along a
given path is given by \eq{pw:3},
\be
W(\Gamma)=-\int_0^tds\dot{H}(s)M(s)
\label{pw:mf:2}
\ee
so the work probability distribution is given by the path integral,
\be
P(W)=\sum_{\Gamma}P(\Gamma)\delta(W(\Gamma)-W)=\int{\cal D}[\vec{\mu}]\delta\Bigl(W+\int_0^tds\dot{H}(s)M(s)\Bigr)
\label{pw:mf:3}
\ee
where we have to integrate over all paths where $\vec{\mu}$ starts at
time 0 in a given equilibrium state up to a final time $t$. To solve
\eq{pw:mf:3} we use the integral representation of the delta function,
\be
\delta(x)=(1/2\pi)\int_{-\infty}^{\infty} d\lambda\exp(-i\lambda
x)~~~. 
\label{pw:mf:3b}
\ee
We also insert the following factor, 
\be 1=\int \frac{{\cal D}[\gamma]{\cal D}[m]}{2\pi}\exp\Bigl(
\frac{i}{\Delta t}\int
\gamma(s)(m(s)-\frac{1}{N}\sum_{i=1}^N\mu_i(s)) \Bigr)
\label{pw:mf:4}
\ee
where $\Delta t$ is the discretization time-step and we have introduced
new scalar fields $\gamma(s),m(s)$. After some manipulations one gets a
closed expression for the work distribution $P(w)$ ($w=W/N$ is the work
per dipole). We quote the final result \cite{Ritort04},
\be
P(w)={\cal N}\int d\lambda {\cal D}[\gamma]{\cal D}[m] \exp\Bigl( Na(w,\lambda,\gamma,m)\Bigr)
\label{pw:mf:5}
\ee
where ${\cal N}$ is a normalization constant and $a$ represents an action given by,
\bea
a(w,\lambda,\gamma,m)=\lambda\Bigl(w+\int_0^tds\dot{H}(s)M(s)\Bigr)+\\
\frac{1}{2}\int_0^tds\Bigl(m(s)(2\dot{\gamma}(s)+c(s))+d(s)
\Bigr)+\log\Bigl(\exp(\gamma(0))k_{\rm up}(H_i)+exp(-\gamma(0)k_{\rm down}(H_i)\Bigr)
\label{pw:mf:6}
\eea
with,
\bea
c(s)=k_{\rm down}(H(s))\Bigl(\exp(-2\gamma(s))-1\Bigr)-k_{\rm
up}(H(s))\Bigl(\exp(2\gamma(s))-1\Bigr)\label{pw:mf:7a}\\
d(s)=k_{\rm down}(H(s))\Bigl(\exp(-2\gamma(s))-1\Bigr)+k_{\rm
up}(H(s))\Bigl(\exp(2\gamma(s))-1\Bigr)\label{pw:mf:7b}
\eea
where the rates $k_{\rm up},k_{\rm down}$ are given in
\eqq{pw:1a}{pw:1b} and we have assumed an initial equilibrium state at
the initial value of the field, $H(0)=H_i$. Equation \eq{pw:mf:6} has to be
solved together with the boundary conditions,
\be
\gamma(t)=0~~~~~;~~~~~m(0)=\tanh\Bigl(\gamma(0)+\frac{\mu H_i}{T}\Bigr)~~~~~.
\label{pw:mf:8}
\ee
Note that these boundary conditions break causality. The function
$\gamma$ has the boundary at the final time $t$ whereas $m$ has the
boundary at the initial time $0$. Causality is broken because by imposing
a fixed value of the work $w$ along the paths we are constraining
the time evolution of the system. 

\begin{figure}
\includegraphics[scale=0.28,angle=-90]{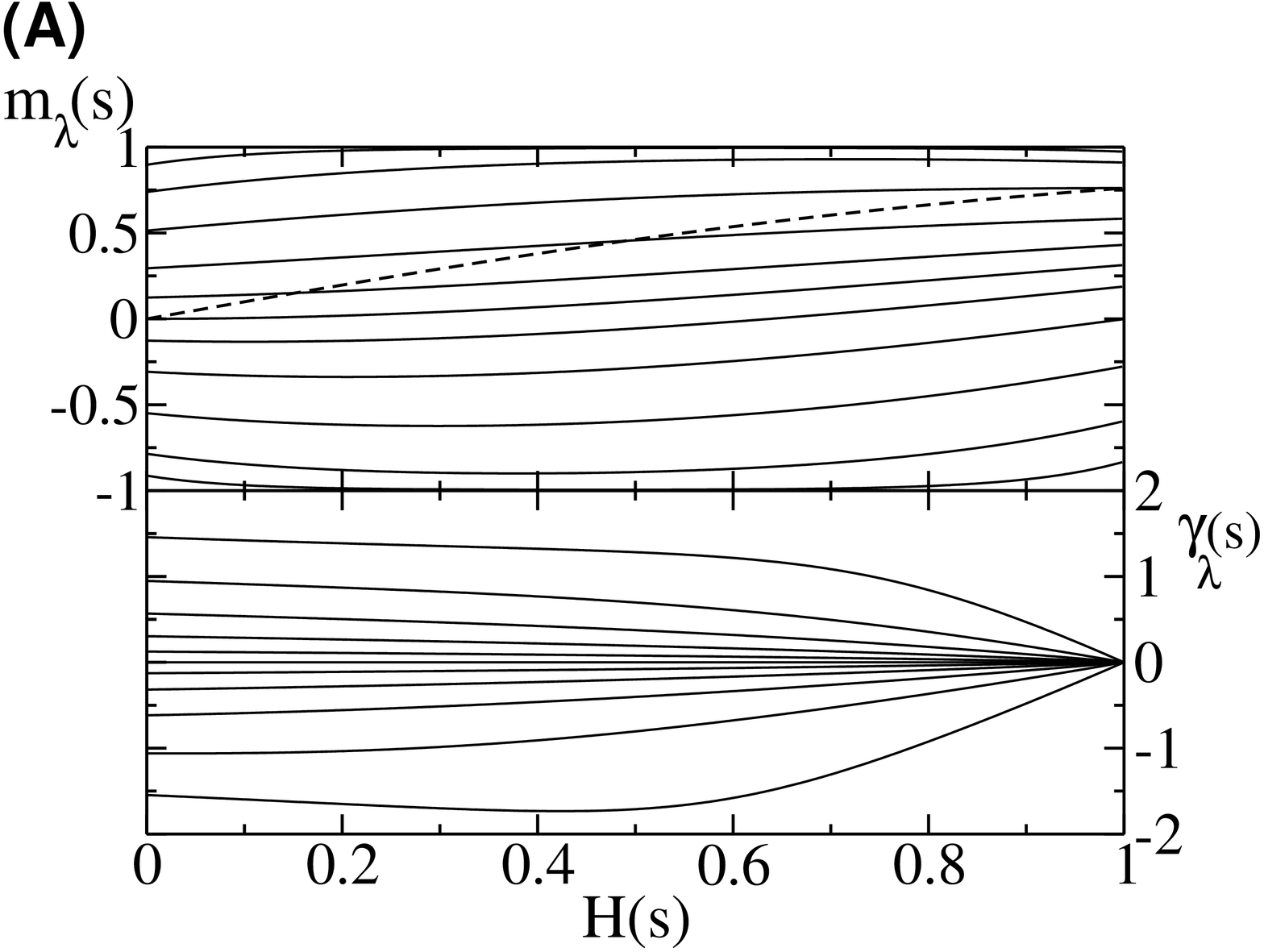}\includegraphics[scale=0.4,angle=-90]{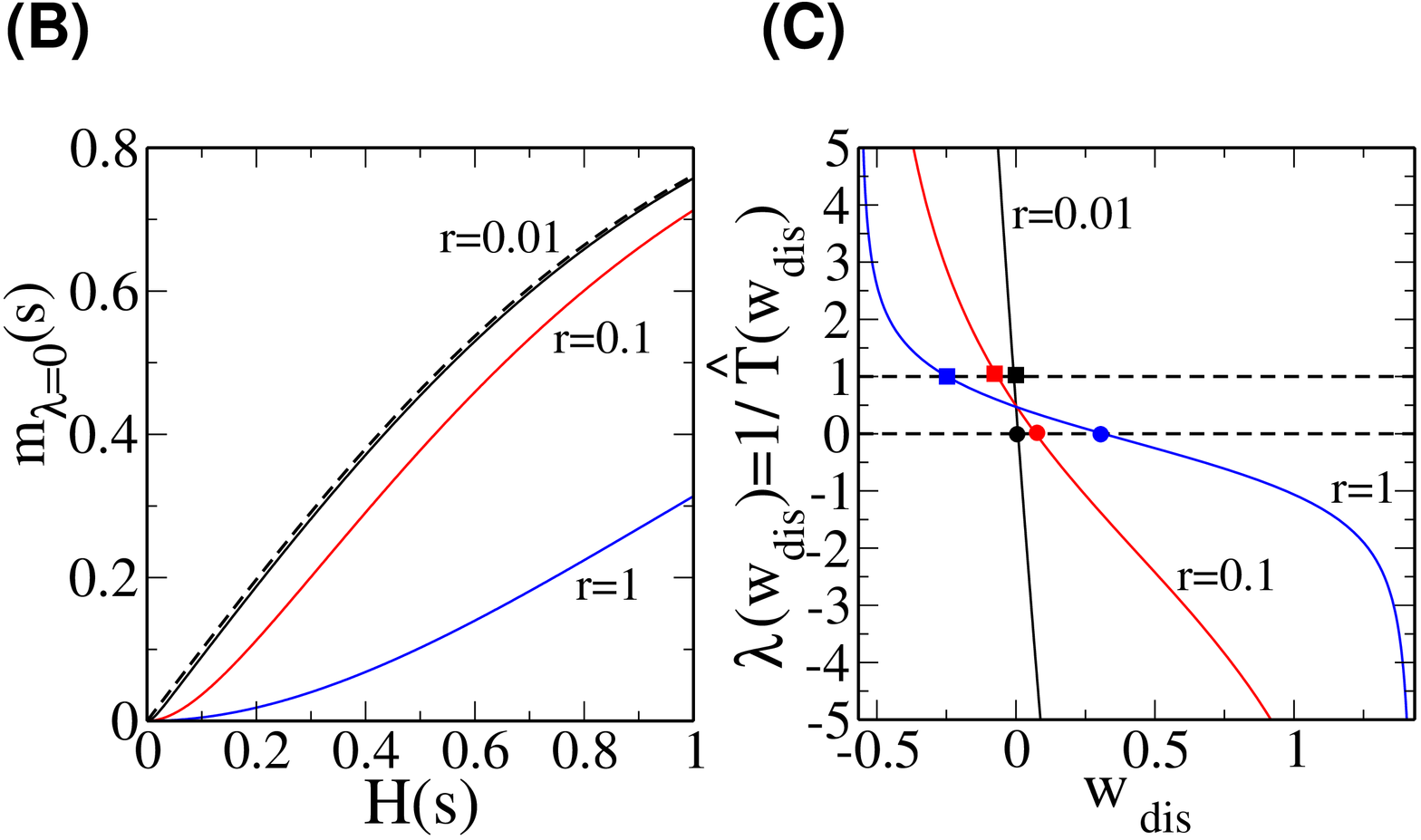}
\begin{center}
\includegraphics[scale=0.45,angle=0]{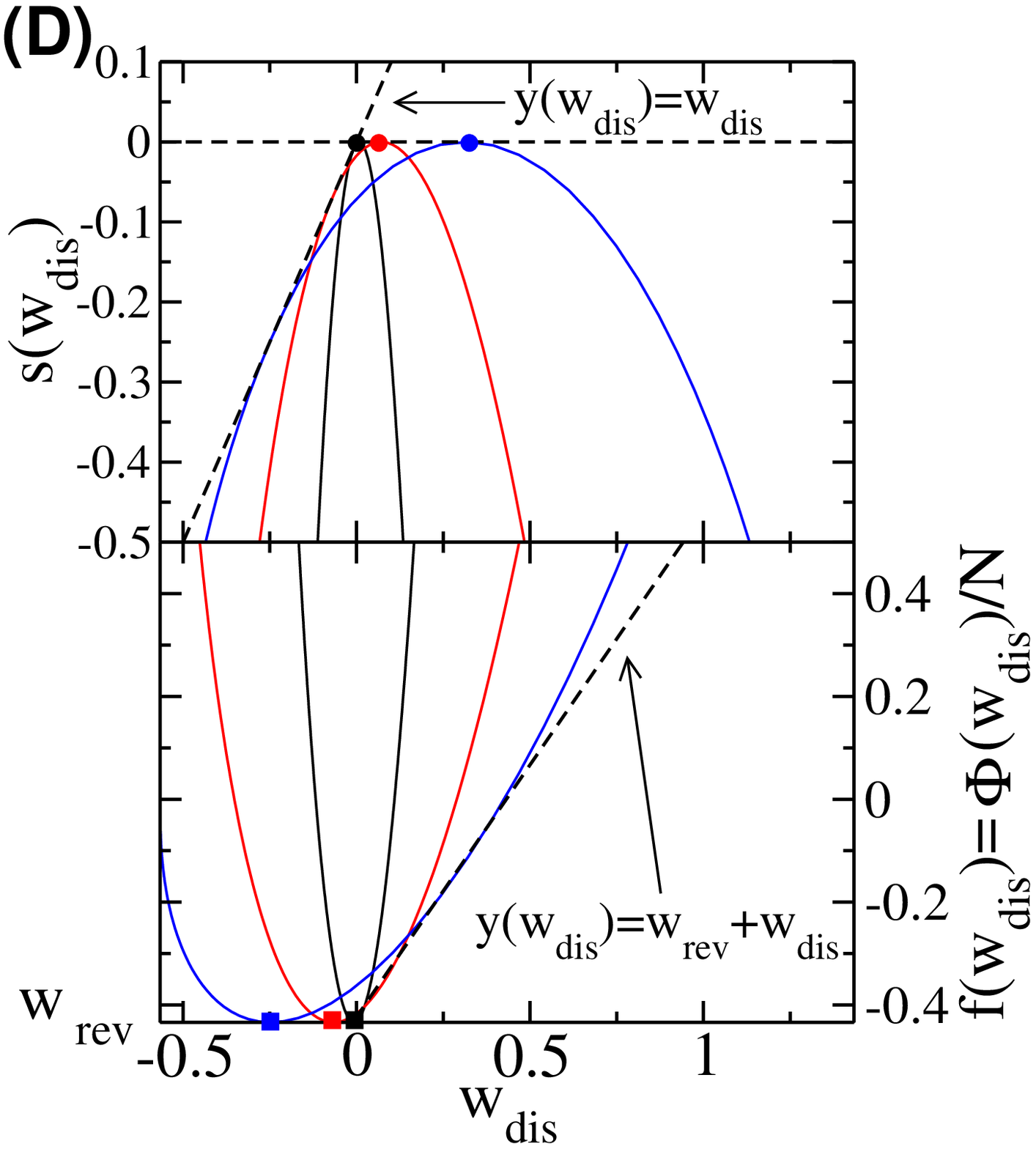}
\end{center}
\vspace{-0.4cm}
\caption{\em Various results for the mean-field solution
\eqqq{pw:mf:10a}{pw:mf:10b}{pw:mf:10c} of a dipole in field that is
ramped from $H_i=0$ to $H_f=1$.  (A) Fields $m_{\lambda}(s)$ and
$\gamma_{\lambda}(s)$ at the ramping speed $r=1$. Curves correspond to
different values of $\lambda$
($\lambda=-5,-2,-1,-0.5,-0.2,0.,0.2,0.5,1,2,5$ from top to bottom in the
upper and lower panel). The dashed line in $m_{\lambda}(s)$ is the
equilibrium solution $m_{\rm eq}(H)=\tanh(H)$ corresponding to the
reversible process $r\to 0$. (B) Magnetization
$m_{\lambda}(s)$ for the most probable path $\lambda=0$. The dashed line
corresponds to the reversible trajectory, $r\to 0$. (C) Lagrange
multiplier $\lambda(w_{\rm dis})$ for three ramping speeds. The
intersection of the different curves with the dashed line $\lambda=0$
gives $w^{\rm mp}$ (filled circles) whereas the intersection with
$\lambda=-1$ gives $w^{\dag}$ (filled squares). The intersection of all
three curves around $\lambda=0.5$ is only accidental (looking at a
larger resolution such crossing is not seen). (D) Path entropy and free
energy corresponding to the solutions shown in (B,C) (larger speeds
correspond to wider distributions). Path entropies are maximum and
equal to zero at $w_{\rm dis}^{\rm mp}=w^{\rm mp}-w_{\rm rev}$ (filled circles)
whereas path free-energies are minimum and equal to $f^{\dag}=w_{\rm
rev}$ at $w_{\rm dis}^{\dag}$ (filled squares).  Figure taken from
\protect\cite{Ritort04}.}
\label{fig14}
\vspace{-0.2cm}
\end{figure}

To compute $P(w)$ we take the large volume limit $N\to\infty$ in
\eq{pw:mf:5}. For a given value of $w$ the probability distribution is
given by,
\be
P(w)\propto \exp(Ns(w))=\exp\Bigl(Na(w,\lambda(w),\gamma_w(s),m_w(s)\Bigr)
\label{pw:mf:9}
\ee
where $s$ is the path entropy \eq{path:3b} and
the functions $\lambda(w),\gamma_w(s),m_w(s)$ are solutions of the
saddle point equations,
\bea
\frac{\delta a}{\delta \lambda}=w+\mu\int_0^tm_w(s)\dot{H}(s)ds=0\label{pw:mf:10a}\\
\frac{\delta a}{\delta \gamma(s)}=\dot{m}_w(s)+m_w(s)(k_{\rm up}(s)+k_{\rm down}(s))-\nonumber\\(k_{\rm
up}(s)-k_{\rm down}(s))+m_w(s)d_w(s)+c_w(s)=0\label{pw:mf:10b}\\
\frac{\delta a}{\delta m(s)}=\dot{\gamma}_w(s)+\lambda(w)\mu \dot{H}(s)+\frac{1}{2}c_w(s)=0~~~.\label{pw:mf:10c}
\eea
These equations must be solved together with the boundary conditions \eq{pw:mf:8}. Note that we use
the subindex (or the argument) $w$ in all fields ($\lambda,m,\gamma$) to
emphasize that there exists a solution of these fields for each value of the work
$w$. From the entropy $s$ in \eq{pw:mf:9} we can evaluate the path free
energy, the path temperature and the values
$w^{\rm mp},w^{\dag}$ introduced in
Sec.~\ref{path:strategy}. We enumerate the different results:

\begin{itemize}

\item{\bf The path entropy $s(w)$.} By inserting \eq{pw:mf:10c} into
\eq{pw:mf:6} we get,
\be
s(w)=\lambda(w) w+\frac{1}{2}\int_0^t d_w(s)ds+\log\Bigl(\exp(\gamma(0))k_{\rm up}(H_i)+\exp(-\gamma(0))k_{\rm down}(H_i)\Bigr)~~.
\label{pw:mf:11}
\ee
From the stationary conditions
\eqqq{pw:mf:10a}{pw:mf:10b}{pw:mf:10c} the path entropy \eq{pw:mf:9} satisfies,
\be
s'(w)=\frac{ds(w)}{dw}=\frac{\partial
a(w,\lambda(w),\gamma_w(s),m_w(s))}{\partial w}=\lambda(w)~~~~.
\ee
The most probable work can
be determined by finding the extremum of the path entropy $s(w)$,
\be 
s'(w^{\rm mp})=\lambda(w^{\rm mp})=0.
\label{pw:mf:12}
\ee
where we used \eq{path:6}. The saddle point equations
\eqqq{pw:mf:10a}{pw:mf:10b}{pw:mf:10c} give $\gamma_{w^{\rm mp}}(s)=c_{w^{\rm mp}}(s)=d_{w^{\rm mp}}(s)=0$ and,
\be
\dot{m}_{w^{\rm mp}}(s)=-m_{w^{\rm mp}}(s)(k_{\rm up}(s)+k_{\rm down}(s))+(k_{\rm up}(s)-k_{\rm down}(s))~~~~.
\label{pw:mf:13}
\ee
which is the solution of the master equation for the magnetisation. The
stationary solution of this equation gives the equilibrium solution
$m^{\rm eq}(s)=\tanh(\mu H(s)/T)$ corresponding to a quasistationary or reversible
process.

\item{\bf The path free energy.} The path
free energy  $f=\Phi/N$ \eq{path:5a} is given by,
\be
f^{\dag}=f(w^{\dag})=\frac{\Delta F}{N}=w_{\rm
rev}=w^{\dag}-Ts(w^{\dag})=\frac{T}{2}\int_0^t  d_{w^{\dag}}(s)ds
\label{pw:mf:14}
\ee
where $w^{\dag}$ is given by,
\be
s'(w^{\dag})=\lambda(w^{\dag})=\frac{1}{T}
\label{pw:mf:15}
\ee
and the path temperature \eq{path:7} satisfies the identity,
\be
\hat{T}(w)=\frac{1}{\lambda(w)}~~~;~~~\hat{T}(w^{\dag})=T
\label{pw:mf:16}
\ee
\end{itemize}
This set of equations can be numerically solved. Figure \ref{fig14} shows some
of the results.

\subsection{Large deviation functions and tails}
\label{pw:tails}
A large deviation function $\hat{P}(x)$ of a function $P_L(x)$ is
defined if the following limit exists,
\be
\hat{P}(x)=\lim_{L\to\infty}\frac{1}{L}\log(P_L(\frac{x}{L}))~~~.
\label{pw:tails:1}
\ee
From this point of view the distribution of the entropy production in
NESS, $P(a)$ \eq{applications:NESS:7}, where $a={\cal S}_p /
\langle{\cal S}_p \rangle$ and the work distribution, $P(W)$
\eq{pw:mf:9}, define large deviation functions. In the first case,
$\lim_{t\to\infty}f_t(a)$ is the large deviation function
(e.g. \eq{example:23}), the average entropy production $\langle {\cal
S}_p\rangle$ being the equivalent of $L$ in \eq{pw:tails:1}. In the
second case the path entropy $s(w)=S(W)/N$ \eqq{path:3b}{pw:mf:9} is a
large deviation function, the
size $N$ being the equivalent of $L$ in \eq{pw:tails:1}.  Large deviation functions are interesting for several
reasons,

\begin{itemize}

\item{\bf Nonequilibrium theory extensions.} By knowing the large
deviation function of an observable in a nonequilibrium system
(e.g. the velocity or position density) we can characterize the
probability of macroscopic fluctuations. For example, by knowing the
function $s(w)$ we can determine the probability of macroscopic work
fluctuations $\delta W\propto N$, where $N$ is the size of the
system. Large deviations (for example, in work) may depend on the
particular details (e.g. the rules) of the nonequilibrium dynamics. In
contrast, small deviations (i.e. $\delta W\propto \sqrt{N}$) are
usually insensitive to the microscopic details of the
dynamics. Nonequilibrium systems are non-universal and often strongly
dependent on the microscopic details of the system. In this regard,
understanding large or macroscopic deviations may be a first step in
establishing a general theory for nonequilibrium systems.

\item{\bf Spectrum of large deviations.} There are few examples where
large deviations can be analytically solved. Over the past years a large
amount of work has been devoted to understand large deviations in some
statistical models such as exclusion processes. General results include
the additivity principle in spatially extended systems
\cite{DerLebSpe01,DerLebSpe02a,DerLebSpe02b} and the existence of
exponential tails in the distributions \cite{GiaKurPel06}. These general
results and the spectrum of large deviations are partially determined by
the validity of the FT \eq{ft13} which imposes a specific relation
between the forward and the reverse work/heat distributions.  For
example, exponential tails in the work distribution $P(W)$ \eq{path:3b}
correspond to a path-entropy $S(W)$ that is linear in $W$. This is the
most natural solution of the FT, see \eq{path:7a}.

\item{\bf Physical interpretation of large deviations.} In small
systems, large deviations are common and have to be considered as
important as small deviations. This means that, in order to understand
the nonequilibrium behavior of small systems, a full treatment of
small and large deviations is necessary. The latter are described
by the shape of the large deviation function. The physical
interpretation of small and large deviations may be different. For
example, if we think of the case of molecular motors, small deviations
(respect to the average value) in the number of mechanochemical cycles may
be responsible of the average speed of a molecular motor whereas large
deviations may be relevant to understand why molecular motors operate so
efficiently along the mechanochemical cycle.

\end{itemize}

\subsubsection{Work and heat tails}
\label{pw:workheat}
Let us consider the case of a NETS that starts initially in equilibrium
and is driven out of equilibrium by some external driving forces.  As we
have seen in \eq{pw:mf:9}, $\frac{1}{N}\log(P(w))=s(w)$ is a large
deviation function. At the same time we could also consider the heat
distribution $P(Q)$ and evaluate its large deviation function
$\frac{1}{N}\log(P(Q))=s(q)$ where $q=Q/N$. Do we expect $s(q)$ and
$s(w)$ to be identical? Heat and work differ by a boundary term, the
energy difference. Yet the energy difference is extensive with $N$,
therefore boundary terms modify the large deviation function so we
expect that $s(q)$
and $s(w)$ are different. An interesting example is the case of
the bead in the harmonic trap discussed in Sec.~\ref{example:bead}. Whereas the
work distribution measured along arbitrary time intervals is always a Gaussian, the heat distribution
is characterized by a Gaussian distribution for small fluctuations
$\delta Q=Q-<Q>\propto\sqrt{t}$, plus exponential tails for large
deviations $\delta Q \propto t$. The difference between the large
deviation function for the heat and the work arises from a boundary
term, the energy difference. Again, in the large $t$ limit, the boundary
term is important for large fluctuations when $a=|Q|/\langle Q \rangle\ge
a^*=1$ \eq{example:23}. Large deviation functions depend always on
boundary terms that often cannot be neglected.

Let us come back now to the example of Sec.~\ref{pw:example} where we
considered work distributions in a system of non-interacting dipoles
driven by an externally varying magnetic field. Again we will focus the
discussion on the particular case where the initial value of the field
is negative and large $H_i=-H_0\to -\infty$ , and the field is ramped at
speed $r$ until reaching the final value $H_f=H_0\to \infty$. In this
case $Q=W$ for individual paths so both large deviation functions
$s(q),s(w)$ are identical. In what follows we will use heat instead of
work for the arguments of all functions. In addition $s_F=s_R$ due to
the time-reversal symmetry of the protocol. Exponential tails are
indicated by a path-temperature $\hat{T}(q)$ \eq{path:7} which is
constant along a finite interval of heat values.

In Sec.~\ref{pw:example} we have evaluated the path entropy $s(q)$
\eq{pw:mf:11} for an individual dipole ($N=1$) in the
approximation of a first order Markov process. The following result has
been obtained \eq{pw:9},
\be
s(q)=-\frac{Tk_0}{2\mu
r}\log(\exp(\frac{q}{T})+1)+\frac{q}{2T}-\log(\cosh(\frac{q}{2T}))+{\rm constant}~~~.
\label{pw:tails:2}
\ee  
For $|q|\to\infty$ we get,
\bea
s(q\to\infty)=-\frac{qk_0}{2\mu r}+{\cal O}\Bigl(\exp(-\frac{q}{T})\Bigr)\label{pw:tails:3a}\\
s(q\to-\infty)=\frac{q}{T}+{\cal O}\Bigl(\exp(\frac{q}{T})\Bigr)
\label{pw:tails:3b}
\eea
The linear dependence of $s(q)$ on $q$ leads to
\bea
\hat{T}(q\to\infty)=T^-=-\frac{2\mu r}{k_0} \label{pw:tails:4a}\\
\hat{T}(q\to-\infty)=T^+=T \label{pw:tails:4b}\\
\eea
where we use the notation $T^+,T^-$ to stress the fact that these path
temperatures are positive and negative respectively. Both path
temperatures are constant and lead to exponential tails for positive and negative work
values. Note that equation \eq{path:7a} reads,
\be
s(q)-s(-q)=\frac{q}{T}\rightarrow
(s)'(q)+(s)'(-q)=\frac{1}{T}\rightarrow \frac{1}{T^+}+\frac{1}{T^-}=\frac{1}{T}
\label{pw:tails:5}
\ee
which is satisfied by \eqq{pw:tails:4a}{pw:tails:4b} up to $1/r$
corrections. 

\begin{figure}
\begin{center}
\includegraphics[scale=0.28,angle=-90]{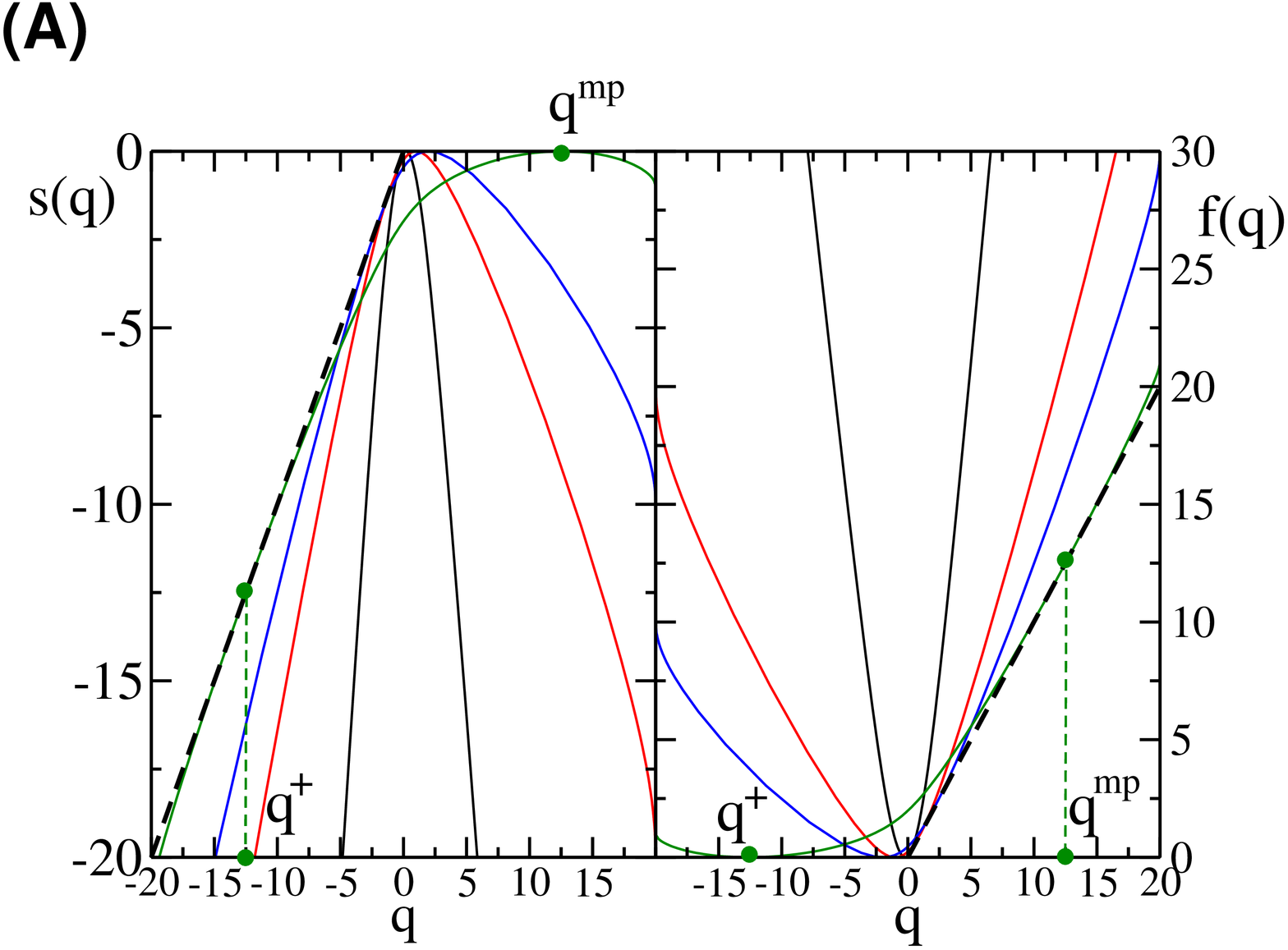}\includegraphics[scale=0.28,angle=-90]{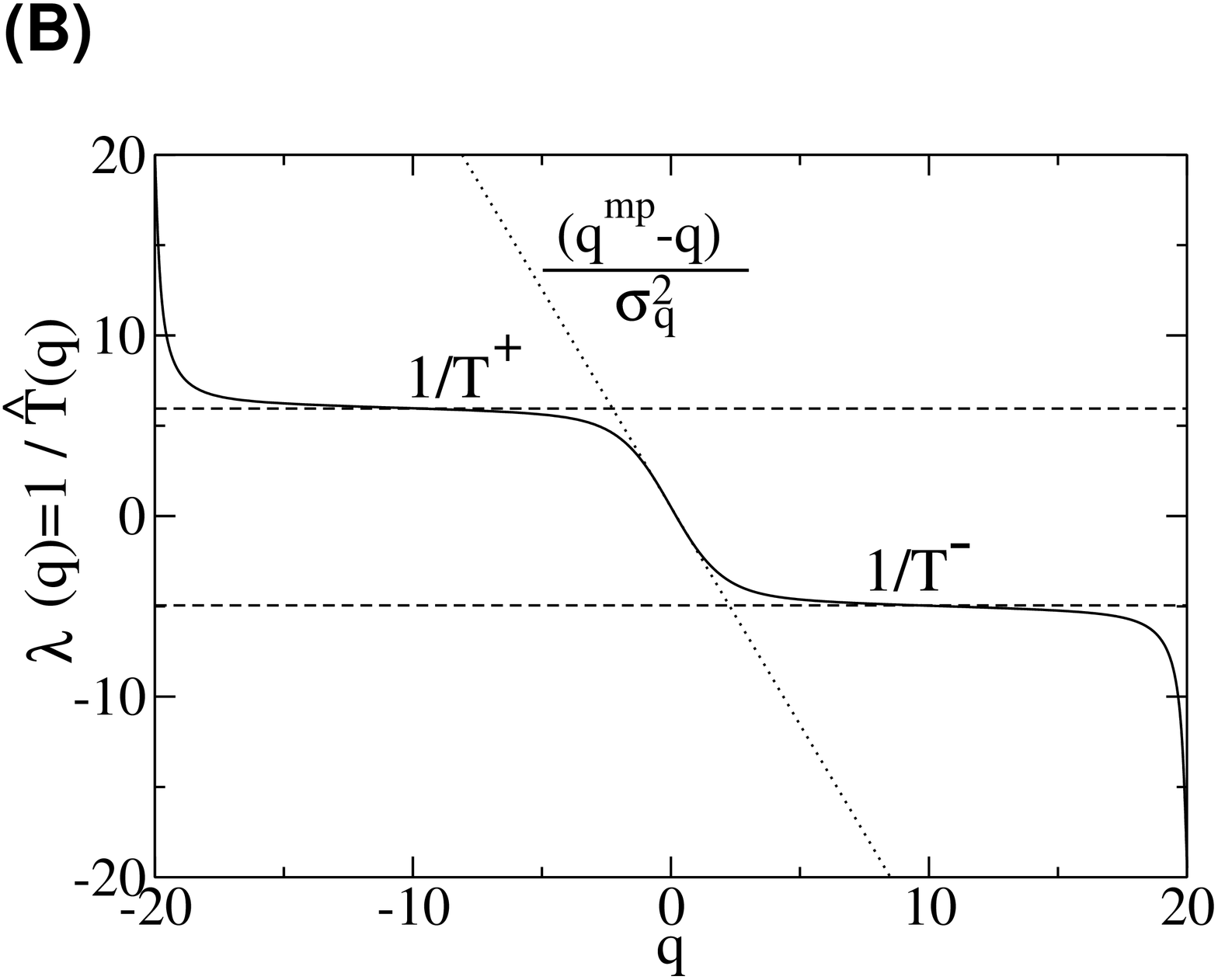}
\end{center}
\vspace{-0.4cm}
\caption{\em (A) Heat distributions (path entropy $s(q)$ and path free
energy $f(q)$) evaluated at four ramping speeds $r=0.1,0.5,1,10$ (from
the most narrower to the most wider distributions). The dashed line in
the left panel is $y(q)=q/T$ (we take $T=1$) and is tangent to $s(q)$ at
$q^{\dag}$ (dots are shown for $r=10$). The dashed line in the right panel
corresponds to $y(q)=q$ and is tangent to the function $f(q)$ at the value $q^{\rm
mp}$ (dots shown for $r=10$). (B) $\lambda(q)$ for the lowest speed
$r=0.1$. It shows a linear behavior for small values of $q$,
$\lambda(q)=(1/\sigma_{q}^2)(q^{\rm mp}-q)$ and two plateaus for $q>>1$
and $q<<-1$. The former contributes as a Gaussian component to the heat
distribution describing the statistics of small deviations respect to
the most probable value (stimulated sector). The latter gives rise to
two exponential tails for the distribution describing the statistics of
rare events (spontaneous sector). Figure adapted from
\protect\cite{Ritort04}.}
\label{fig15}
\vspace{-0.2cm}
\end{figure}

Another interesting limit is the quasistatic limit $r\to 0$. Based on
the numerical solution of the saddle point equations
\eqqq{pw:mf:10a}{pw:mf:10b}{pw:mf:10c}, it was suggested in
\cite{Ritort04} that $\hat{T}(q)$ converged to a constant value over a
finite range of work values. Figure \ref{fig15}A shows the results obtained
for the heat distributions whereas the path temperature is shown in 
Figure \ref{fig15}B. A more detailed analysis \cite{ImpPel05c} has shown that a
plateau is never fully reached for a finite interval of heat values when
$r\to 0$. The presence of a plateau has been interpreted as the
occurrence of a first order phase transition in the path entropy $s(q)$
\cite{ImpPel05c}.

There is an interesting analogy between the different type of
work/heat fluctuations and the emission of light radiation by atoms in
a cavity. Atoms can absorb and reemit photons
following two different mechanisms. One type of radiative mechanism is
called stimulated because it depends on the density of blackbody
radiation in the cavity (directly related to the temperature of the
cavity). The other radiative mechanism is called spontaneous and is
independent on the density of radiation in the cavity (i.e. it does
not depend on its temperature). The stimulated process contributes to
the adsorption and emission of radiation by atoms. The spontaneous
process only contributes to the emitted radiation.  In general, the
path entropy $s(w)$ contains two sectors reminiscent of the stimulated
and spontaneous processes in the blackbody radiation,
\begin{itemize}

\item{\bf The FDT or stimulated sector.} This sector is described by Gaussian
work fluctuations \eq{path:8} leading to $s(q)=-(q-q^{\rm
mp})^2/(2\sigma_q^2)+{\rm constant}$. Therefore we get \eq{path:9},
\be
\lambda(q)=\frac{1}{\hat{T}(q)}=-\frac{q-q^{\rm
mp}}{\sigma_q^2}
\label{pw:tails:6}
\ee
which behaves linearly in $q$ for small deviations around $q^{\rm
mp}$. Note that $\hat{T}(q)$ satisfies \eq{pw:tails:5} and therefore
\be
\sigma_q^2=2Tq^{\rm mp},
\label{pw:tails:6b}
\ee
leading to a fluctuation-dissipation parameter $R=1$, a result
equivalent to the validity of the fluctuation-dissipation theorem
(FDT). This sector we call stimulated because work fluctuations \eq{pw:tails:6b} depend
directly on the temperature of the bath.

\item{\bf The large deviation or spontaneous sector.} Under some conditions this
sector is well reproduced by exponential
tails in the work distribution describing large or macroscopic deviations. In this sector,
\be
\frac{1}{\hat{T}(q)}-\frac{1}{\hat{T}(-q)}=\frac{1}{T}\rightarrow \frac{1}{T^+}+\frac{1}{T^-}=\frac{1}{T}~~~~.
\label{pw:tails:7}
\ee
The physical interpretation of $T^+,T^-$ is as follows. Because $T^-$ is
negative then $T^-$ describes fluctuations where net heat is released to
the bath, whereas $T^+$ is positive and describes fluctuations where net
heat is absorbed from the bath. Equation \eq{pw:tails:7} imposes $T^+ <
|T^-|$ implying that large deviations also satisfy the second law: the
average net amount of heat supplied to the bath ($\propto |T^-|$) is
always larger that the average net heat absorbed from the bath ($\propto
T^+$). In the previous example \eqq{pw:tails:4a}{pw:tails:4b}, $T^+$
converges to the bath temperature whereas $T^-$ diverges to $-\infty$
when $r\to\infty$. We call this sector spontaneous because the energy
fluctuations mainly depend on the nonequilibrium protocol (in the
current example, such dependence is contained in the $r$-dependence of
$T^-$, \eq{pw:tails:4a}).

\end{itemize}

\subsubsection{The bias as a large deviation function}
\label{pw:bias}
The bias defined in \eq{bio:sm:7} is still another example of a large deviation
function. Let us define the variable,
\be
X=\sum_{i=1}^N\exp(-\frac{W_i}{T})
\label{pw:bias:0a}
\ee
where $W_i\to W_i-\Delta F$ stands for the dissipated work. The free-energy estimate \eq{bio:sm:7} satisfies the relation,
\be
x=\exp(-\frac{F^{\rm JE}-\Delta F}{T})~~~;~~~x=\frac{X}{N}=\frac{1}{N}\sum_{i=1}^N\exp(-W_i)
\label{pw:bias:1}
\ee
where $N$ is the total number of experiments. The $N$ values $W_i$ are
extracted from a distribution $P(W)$ that satisfies the relations,
\be
\langle1\rangle=\int_{-\infty}^{\infty}P(W)dW=1~~~;~~~\langle\exp(-W)\rangle=\int_{-\infty}^{\infty}\exp(-W)P(W)dW=1
~~~.\label{pw:bias:2}
\ee
We follow the same procedure as in Sec.~\ref{bio:free} and extract $N$
different values $W_i$ to obtain a single $x$ using \eq{pw:bias:1}. By repeating this
procedure a large number of times, $M$, we generate the probability distribution of $x$,
that we will call ${\cal P}_N(x)$, in the limit $M\to \infty$. The
bias \eq{bio:sm:8} is defined by,
\be
B(N)=-T\langle \log(x) \rangle=-T\int_{-\infty}^{\infty} \log(x) {\cal
P}_N(x) dx~~~~.
\label{pw:bias:2b}
\ee
In the following we show that ${\cal
P}_N(x)$ defines a large deviation function in the limit
$N\to\infty$. We write, 
\bea
{\cal P}_N(x)=\int \prod_{i=1}^NdW_iP(W_i)\delta \Bigl(
x-\frac{1}{N}\sum_{i=1}^N\exp(-W_i)\Bigr)=\nonumber\\
=\frac{1}{2\pi i}\int_{-i\infty}^{i\infty} d\mu \exp\Bigl(\mu
x-\frac{\mu}{N}\sum_{i=1}^N\exp(-W_i)\Bigr)\prod_{i=1}^N P(W_i)dW_i=\nonumber\\
=\frac{N}{2\pi
i}\int_{-i\infty}^{i\infty}d\hat{\mu}\exp\Bigl(N\hat{\mu}x+N\log(\int
dW P(W)\exp(-\hat{\mu}\exp(-W)))\Bigr)=\nonumber\\=\frac{N}{2\pi
i}\int_{-i\infty}^{i\infty}d\hat{\mu}\exp\Bigl(Ng(\hat{\mu},x)\Bigr)\sim_{N\to\infty} \exp\Bigl(Ng(\hat{\mu}^*,x)\Bigr)
\label{pw:bias:3}
\eea
where in the second line we used the integral representation of the
delta function \eq{pw:mf:3b}; in the third line we separate
the integrals and independently integrate the contribution of each variable $W_i$; in the last line we
apply the saddle point integration method to
the function $g(\hat{\mu},x)$ defined as,
\be
g(\hat{\mu},x)=\hat{\mu}x+\log\Bigl(\int_{-\infty}^{\infty}\exp(-\hat{\mu}\exp(-W)) \Bigr)
\label{pw:bias:4}
\ee
where $\hat{\mu}^*$ is equal to the absolute maximum of
$g(\hat{\mu},x)$,
\be \Bigl(\frac{\partial g(\hat{\mu},x)}{\partial
\hat{\mu}}\Bigr)_{\hat{\mu}=\hat{\mu}^*}=0\rightarrow
x=<<\exp(-W)>>_{\hat{\mu}^*}
\label{pw:bias:5}
\ee
with,
\be
<<..>>_{\hat{\mu}}=\frac{\int_{-\infty}^{\infty}\exp(-W)\exp\Bigl(-\hat{\mu}\exp(-W)\Bigr)P(W)dW}{\int_{-\infty}^{\infty}\exp\Bigl(-\hat{\mu}\exp(-W)\Bigr)P(W)dW}~~~.
\label{pw:bias:6}
\ee
The function $g(\hat{\mu},x)$ evaluated at $\hat{\mu}=\hat{\mu}^*$
defines a large deviation function \eq{pw:tails:1},
\be
g^*(x)=g(\hat{\mu}^*(x),x)=\lim_{N\to\infty}\frac{1}{N}\log\Bigl(
{\cal P}_N(\frac{X}{N})\Bigr)=\lim_{N\to\infty}\frac{1}{N}\log\Bigl(
{\cal P}_N(x)\Bigr)~~~.
\label{pw:bias:7}
\ee
Using \eq{pw:bias:7} we can write for the bias \eq{pw:bias:2b} in the large $N$ limit,
\be
B(N)=-T\frac{\int dx \log(x)\exp\Bigl(Ng^*(x)\Bigr)}{\int dx \exp\Bigl(Ng^*(x)\Bigr)}~~~.
\label{pw:bias:8}
\ee
The integrals in the numerator and denominator can be estimated by using
the saddle point method again. By expanding $g^*(x)$ around the maximum
contribution at $x^{\rm max}$ we get, up to second order, 
\be
g^*(x)=g^*(x^{\rm max})+\frac{1}{2}(g^*)''(x^{\rm max})(x-x^{\rm max})^2~~~.
\label{pw:bias:9}
\ee
To determine $x^{\rm max}$ we compute first,
\bea
(g^*)'(x)=\Bigl(\frac{\partial g(\hat{\mu},x)}{\partial
{\hat\mu}}\Bigr)_{\hat{\mu}=\hat{\mu}^*(x)}\Bigl(\frac{d\hat{\mu}^*(x)}{dx}
\Bigr)+
\Bigl(\frac{\partial g(\hat{\mu}^*,x)}{\partial x}\Bigr)=\nonumber\\
=\Bigl(\frac{\partial g(\hat{\mu}^*,x)}{\partial x}\Bigr)=\hat{\mu}^*(x)
\label{pw:bias:10}
\eea
where we have used \eqq{pw:bias:4}{pw:bias:5}. The value $x^{\rm max}$
satisfies, 
\be
(g^*)'(x^{\rm max})=\hat{\mu}^*(x^{\rm max})=0~~~.
\label{pw:bias:11}
\ee
Inspection of \eqq{pw:bias:5}{pw:bias:6} shows that  $x^{\rm max}=1$. The second term in
the rhs of \eq{pw:bias:9} is then given by,
\be
(g^*)''(x=1)=(\hat{\mu}^*)'(x=1)=\frac{1}{1-\langle\exp(-2W)\rangle}
\label{pw:bias:12}
\ee
where $\langle...\rangle$ denotes the average over the distribution $P(W)$ \eq{pw:bias:2}.
Using \eq{pw:bias:12} and inserting \eq{pw:bias:9} into \eq{pw:bias:8}
we finally obtain,
\be
B(N)=T\frac{\langle\exp(-2W)\rangle-1}{2N}+{\cal O}\Bigl(\frac{1}{N^2}\Bigr)
\label{pw:bias:13}
\ee
For a Gaussian distribution we get,
$B(N)=T\exp(\sigma_W^2-1)/(2N)$. Equation \eq{pw:bias:13} was derived in \cite{WooMuhTho91}.
For intermediate values of $N$ (i.e. for values of $N$ where $B(N)>1$)
other theoretical methods are necessary.

\section{Glassy dynamics}
\label{glassy}
Understanding glassy systems (see Sec. \ref{small:physics}) is a major
goal in modern condensed matter physics
\cite{Jackle86,Angell95,EdiAngNag96,Trieste99}. Glasses represent an
intermediate state of matter sharing some properties of solids and
liquids. Glasses are produced by fast cooling of a liquid when the
crystallization transition is avoided and the liquid enters the
metastable supercooled region. The relaxation of the glass to the
supercooled state proceeds by reorganization of molecular clusters
inside the liquid, a process that is thermally activated and strongly
dependent on the temperature. The relaxation of the supercooled liquid
is a nonequilibrium process that can be extremely slow leading to
aging. The glass analogy is very fruitful to describe the nonequilibrium
behavior of a large variety of systems in condensed matter physics, all
them showing a related phenomenology.

The nonequilibrium aging state (NEAS, see Sec.\ref{stoctherm:noneq}) is
a non-stationary state characterized by slow relaxation and a very low rate of energy
dissipation to the surroundings. Aging systems fail to reach equilibrium unless one waits an
exceedingly large amount of time. For this reason, the
NEAS is very different from either the nonequilibrium transient state
(NETS) or the nonequilibrium steady state (NESS). 

What do aging systems have in common with the nonequilibrium behavior of
small systems? Relaxation in aging systems is
driven by fluctuations of a small number of molecules that relax
by releasing a small amount of stress energy to the surroundings. These
molecules are grouped into clusters often called cooperatively
rearranging regions (CRRs). 
A few observations support this interpretation,
\begin{itemize}

\item{\bf Experimental facts.} Traditionally, the glass transition has
been studied with bulk methods such as calorimetry or light
scattering. These measurements perform an average over all mesoscopic
regions in the sample, but are not suitable to follow the motion of
individual clusters of a few nanometers in extension.  The few direct
evidences we have on aging as driven by the rearrangement of small
regions comes from AFM measurements on glass surfaces, confocal
microscopy of colloids and the direct observation of molecular motion
(NMR resonance and photobleaching tests) \cite{CipRam05}. More indirect
evidence is obtained from the heterogeneous character of the dynamics,
i.e. the presence of different regions in the system that show a great
disparity of relaxation times \cite{Ediger00}. The observation of strong
intermittent signals \cite{BuiBelCil03} in Nyquist noise measurements
while the system ages has been interpreted as the result of CRRs,
i.e. events corresponding to the rearrangement of molecular
clusters. Finally, the direct measure of a correlation length in
colloidal glasses hints at the existence of CRRs \cite{BerBirBouCipMasHotLadPie05}.  Future
developments in this area are expected to come from developments in
micromanipulation and nanotechnology applied to the direct experimental observation of molecular clusters.

\item{\bf Numerical facts.} Numerical simulations are a very useful
approach to examine our understanding of the NEAS
\cite{CriRit04}. Numerical simulations allow to measure correlation
functions and other observables which are hardly accessible in
experiments. Susceptibilities in glasses are usually defined in terms of
four-point correlation functions (two-point in space and two-point in
time) which give information about how spatially separated regions are
correlated in time \cite{BirBou04}. A characteristic quantity is the
typical length of such regions.  Numerical simulations of glasses show
that the maximum length of spatially correlated regions is small, just a
few nanometers in molecular glasses or a few radii in colloidal
systems. Its growth in time is also exceedingly slow (logarithmic in
time) suggesting that the correlation length is small for the experimentally
accessible timescales.

\item{\bf Theoretical facts.} There are several aspects that suggest
that glassy dynamics must be understood as a result of the relaxation of
CRRs. Important advances in the understanding of glass phenomena come
from spin glass theory \cite{MezParVir87,Young98}. Historically, this
theory was proposed to study disordered magnetic alloys which show
nonequilibrium phenomena (e.g. aging) below the spin-glass
transition temperature. However, it has been shown later how spin glass theory
provides a consistent picture of the NEAS in structural glass models
that do not explicitly contain quenched disorder in the Hamiltonian
\cite{BouMez94,MarParRit94a,CugKurParRit95,FraHer95}. Most of the
progress in this area comes from the study of mean-field models,
i.e. systems with long-range interactions. The success of mean-field
theory to qualitatively reproduce most of the observed phenomenology in glasses
suggests that NEAS are determined by the relaxation of mean-field like
regions, maybe the largest CRRs in the system. Based on this analogy several mean-field based
phenomenological approaches have been proposed \cite{XiaWol00,CriRit01,GarCha03,BirBou04b,GarRit05}.

\end{itemize}

In the next sections I briefly discuss some of the theoretical concepts
important to understand the glass state and the nonequilibrium aging dynamics.

\subsection{A phenomenological model}
\label{glassy:pheno}
To better understand why CRRs are predominantly small we introduce here
a simple phenomenological aging model inspired by mean-field theory
\cite{CriRit01}. The model consists of a set of regions or domains of
different sizes $s$. A region of size $s$ is just a molecular cluster
(colloidal cluster), containing $s$ molecules (or $s$ colloidal
particles). The system is prepared in an initial high energy
configuration where spatially localized regions in the system contain
some stress energy. That energy can be irreversibly released to the
environment if a cooperative rearrangement of that region takes place. The release occurs when
some correlated structures are built inside the region by a cooperative
or anchorage mechanism. Anchorage occurs when all $s$ molecules in that
region move to collectively find a transition state that gives access to
the {\em release pathway}, i.e. a path in configurational space that
activates the rearrangement process. Because the cooperative process
involves $s$ particles, the characteristic time to anchor the transition
state is given by,
\be
\frac{\tau_s}{\tau_0}\propto\Bigl(\frac{\tau^*}{\tau_0}\Bigr)^s=\exp\Bigl(\frac{Bs}{T} \Bigr)
\label{glassy:pheno:1}
\ee
where $\tau^*=\tau_0\exp(B/T)$ is the activated time required to anchor
one molecule, $\tau_0$ is a microscopic time and $B$ is the activation
barrier which is equal to the energy of the transition state. How CRRs exchange energy with the
environment? Once relaxation starts, regions of all sizes contain some
amount of stress energy ready to be released to the environment in the
form of heat. The first time a given region rearranges it typically releases an
amount of heat $\overline{Q}$ that does not scale with the size of the
region. After the first rearrangement has taken place the region
immediately equilibrates with its environment. Subsequent rearrangement
events in that same region do not release more stress energy to the
environment. These regions can either absorb or release heat from/to the
environment as if they were thermally equilibrated with the bath, the net average heat
exchanged with the environment being equal to 0. The release of the stored
stress energy in the system proceeds in a hierarchical fashion. At a
given age $t$ (the time elapsed after relaxation starts, also called
waiting time) only the CRRs of size $s^*$ have some stress
energy $\overline{Q}$ available to be released to the
environment. Smaller regions with $s<s^*$ already released their stress
energy sometime in the past being now in thermal
equilibrium with the environment. Larger regions with $s>s^*$ have not
had yet enough time to release their stress energy.
Only the CRRs with $s$ in the vicinity of $s^*$ contribute to the
overall relaxation of the glass toward the supercooled state. That size
$s^*$ depends on the waiting time or time elapsed since the relaxation started.

\begin{figure}
\begin{center}
\includegraphics[scale=0.3,angle=-90]{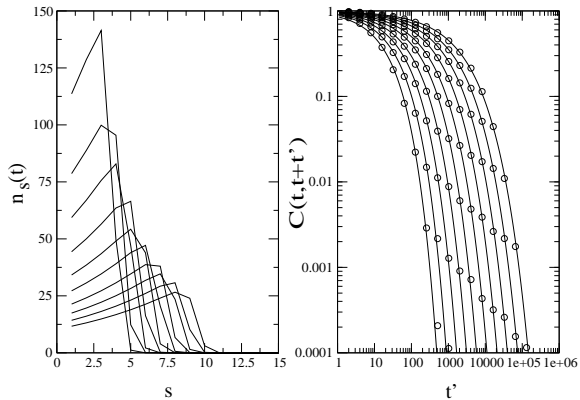}
\end{center}
\vspace{-0.4cm}
\caption{\em $n_s(t)$ (left panel) and $C(t,t+t')$ (right panel) for different waiting times
$t=10^{14}-10^{33}$ for the numerical
solution of \eq{glassy:pheno:3} with $l=8,g=1$ and
$T=0.45$. The relaxation time and the stretching exponent are very
well fitted by $\tau_t=2.2t^{0.35}$,
$\beta_s(t)=0.34+0.45t^{-0.06}$. Figure
taken from \protect\cite{CriRit01}.}
\label{fig16}
\vspace{-0.2cm}
\end{figure}

Let $n_s(t)$ be the number of CRRs of size $s$ at time $t$. At a given
time the system is made out of non-overlapping regions in the system
that randomly rearrange according to \eq{glassy:pheno:1}.  After a
rearrangement occurs CRRs destabilize probably breaking up into smaller
regions. In the simplest description we can assume that regions can just
gain or loose one particle from the environment with respective
(gain,loose) rates $k^g_s,k^l_s$ with $k^g_s+k^l_s=k_s$. $k_s$, the rate
of rearrangement, is proportional to $1/\tau_s$ where $\tau_s$ is
given in \eq{glassy:pheno:1}. To further simplify the description we
just take $k^g_s=gk_s,k^l_s=lk_s$ with $g+l=1$. Consequently, the
balance equations involve the following steps:
\be
{\cal D}_s \rightarrow {\cal D}_{s-1}+ p~~~;~~~~{\cal D}_s+p \rightarrow {\cal
D}_{s+1}
\label{glassy:pheno:2}
\ee
with rates $k^l_s,k^g_s$, where ${\cal D}_s$ denotes a region of size
$s$ and $p$ a particle (an individual molecule or a colloidal particle)
in the system. The balance equations for the
occupation probabilities read ($s\ge 2$),
\be
\frac{\partial n_s(t)}{\partial
t}=k^l_{s+1}n_{s+1}(t)+k^g_{s-1}n_{s-1}(t)-k_sn_s(t)~~~.
\label{glassy:pheno:3}
\ee
This set of equations must be solved together with mass conservation
$\sum_{s=1}^{\infty}sn_s(t)={\rm const}$. The equations can be numerically
solved for all parameters of the model. Particularly interesting results
are found for $g\ll l$. Physically this means that, after rearranging,
regions are more prone to loose molecules than to capture them, a
reasonable assumption if a cooperative rearrangement leads to
a destabilization of the region.  A few remarkable results can be inferred
from this simple model,

\begin{itemize}
\item{\bf Time dependent correlation length.} In Figure \ref{fig16}
(left panel) we show the time evolution for $n_s(t)$. At any time it
displays a well defined time-dependent cutoff value $s^*(t)$ above which
$n_s(t)$ abruptly drops to zero. The distribution of the sizes of the
CRRs scales like $n_s(t)=(1/s^*)\hat{n}(s/s^*)$ where $s^*$ is a waiting
time dependent cutoff size (data not shown).  The NEAS can be
parametrized by either the waiting time or the size of the region
$s^*(t)$. Relaxation to equilibrium is driven by the growth of $s^*(t)$
and its eventual convergence to the stationary solution of
\eq{glassy:pheno:3}. The size $s^*(t)$ defines a characteristic growing
correlation length, $\xi(t)=(s^*(t))^{\frac{1}{d}}$, where $d$ is the
dimensionality of the system. Because $s^*(t)$ grows logarithmically in
time (see \eq{glassy:pheno:1}) sizes as small as $\simeq 10$ already require
$10^{33}$ iteration steps. Small CRRs govern the relaxation of the
system even for exceedingly long times.

\item{\bf Logarithmic energy decay.}  The release of stress energy to
the environment occurs when the regions of size $s^*$ rearrange
for the first time.  The advance of the {\it front} in $n_s(t)$ located
at $s=s^*$ is the leading source of energy dissipation. Cooperative
rearrangements of regions of size smaller than $s^*$ have already
occurred several times in the past and do not yield a net thermal heat
flow to the bath whereas regions of size larger than $s^*$ have not yet
released their stress energy.  The supercooled state is reached when the cutoff size
$s^*$ saturates to the stationary solution of \eq{glassy:pheno:3} and the net energy flow
between the glass and the bath vanishes. The rate of energy decay in the
system is then given by the stress energy $\overline{Q}$ released by regions of size
$s^*(t)$ times their number $n_{s^*}(t)=\hat{n}(1)/s^*$, divided by the
activated time \eq{glassy:pheno:1} (equal to the waiting time $t\sim\exp(Bs^*/T)$),
\be
\frac{\partial E}{\partial t}\sim \frac{\overline{Q} n_{s^*}(t)}{t}\sim \frac{\overline{Q}}{s^* t}~~~~.
\label{glassy:pheno:4}
\ee
Because $s^*(t)\sim T\log(t)$ then the energy decays logarithmically
with time, $E(t)\sim 1/\log(t)$.

\item{\bf Aging.} If we assume independent exponential relaxations for the CRRs we 
obtain the following expression for the two-times correlation function,
\be
C(t,t+t')=\sum_{s\ge 1} s n_s(t)\exp(-t'/\tau_s)
\label{glassy:pheno:5}
\ee 
where $t$ denotes the waiting time after the initiation of the
relaxation and $\tau_s$ is given by \eq{glassy:pheno:1}. In the right
panel of Figure
\ref{fig16} we show the correlation function \eq{glassy:pheno:5} for
different values of $t$ (empty circles in the figure). Correlations
\eq{glassy:pheno:5} are excellently fitted by a stretched exponential
with a $t$ dependent stretching exponent $\beta_s$,
\be
C(t,t+t')\equiv C_{t}(t')=\exp\Bigl(-\Bigl(\frac{t'}{\tau_t}\Bigr)^{\beta_s(t)}\Bigr)~~~~.
\label{glassy:pheno:5b}
\ee 
In the right panel of
Figure \ref{fig16} we also show the best fits (continuous lines).  Correlation functions show simple aging and scale like
$t'/t$ with $t=\exp(s^*/T)$ where $s^*$ is the waiting-time
dependent cutoff size.

\item{\bf Configurational entropy and effective temperature.} An
important concept in the glass literature that goes back to Adam and
Gibbs in the Fifties \cite{GibDim58,AdaGib65} is the configurational
entropy, also called complexity and denoted by ${\cal S}_c$
\cite{CriRit04}. It is proportional to the logarithm of the number of
cooperative regions with a given free-energy $F$, $\Omega(F)$
\be
{\cal S}_c(F)=\log\Bigl(\Omega(F)\Bigr)~~~~.
\label{glassy:pheno:6}
\ee
At a given time $t$ after relaxation starts, the regions of size
$s^*$ contain a characteristic free energy $F^*$. Fluctuations in these
regions lead to rearrangements that release a net amount of heat to the
environment \eq{glassy:pheno:4}. Local detailed balance implies that,
after a rearrangement takes place, new regions with free energies around $F^*$ are
generated with identical probability. Therefore,
\be
\frac{{\cal W}(F\to F')}{{\cal W}(F'\to
F)}=\frac{\Omega(F')}{\Omega(F)}=\exp\Bigl({\cal S}_c(F')-{\cal S}_c(F)
\Bigr)
\label{glassy:pheno:7}
\ee
where ${\cal W}(F\to F')$ is the rate to create a region of free energy $F'$
after rearranging a region of free energy $F$. Note the similarity between
\eq{glassy:pheno:7} and \eq{demo5}. If $\Delta F'=F'-F$ is much smaller
than ${\cal S}_c(F)$ we can expand the difference in the configurational
entropy in \eq{glassy:pheno:7} and write,
\be
\frac{{\cal W}(\Delta F)}{{\cal W}(-\Delta F)}=\exp\Bigl(\Bigl(\frac{\partial {\cal
S}_c(F)}{\partial F}\Bigr)_{F=F^*}\Delta
F\Bigr)=\exp\Bigl(\frac{\Delta F}{T_{\rm eff}(F^*)}\Bigr)
\label{glassy:pheno:8}
\ee
with the shorthand notation ${\cal W}(\Delta F)={\cal W}(F\to F')$ and
the time-dependent effective temperature $T_{\rm eff}(F^*)$ defined as,
\be
\frac{1}{T_{\rm eff}(F^*)}=\Bigl(\frac{\partial {\cal S}_c(F)}{\partial F}\Bigr)_{F=F^*}~~~~.
\label{glassy:pheno:9}
\ee
In the present phenomenological model only regions
that have not yet equilibrated (i.e. of size $s\ge s^*(t)$) can release stress
energy in the form of a net amount of heat to the
surroundings. This means that only transitions with $\Delta F<0$
contribute to the overall relaxation toward equilibrium.

Therefore, the rate of energy dissipated by the system can
be written as,
\be
\frac{\partial E}{\partial t}\propto
\frac{1}{t}\frac{\int_{-\infty}^0dx x {\cal W}(x)}{\int_{-\infty}^0dx
{\cal W}(x)}=\frac{2T_{\rm eff}(F^*)}{t}~~~~.
\label{glassy:pheno:10}
\ee
where we take 
\be
{\cal W}(\Delta F)\propto \exp\Bigl(\frac{\Delta F}{2T_{\rm eff}(F^*)}\Bigr)
\label{glassy:pheno:10b}
\ee
as the solution of \eq{glassy:pheno:8}. Identifying \eq{glassy:pheno:10}
and \eq{glassy:pheno:4} we get,
\be
T_{\rm eff}(F^*)=\frac{\overline{Q}}{2s^*}~~~.
\label{glassy:pheno:11}
\ee
The time dependence of $s^*$ derived in \eq{glassy:pheno:4} shows that
the effective temperature decreases logarithmically in time.

\end{itemize}

\subsection{Nonequilibrium temperatures}
\label{glassy:noneqtemp}
The concept of a nonequilibrium temperature has stimulated a lot the
research in the area of glasses. This concept has been promoted by
Cugliandolo and Kurchan in the study of mean-field models of spin
glasses \cite{CugKur93,CugKur95b} that show violations of the fluctuation-dissipation theorem
(FDT) in the NEAS.  The main result in the theory is that two-time
correlations $C(t,t_w)$ and responses $R(t,t_w)$ satisfy a a modified
version of the FDT. It is customary to introduce the
effective temperature through the fluctuation-dissipation ratio (FDR) \cite{CugKurPel97} defined as,
\be
T_{\rm eff}(t_w)=\frac{\frac{\partial C(t,t_w)}{\partial t_w}}{R(t,t_w)}
\label{glassy:noneqtemp:1}
\ee
in the limit where $t-t_w\gg t_w$. In contrast, in the limit $t-t_w\ll
t_w$ local equilibrium holds and $T_{\rm eff}(t_w)=T$. In general
$T_{\rm eff}(t_w)\ge T$, although there are exceptions to this rule and
even negative effective temperatures have been found \cite{Sollich06}.
These predictions have been tested in many exactly solvable models and
numerical simulations of glass formers \cite{CriRit04}.  In what follows
we try to emphasize how the concept of the effective temperature $T_{\rm
eff}(t_w)$ contributes to our understanding of nonequilibrium fluctuations
in small systems.

Particularly illuminating in this direction has been the study of mean-field
spin glasses. These models have the advantage that can be analytically
solved in the large volume limit. At the same time, numerical
simulations allow to investigate finite-size effects in
detail. Theoretical calculations in mean-field spin glasses are usually
carried out by first taking the infinite-size limit and later the long
time limit. Due to the infinite range nature of the interactions this
order of limits introduces pathologies in the dynamical solutions and
excludes a large spectrum of fluctuations that are relevant in real
systems. The infinite-size limit in mean-field models, albeit physically
dubious, is mathematically convenient. Because analytical computations
for finite-size systems are not available we can resort to numerical
simulations in order to understand the role of finite-size effects in
the NEAS. A spin glass model that has been extensively studied is the
random orthogonal model (ROM) \cite{MarParRit94b}, a variant of the
Sherrington-Kirkpatrick model \cite{SheKir75}, known to reproduce the
ideal mode coupling theory \cite{GotSjo92}. The model is defined in
terms of the following energy function,
\be
{\cal H}=-\sum_{(i,j)}J_{ij}\s_i\s_j
\label{glassy:noneqtemp:2}
\ee
where the $\sigma_i$ are $N$ Ising spin variables ($\sigma= \pm 1$) and
$J_{ij}$ is a random $N\times N$ symmetric orthogonal matrix with zero
diagonal elements. In the limit $N\to\infty$ this model has the same
thermodynamic properties as the random-energy model of Derrida
\cite{Derrida80,Derrida81} or the $p$-spin model \cite{Gardner85} in the
large $p$ limit \cite{Campellone,ParRitSan95}. The ROM shows a dynamical
transition at a characteristic temperature $T_{\rm dyn}$ (that
corresponds to the mode coupling temperature $T_{\rm MCT}$ in mode
coupling theories for the glass transition \cite{BouCugKurMez96}). Below
that temperature ergodicity is broken and the phase space splits up into
disconnected regions that are separated by infinitely high energy
barriers. For finite $N$ the dynamics is different and the dynamical
transition is smeared out. The scenario is then much reminiscent of the
phenomenological model we discussed in
Sec.~\ref{glassy:pheno}. Different sets of spins collectively relax in
finite timescales, each one representing a CRR. There are two
important and useful concepts in this regard,

\begin{itemize}

\item{\bf The free energy landscape.} An interesting approach to
identify CRRs in glassy systems is the study of the topological
properties of the potential energy landscape \cite{Goldstein69}. The
slow dynamics observed in glassy systems in the NEAS is attributed to
the presence of minima, maxima and saddles in the potential energy
surface. Pathways connecting minima are often separated by large energy
barriers that slow down the relaxation. Stillinger and Weber have
proposed to identify phase space regions with the so-called inherent
structures (IS) \cite{StiWeb82,Stillinger95b}. The inherent structure of
a region in phase space is the configuration that can be reached by
energy minimization starting from any configuration contained in the
region. Inherent structures are used as labels for regions in phase
space. Figure \ref{fig17} (left panel) shows a schematic representation
of this concept. Figure \ref{fig17} (right panel) shows the relaxation
of the energy of the inherent structure energy starting from a high
energy initial nonequilibrium state
\cite{CriRit00,CriRit00b,CriRit00c}. Inherent structures are a useful
way to keep track of all cooperative rearrangements that occur during
the aging process \cite{CriRit02}.

\begin{figure}
\begin{center}
\includegraphics[scale=0.7,angle=0]{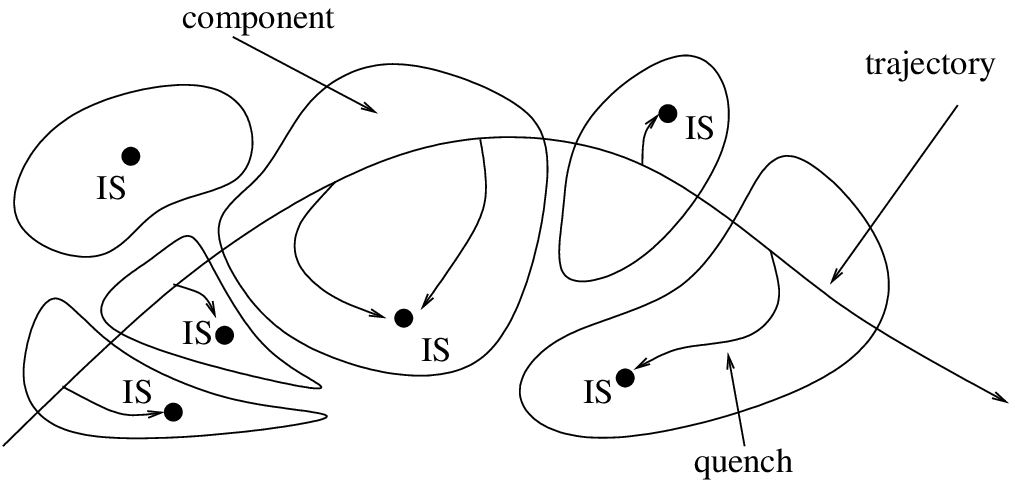}
\includegraphics[scale=0.7,angle=0]{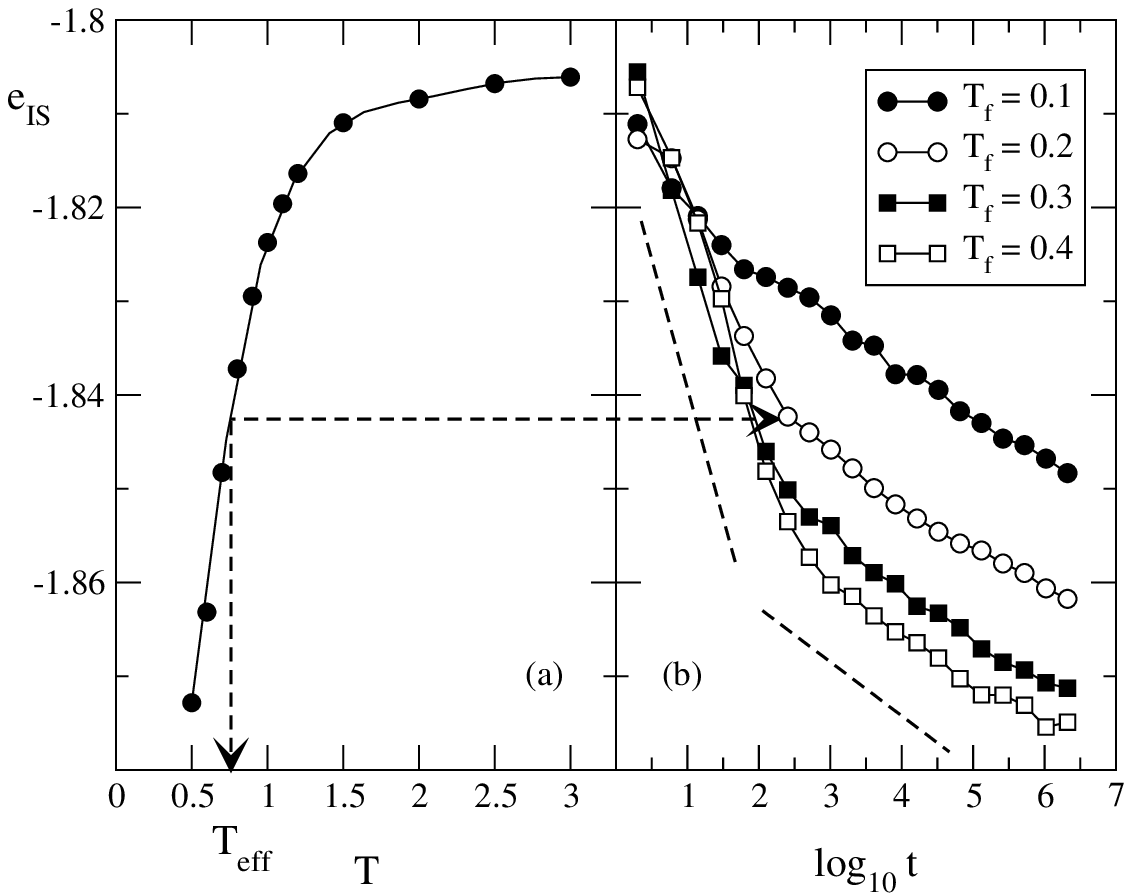}
\end{center}
\vspace{-0.4cm}
\caption{\em (Left) Stillinger and Weber decomposition. Schematic
picture showing regions or components in phase space that are labeled
by a given IS. (Right) Relaxation in the ROM. Panel (a): Equilibrium
average $e_{\rm IS}$ a function of temperature. The arrows indicate the
construction of the effective temperature $T_{\rm eff}(t_w)$
\eq{glassy:noneqtemp:1}. Panel (b): Average inherent structure energy
as function of time for the initial equilibrium temperature
$T_{\rm i}= 3.0$ and final quench temperatures $T_{\rm f}= 0.1$, $0.2$,
$0.3$ and $0.4$.  The average is over $300$ initial configurations. The
system size is $N=300$.  Left figure taken from \protect\cite{CriRit02}. Right
figure taken from \protect\cite{CriRit00b,CriRit00c}}
\label{fig17}
\vspace{-0.2cm}
\end{figure}

\item{\bf FD plots.} Numerical tests of the validity of the FDR
\eq{glassy:noneqtemp:1} use fluctuation-dissipation plots (FD plots) to
represent the integrated response as a function of the correlation. The
integrated version of relation \eq{glassy:noneqtemp:1} is expressed in
terms of the susceptibility,
\be
\chi(t,t_w)=\int_{t_w}^tdt'R(t,t')~~~.
\label{glassy:noneqtemp:3}
\ee
By introducing \eq{glassy:noneqtemp:3} in \eq{glassy:noneqtemp:1} we
obtain,
\be
\chi(t,t_w)=\int_{t_w}^t dt'\frac{1}{T_{\rm eff}(t')}\frac{\partial
C(t,t')}{\partial t'}=\frac{1}{T_{\rm eff}(t_w)}\Bigl(C(t,t)-C(t,t_w)\Bigr)
\label{glassy:noneqtemp:4}
\ee
where we have approximated $T_{\rm eff}(t')$ by $T_{\rm eff}(t_w)$. By
measuring the susceptibility and the correlation function for a fixed
value of $t_w$ and plotting one respect to the other, the slope of the
curve $\chi$ respect to $C$ gives the effective temperature. This
result follows naturally from \eq{glassy:noneqtemp:4} if we take
$C(t,t)$ time independent (which is the case for spin systems). If
not, proper normalization of the susceptibility and correlations by
$C(t,t)$ is required and a similar result is obtained \cite{FieSol02}.
A numerical test of these relations in the ROM and in binary
Lennard-Jones mixtures is shown in Figure \ref{fig18}. We stress that
these results have been obtained in finite-size systems. As the system
becomes larger the timescales required to see rearrangements events
become prohibitively longer and the relaxation of the system toward
equilibrium drastically slows down.

\end{itemize}

\begin{figure}
\begin{center}
\includegraphics[scale=0.62,angle=0]{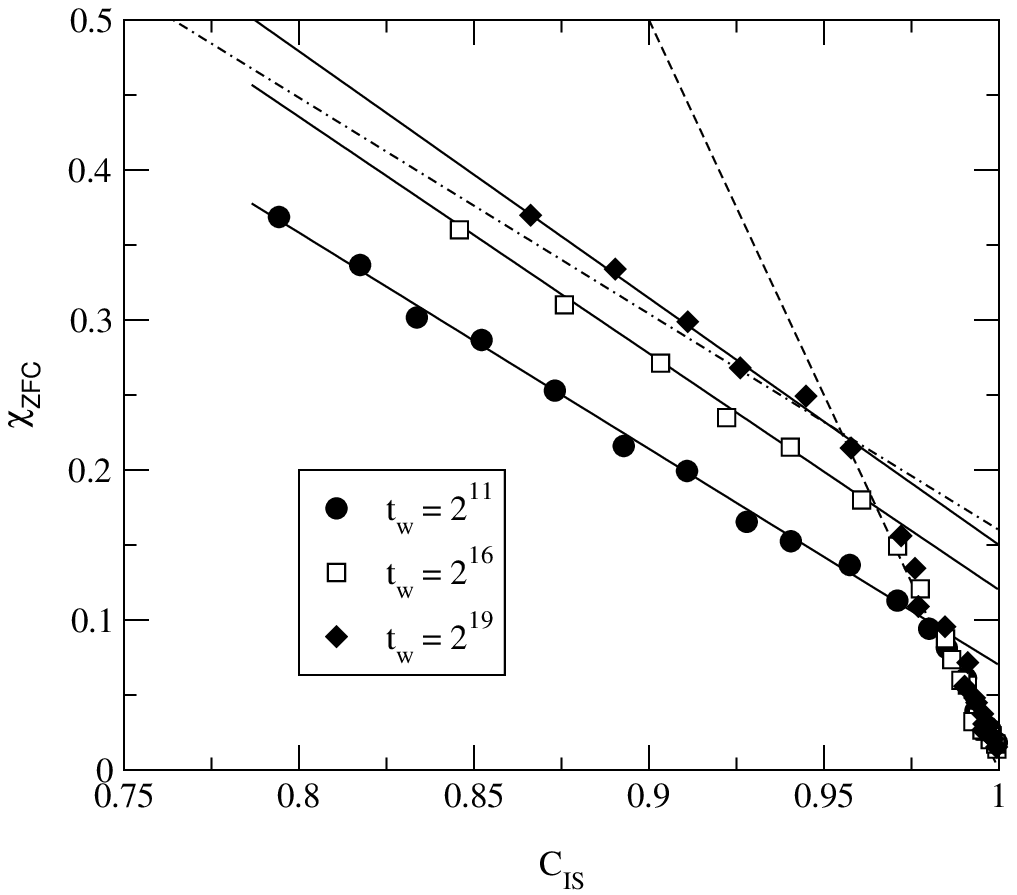}\includegraphics[scale=0.75,angle=0]{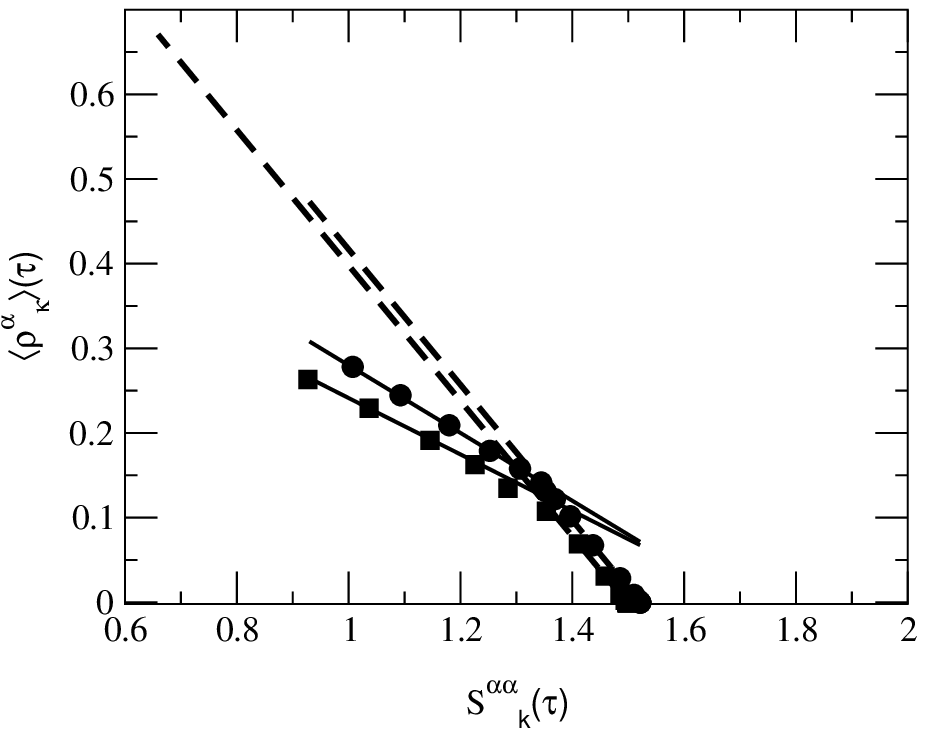}
\end{center}
\vspace{-0.4cm}
\caption{\em (Left panel) Integrated response function as a
function of IS correlation function, i.e, the correlation between
different IS configurations, in the ROM.  The dashed line has slope
$T_{\rm f}^{-1} = 5.0$, where $T_{\rm f}$ is the final quench
temperature, whereas the full lines are the prediction from
\protect\eq{glassy:pheno:9} and $F^*=E_{\rm IS}(t_w)$: $T_{\rm
eff}(2^{11})\simeq 0.694$, $T_{\rm eff}(2^{16})\simeq 0.634$ and $T_{\rm
eff}(2^{19})\simeq 0.608$.  The dot-dashed line has slope $1/T_{\rm eff}(t_w)$
for $t_{\rm w}=2^{11}$ and is drawn for comparison. Figure taken from
\protect\cite{CriRit00b}. (Right panel) Integrated response versus the dynamical structure factor for
the binary mixture Lennard-Jones particles system in a quench from the
initial temperature $ T_{\rm i} = 0.8$ to a final temperature $T_{\rm
f}=0.25$ and two waiting times $t_w=1024$ (square) and $t_w=16384$
(circle).  Dashed lines have slope $1/T_{\rm f}$ while thick lines have
slope $1/T_{\rm eff}(t_w)$.  Figure taken from
\protect\cite{SciTar01}. }
\label{fig18}
\vspace{-0.2cm}
\end{figure}

\subsection{Intermittency}
\label{glassy:inter}
Indirect evidence of nonequilibrium fluctuations due to CRRs in
structural glasses has been obtained in Nyquist noise experiments by
Ciliberto and coworkers. In these experiments a polycarbonate glass is
placed between the plates of a condenser and quenched at temperatures
below the glass transition temperature.  Voltage fluctuations are then
recorded as a function of time during the relaxation process and the
effective temperature is measured,
\be
T_{\rm eff}(\omega,t_w)=\frac{S_Z(\omega,t_w)}{4{\cal R}(Z(\omega,t_w))}
\label{glassy:inter:1}
\ee
where ${\cal R}(Z(\omega,t_w))$ is the real part of the impedance of the
system and $S_Z(\omega,t_w)$ is the noise spectrum of the impedance that
can be measured from the voltage noise \cite{BuiBelCil03}.

Experimental data shows a strong variation of the effective temperature
with the waiting time by several orders of magnitude. The voltage signal
is also intermittent with strong voltage spikes at random times. The
distribution of the times between spikes follows a power law
characteristic of trap models. These results point to the fact that the
observed voltage spikes correspond to CRRs occurring in the polycarbonate
sample. Finally, the probability distribution function (PDF) of the
voltage signal strongly depends on the cooling rate in the glass
suggesting that relaxational pathways in glasses are very sensitive to
temperature changes. A related effect that goes under the name of Kovacs
effect has been also observed in calorimetry experiments, numerical
simulations and exactly solvable models
\cite{MosSci04,NavMosSci02,BerBouDroGod03}.

\begin{figure}
\begin{center}
\includegraphics[scale=0.5,angle=-90]{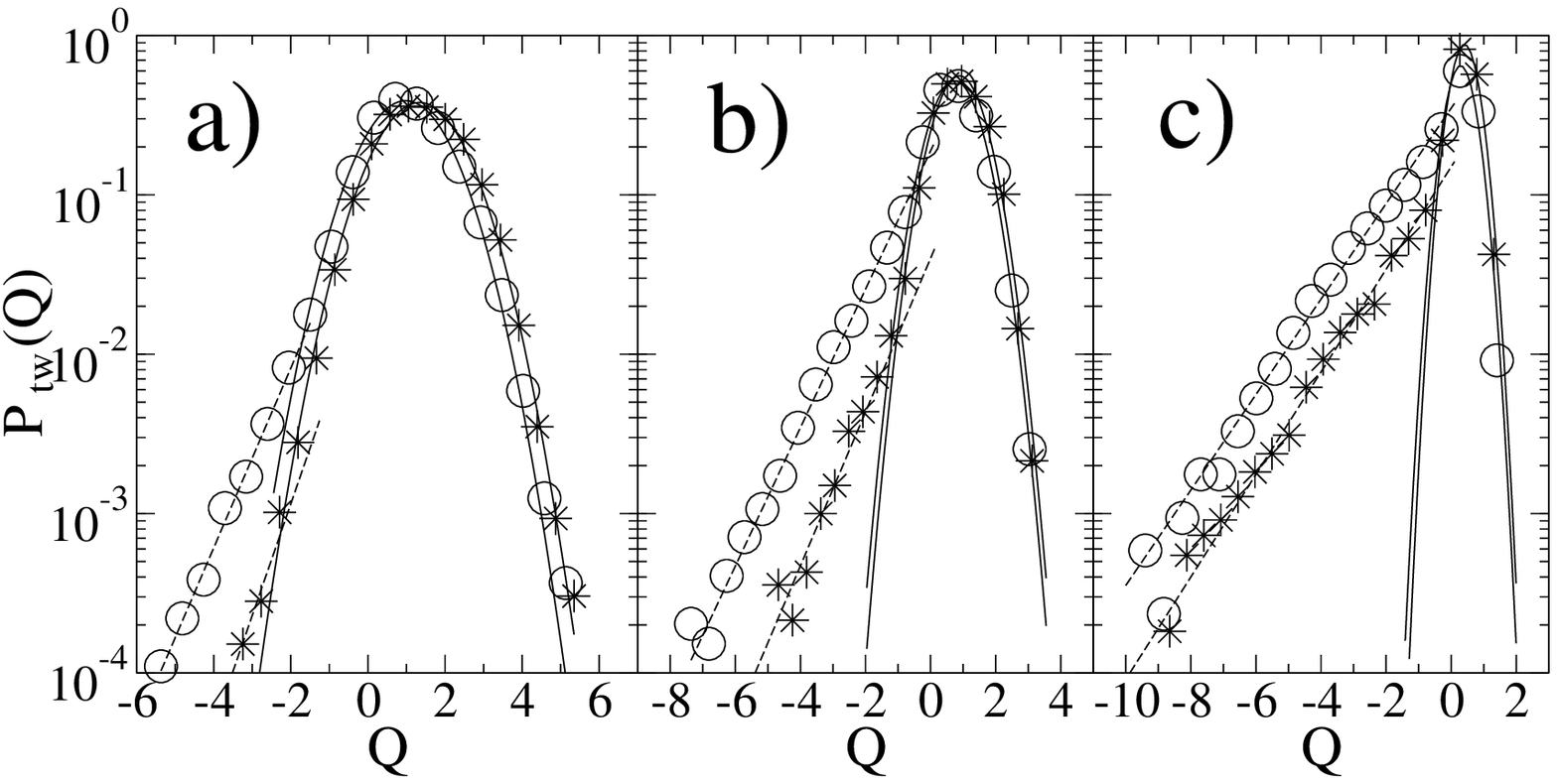}
\includegraphics[scale=.5,angle=-90]{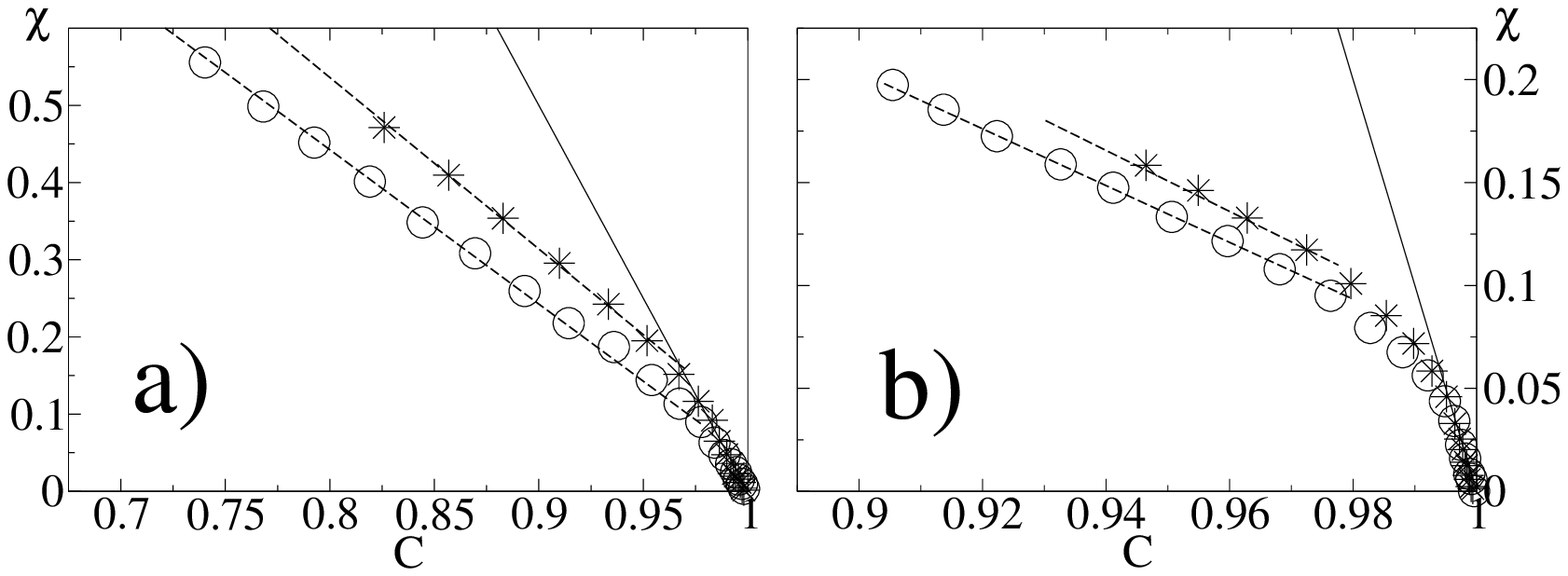}
\end{center}
\vspace{-0.4cm}
\caption{\em Heat exchange PDFs for $T=0.3$ (a),
$T=0.2$ (b), $T=0.1$ (c). Circles are for $t_w=2^{10}$ and
asterisks for $t_w=2^{15}$. The continuous lines are Gaussian fits to the stimulated sector,the
dashed lines are the exponential fits to the spontaneous sector. Figure
taken from \protect\cite{CriRit04b}.}
\label{fig20}
\vspace{-0.2cm}
\end{figure}

A physical interpretation of the intermittency found in aging systems
has been put forward based on exactly solvable models of glasses
\cite{CriRit04b,Ritort04b,Ritort04c}. According to this, energy
relaxation in glassy systems follows two different mechanisms (see
Sec.~\ref{pw:workheat}): stimulated relaxation and spontaneous
relaxation. In the NEAS the system does not do work but exchanges heat
with the environment. At difference with previous Sections here we adopt
the following convention: $Q>0$ ($Q<0$) denotes heat absorbed (released)
by (from) the system from (to) the environment. Energy conservation then
reads $\Delta E=Q+W$ where $W$ is the worked done on the system. In the
NEAS, $W=0$ and $\Delta E=Q$, the energy released by the system is
dissipated in the form of heat. In the phenomenological model put
forward in Sec.~\ref{glassy:pheno} different CRRs can exchange (adsorb
or release) heat to the environment. The regions that cooperatively
rearrange for the first time release stress energy to the environment
and contribute to the net energy dissipation of the glass. We call this
mechanism the {\em spontaneous relaxation}.  Regions that have already
rearranged for the first time can absorb or release energy from/to the
bath several times but do not contribute to the net heat exchanged
between the system and the bath. We call this mechanism the {\em
stimulated relaxation}. There are several aspects worth mentioning,

\begin{itemize}

\item{\bf Heat distribution.} The distribution of heat exchanges
$Q=E(t_w)-E(t)$ for the stimulated process is a Gaussian distribution
with zero mean and finite variance. This process corresponds to the
heat exchange distribution of the system in equilibrium at the
quenching temperature. In contrast, in the spontaneous process a net
amount of heat is released to the bath. {\em Spontaneous heat} arises
from the fact that the system has been prepared in a nonequilibrium
high-energy state.  Let us consider a glass that has been quenched at
a temperature $T$ and has relaxed for an age $t_w$. During aging, CRRs
that release stress energy (in the form of heat $Q<0$) to the
environment satisfy the relation \eq{glassy:pheno:8},
\be
\frac{P^{\rm sp}(Q)}{P^{\rm sp}(-Q)}=\exp\Bigl(\frac{Q}{T_{\rm eff}(F^*)}\Bigr)
\label{glassy:inter:2}
\ee
Therefore, as in the phenomenological model \eq{glassy:pheno:10b}, we expect,
\be
P^{\rm sp}(Q)\propto \exp\Bigl(\frac{Q}{2T_{\rm eff}(F^*)}\Bigr)~~{\rm for}~~~Q<0~~~.
\label{glassy:inter:4}
\ee
Note that $T_{\rm eff}(F^*)$ depends on the age of the system through the
value of the typical free-energy of the CRRs that release their stress
energy at $t_w$, $F^*(t_w)$.  This relation has been numerically tested in the ROM
\eq{glassy:noneqtemp:2} by carrying out aging simulations at different
temperatures and small sizes $N$ \cite{CriRit04b} (see the next item). 

\item{\bf Numerical tests.} How to measure the heat distribution
\eq{glassy:inter:4} in numerical simulations of NEAS? A useful
procedure uses the concept of inherent structures and goes as
follows. The heat exchanged during the time interval $[t_w,t]$ ($t>t_w$)
has to be averaged over many aging paths (ideally an infinite number of
paths). Along each aging path many rearrangement events occur between
$t_w$ and $t$. Most of them are stimulated, a few of them are
spontaneous. In fact, because the spontaneous process gets contribution
only from those cooperative regions that rearrange for the first time,
its PDF signal gets masked by the much larger one coming from the
stimulated component where rearrangements events from a single region
contribute more than once. To better disentangle both processes we
measure, for a given aging path, the heat exchange corresponding to the
first rearrangement event observed after $t_w$. To identify a
rearrangement event we keep track of the IS corresponding to the run
time configuration. Following the IS is an indirect way of catching
rearranging events due to CRRs. Only
when the system changes IS we know that a cooperative rearrangement
event has taken place. Rearrangement events take place at different
times $t$ after $t_w$, therefore the heat distribution $P^{\rm sp}(Q)$
is measured along an heterogeneous set of time intervals.  The results
for the heat distributions at various ages $t_w$ are shown in Figure
\ref{fig20}.  We notice the presence of two well defined contributions
to the heat PDFs: A Gaussian central component plus additional exponential tails at
large and negative values of $Q$. The Gaussian component corresponds to the
stimulated process, however its mean is different from zero. The reason
of this apparent discrepancy lies in the numerical procedure used to measure the
heat PDF: the average {\em stimulated heat} is not equal to the net
exchanged heat (which should be equal to 0) because different aging paths contribute to the heat
exchange along different time intervals. The Gaussian component should be
equal to the heat PDF for the system in thermal equilibrium at the same
temperature and therefore independent of $t_w$. Indeed the variance of the Gaussian distribution is found
to be independent of $t_w$ \cite{CriRit04b}.

\item{\bf Spontaneous events release stress energy.}
One striking aspect of the spontaneous process is that, according to
\eq{glassy:inter:2}, the probability of heat absorption ($Q>0$) should
be much larger than the probability of heat release ($Q<0$). However,
this is not observed in the numerical results of Figure \ref{fig20} where
the exponential tail is restricted to the region $Q<0$. Why spontaneous
events are not observed for $Q>0$?  The reason is that spontaneous events
can only release and not absorb energy from the environment, see \eq{glassy:inter:4}. 
This is in the line of the argumentation put forward in
Sec.~\ref{glassy:pheno} where it was said that the first time that cooperative
regions release the stress energy, this gets irreversibly lost as
heat released to the environment. As the number of stressed
regions monotonically decreases as a function of time the weight of the
heat exponential tails decreases with the age of the system as observed in Figure
\ref{fig20}. The idea that only energy decreasing events contribute to
the effective temperature \eq{glassy:inter:4} makes it possible to
define a time-dependent configurational entropy \cite{BirKur01}.

\item{\bf Zero-temperature relaxation.}  This interpretation
rationalizes the aging behavior found in exactly solvable entropy
barrier models that relax to the ground state and show aging at zero
temperature \cite{Ritort95,BonPadRit98}. At $T=0$, the stimulated
process is suppressed (microscopic reversibility \eq{demo5} does not
hold), and \eq{glassy:pheno:8} holds by replacing the free energy of a
CRR by its energy, $F=E$. In such models a region corresponds to just a
configuration in phase space and relaxation occurs through
spontaneous rearrangements where configurations are visited only once.  In
entropic barrier models the effective temperature \eq{glassy:pheno:9}
still governs aging at $T=0$. Because the energy is a monotonically
decreasing quantity for all aging paths, \eq{glassy:pheno:8} does not
strictly hold as ${\cal W}(\Delta E>0)=0$. Yet, the effective
temperature obtained from \eq{glassy:pheno:8} has been shown to coincide
with that derived from the FDR \eq{glassy:noneqtemp:1}
\cite{Ritort04b,Ritort04c}.

\end{itemize}

\section{Conclusions and outlook}
\label{conclusions}

We have presented a general overview of several topics related to the
nonequilibrium behavior of small systems: from fluctuations in
mesoscopic systems such as small beads in optical traps up to molecular
machines and biomolecules. The main common theme being that, under
appropriate conditions, physical systems exchange small amounts of
energy with the environment leading to large fluctuations and strong
deviations from the average behavior. We call such systems {\em small}
because of their properties and behavior are markedly different from
macroscopic systems. We started our tour by stressing the similarities
between colloidal systems and molecular machines: intermittency and
nonequilibrium behavior are common aspects there. We have then discussed
fluctuation theorems (FTs) in detail, and focused our discussion in two
well studied systems: the bead in a trap and single molecule force
experiments.  Experimental results in such systems show the presence of
large tails in heat and work distributions in marked contrast to
Gaussian distributions, characteristic of macroscopic systems. Such
behavior can be rationalized by introducing a path formalism that
quantifies work/heat distributions. Finally, we revised some of the main
concepts in glassy dynamics where small energy fluctuations appear an
essential underlying ingredient of the observed slow relaxation. Yet, we
still lack a clear understanding of the right theory that unifies all
phenomena and a clear and direct observation of the postulated small and
cooperatively rearranging regions remains an experimental challenge. We
envision three future lines of research,

\begin{itemize}

\item{\bf Developments in FTs.} FTs are simple results that provide a
new view to better understand issues related to irreversibility and
the second law of thermodynamics. The main assumption of FTs is
microscopic reversibility or local equilibrium, an assumption that has
received some criticism
\cite{CohMau04,Jarzynski04,Astumian05}. Establishing limitations to
the validity of FTs is the next task for the future. At present no
experimental result contradicts any of the FTs, mainly because the
underlying assumptions are respected in the experiments or because
current experimental techniques are not accurate enough to detect systematic
discrepancies. Under some experimental conditions, we might discover
that microscopic reversibility breaks down and a more refined and
fundamental description of the relevant degrees of freedom in the
system becomes necessary. Validation of FTs under different and far
from equilibrium conditions will be useful to test the main
assumptions.

\item{\bf Large deviation functions.} The presence of large tails can be
investigated in statistical mechanics theories by exact analytical solutions of
simple models, by introducing simplified theoretical approaches or even
by designing smart and efficient algorithms. In all cases, we
expect to obtain a good theoretical understanding of the relation
between large deviations and nonequilibrium processes.  Ultimately, this
understanding can be very important in biological systems where
nonequilibrium fluctuations and biological function may have gone hand
by hand during biological evolution on earth over the past 4.5 billion
years.  A very promising line of research in this area will be
the study of molecular motors where the large efficiency observed at the
level of a single mechanochemical cycle might be due to a very specific
adaptation of the molecular structure of the enzyme to
its environment. This fact may have
important implications at the level of single molecules and cellular structures.

\item{\bf Glassy systems.} We still need to have direct and clear
experimental evidence of the existence of the cooperatively rearranging
regions, responsible for most of the observed nonequilibrium
relaxational properties in glasses. However, the direct observation of
these regions will not be enough. It will be also necessary to have a
clear idea of how to identify them in order to extract useful
statistical information that can be interpreted in the framework of a
predictive theory. Numerical simulations will be very helpful in this
regard. If the concept of nonequilibrium temperature has to survive the
time then it will be necessary also to provide accurate experimental measurements
at the level of what we can now get from numerical simulations.

\end{itemize} 

Since the discovery of Brownian motion in 1827 by the biologist Robert
Brown and the later development of the theory for the Brownian motion in
1905, science is witnessing an unprecedented convergence of physics
toward biology. This was anticipated several decades ago by Erwin
Schr\"odinger who, in his famous monography published in 1944 and entitled {\em What is life?}
\cite{Sch44}, wrote when talking about the motion
of a clock: ``The true physical picture includes the possibility that
even a regularly going clock should all at once invert its motion and,
working backward, rewind its own spring -at the expense of the heat of
the environment. The event is just 'still a little less likely' than a
'Brownian fit' of a clock without driving mechanism''.  Biological
systems seem to have exploited thermal fluctuations to built new
molecular designs and structures that efficiently operate out of
equilibrium at the molecular and cellular level
\cite{Deduve95,Loewe99,Harold01,Houches02}. The synergy between structure and
function is most strong in living systems where nonequilibrium
fluctuations are at the root of their amazing and rich behavior.

{\bf Acknowledgments:} I am grateful to all my collaborators and
colleagues, too numerous to mention, from whom I learned in the past and
who have made possible the writing on the many of the topics covered in
this paper. I acknowledge financial support from the Ministerio de
Eduaci\'on y Ciencia (Grant FIS2004-3454 and NAN2004-09348), the Catalan
government (Distinci\'o de la Generalitat 2001-2005, Grant SGR05-00688),
the European community (STIPCO network) and the SPHINX ESF program.

\section{List of abbreviations}

\begin{tabbing}
KWW:zz\=\kill

CFT 	\>Crooks fluctuation theorem\\
JE 	\>Jarzynski equality\\
FDT	\>fluctuation-dissipation theorem\\
FEC 	\>Force-extension curve\\
FT 	\>Fluctuation theorem\\
IS	\>inherent structure\\
PDF 	\>probability distribution function\\
\end{tabbing}

\bibliographystyle{unsrt}
\bibliography{ACPWILEY.FRitortv2}

\end{document}